\documentclass[aps,prb,twocolumn,amsmath,amssymb,showpacs,nobibnotes,floatfix,superscriptaddress,nolongbibliography,noeprint,citeautoscriptm]{revtex4-2}

\usepackage{graphicx}
\usepackage{dcolumn}
\usepackage{bm}
\usepackage{lipsum}
\usepackage{multirow}
\usepackage{hyperref}
\appdef \turnpage {%
  \AddToHookNext{shipout/after}{%
    \global\pdfpageattr\expandafter{\the\pdfpageattr/Rotate 90}%
    \AddToHookNext{shipout/after}{%
    }%
  }%
}

\usepackage{soul}
\usepackage{color}

\makeatletter
\renewcommand*{\p@subsection}{}
\renewcommand*{\p@subsubsection}{}

\renewcommand*{\fnum@figure}{{\normalfont\bfseries \figurename~\thefigure}}
\renewcommand*{\fnum@table}{{\normalfont\bfseries \tablename~\thetable}}
\makeatother

\begin{document}

\setcitestyle{numbers,square}
\setcitestyle{super}

\title{Reducing Self-Interaction Error in Transition-Metal Oxides with
Different Exact-Exchange Fractions for Energy and Density}

\author{Harshan Reddy Gopidi}
\affiliation{%
 Department of Electrical and Computer Engineering, University of Houston, Houston, TX 77204, USA
}%
\affiliation{
Texas Center for Superconductivity, University of Houston, Houston, TX 77204, USA
}%

\author{Ruiqi Zhang}
\affiliation{
Department of Physics and Engineering Physics, Tulane University, New Orleans, LA 70118, USA
}%

\author{Yanyong Wang}
\affiliation{
Department of Physics and Engineering Physics, Tulane University, New Orleans, LA 70118, USA
}%

\author{Abhirup Patra}
\affiliation{
Shell International Exploration and Production Inc., Houston, TX 77082, USA
}%

\author{Jianwei Sun}
\affiliation{
Department of Physics and Engineering Physics, Tulane University, New Orleans, LA 70118, USA
}%

\author{Adrienn
Ruzsinszky}
\affiliation{
Department of Physics and Engineering Physics, Tulane University, New Orleans, LA 70118, USA
}%

\author{John P. Perdew}
\email{perdew@tulane.edu}
\affiliation{
Department of Physics and Engineering Physics, Tulane University, New Orleans, LA 70118, USA
}%

\author{Pieremanuele Canepa}
\email{pcanepa@uh.edu}
\affiliation{%
 Department of Electrical and Computer Engineering, University of Houston, Houston, TX 77204, USA
}%
\affiliation{
Texas Center for Superconductivity, University of Houston, Houston, TX 77204, USA
}%

\begin{abstract}
\noindent Density functional theory (DFT) in chemistry and materials science aims for "chemical accuracy," but this goal is challenged by the need to approximate the exact exchange-correlation (XC) energy functional. The restored-regularized strongly constrained and
appropriately normed (r\textsuperscript{2}SCAN), meta-generalized gradient approximation to the 
XC functional fulfills 17 exact constraints of the XC energy, and has significantly
boosted prediction accuracy for molecules and materials.
However, r\textsuperscript{2}SCAN remains inadequate at predicting properties of open \textit{d} and \textit{f} transition-metal strongly correlated compounds, such as band gaps, magnetic moments, and oxidation energies. Prediction inaccuracies of
r\textsuperscript{2}SCAN energies arise from functional and density-driven
errors, mainly resulting from the DFT self-interaction error. Here, we
propose a novel method termed
r\textsuperscript{2}SCANY@r\textsuperscript{2}SCANX to mitigate the self-interaction error of XC functionals for the accurate
simulations of electronic, magnetic, and thermochemical properties of
transition metal oxides.
r\textsuperscript{2}SCANY@r\textsuperscript{2}SCANX uses different fractions of exact Hartree-Fock exchange: X for the electronic density and Y for the density functional approximation of the total energy, thereby simultaneously addressing functional-driven and density-driven inaccuracies. Building
just on 1 (or maximum 2) parameters that apply unchanged to \emph{s-p}-bonded systems, we demonstrate that, 
r\textsuperscript{2}SCANY@r\textsuperscript{2}SCANX improves upon 
the r\textsuperscript{2}SCAN predictions for 20 highly correlated oxides and
even outperforms the highly parameterized
DFT(r\textsuperscript{2}SCAN)+\emph{U} method ---the state-of-the-art
approach to predict strongly correlated materials. 
Prediction uncertainties for oxidation energies and magnetic moments of
transition metal oxides are significantly reduced by
r\textsuperscript{2}SCAN10@r\textsuperscript{2}SCAN50 and band gaps with
r\textsuperscript{2}SCAN10@r\textsuperscript{2}SCAN.
r\textsuperscript{2}SCAN10@r\textsuperscript{2}SCAN50 diminishes the density-driven error of the energy in r\textsuperscript{2}SCAN and
r\textsuperscript{2}SCAN10. We demonstrate that the computationally
efficient r\textsuperscript{2}SCAN10@r\textsuperscript{2}SCAN is nearly
as accurate as the global hybrid r\textsuperscript{2}SCAN10 for
oxidation energies. This indicates that accurate energy differences can be obtained through rate-limiting self-consistent iterations and geometry optimizations with the efficient r\textsuperscript{2}SCAN.
Subsequently, a more expensive nonlocal functional, such as a hybrid or
self-interaction correction, can be applied in a fast, single
post-self-consistent calculation, as in
r\textsuperscript{2}SCAN10@r\textsuperscript{2}SCAN. 

\end{abstract}

\maketitle

\section{Introduction}\label{sec:introduction}
Electronic structure methods, especially density functional theory
(DFT),\cite{kohn_self-consistent_1965} have become essential for materials discovery,
enabling the prediction of properties and behaviors of technologically
relevant materials.\cite{saal_materials_2013,jain_commentary_2013,zunger_inverse_2018,park_developing_2020,Cheetham2022} This has resulted in the
widespread use of DFT for extensive materials databases, such as the
Alexandria Materials Database,\cite{schmidt_dataset_2022}
AFLOW,\cite{curtarolo_aflow_2012} GNoME,\cite{merchant_scaling_2023} the Materials
Project,\cite{jain_commentary_2013} the NREL materials
databases,\cite{stevanovic_correcting_2012} and OQMD.\cite{saal_materials_2013} These
databases provide a critical foundation for materials science research,
enabling direct comparison of computed properties with experimental
data, developing predictive models, and lately, foundational training
sets for machine learning potentials,\cite{chen_universal_2022,deng_chgnet_2023,batatia_mace_2023,Kanungo2025} thus
necessitating highly accurate datasets approaching ``chemical
accuracy''.

Nevertheless, DFT in the Kohn-Sham formulation requires approximations
for the for the exact exchange and correlation (XC) functional, which has been
a matter of intense research for the past 60
years.\cite{parr_density-functional_1995,koch_chemists_2001,cohen_insights_2008} Among the density functional
approximations (DFAs), the local spin density approximation
(LSDA),\cite{kohn_self-consistent_1965,perdew_accurate_1992} and the generalized gradient approximation
(GGA)\cite{perdew_generalized_1996} exhibit inaccuracies in predicting
structural parameters, energetics, and electronic properties, especially
for strongly correlated systems. DFA and XC are synonyms and will be
used interchangeably.

The strongly constrained and appropriately normed approximation (SCAN)\cite{sun_strongly_2015} and its regularized and computationally
efficient version, r\textsuperscript{2}SCAN,\cite{furness_accurate_2018} are
highly accurate meta-GGA functionals. SCAN and
r\textsuperscript{2}SCAN satisfy 17 known exact constraints for constructing XC functionals, ensuring a well-balanced description of XC 
effects for a wide range of systems.\cite{kaplan_predictive_2023} LAK is a constraint-based meta-GGA that accurately describes electronic bonding and band gaps of main-group molecules and semiconductors.\cite{lebeda_balancing_2024} SCAN and r\textsuperscript{2}SCAN have done admirably well in improving the
quality of predictions of molecules and materials relative to
GGA.\cite{kitchaev_energetics_2016,zhang_symmetry-breaking_2020,sai_gautam_evaluating_2018,long_evaluating_2020,artrith_data-driven_2022,bartel_role_2019}  Most materials databases depend on GGA's predictive power, using \emph{ad hoc} corrections, \emph{e.g.}, GGA+\emph{U},{\cite{wang_oxidation_2006,jain_formation_2011,stevanovic_correcting_2012}} or newer DFA, \emph{e.g.},  r{\textsuperscript{2}}SCAN.\cite{kingsbury_flexible_2022}

Challenges persist when SCAN and r\textsuperscript{2}SCAN are used to
predict several valuable material properties of open \emph{d} (\emph{f}) transition-metal compounds, including band gaps, magnetic
moments, and oxidation (reduction) energies. While
r\textsuperscript{2}SCAN (SCAN) reduces the self-interaction
error (SIE) magnitude compared to standard semi-local GGA XC
functionals, this pernicious error
remains.\cite{kitchaev_energetics_2016,zhang_symmetry-breaking_2020,Sun2016,zhang_comparative_2017,zhang_subtlety_2019,zhang_critical_2022,ning_reliable_2022,furness_accurate_2018}
 
Pragmatic but material- and property-dependent solutions to address the
SIE in GGAs and meta-GGAs involve parametrizing \emph{ad
hoc} on-site Hubbard \emph{U} corrections, such as
GGA(LSDA)+\emph{U},\cite{dudarev_electron-energy-loss_1998} and
r\textsuperscript{2}SCAN(SCAN)+\emph{U}.\cite{sai_gautam_evaluating_2018,long_evaluating_2020,artrith_data-driven_2022,wang_oxidation_2006,jain_formation_2011,tekliye_accuracy_2024,swathilakshmi_performance_2023,devi_effect_2022}
Although numerically accurate and affordable, these +\emph{U} approaches often lack transferability because the \emph{U} parameters depend on material chemistry, dimensionality, and the oxidation states of transition metals. Strategies exist to
fit the \emph{U} values to the thermochemical data of redox
reactions,\cite{sai_gautam_evaluating_2018,long_evaluating_2020,artrith_data-driven_2022,wang_oxidation_2006,jain_formation_2011,swathilakshmi_performance_2023,devi_effect_2022} or band
gaps.\cite{dudarev_electron-energy-loss_1998,kirchner-hall_extensive_2021} \emph{U} values can also be derived
from linear response,\cite{cococcioni_linear_2005,timrov_hubbard_2018} or  with machine
learning models.\cite{yu_machine_2020,uhrin_machine_2025} A persistent issue is that applying a \emph{U}-specific value to oxidation energies doesn't ensure the same \emph{U} will accurately replicate band gaps or magnetic moments in similar materials.\cite{wang_oxidation_2006,jain_formation_2011}

Hybrid XC functionals are a universal approach to address SIE in materials,{\cite{illas_magnetic_1998,de_p_r_moreira_effect_2002,kaltsoyannis_performance_2004,alfredsson_electronic_2004,alfredsson_structural_2005,illas_spin_2006,marsman_hybrid_2008,da_silva_hybrid_2007,janesko_screened_2009,paier_dielectric_2008,chevrier_hybrid_2010,canepa_comparison_2011,yang_range-separated_2023}} replacing part of the DFA exchange with a fraction of exact HF exchange. 
 Global and range-separated hybrids are successfully applied to transition-metal oxides
(M\textsubscript{i}O\textsubscript{j}s).\cite{illas_magnetic_1998,de_p_r_moreira_effect_2002,kaltsoyannis_performance_2004,alfredsson_electronic_2004,alfredsson_structural_2005,illas_spin_2006,marsman_hybrid_2008,da_silva_hybrid_2007,janesko_screened_2009,paier_dielectric_2008,chevrier_hybrid_2010,canepa_comparison_2011,yang_range-separated_2023,tran_hybrid_2006,graciani_comparative_2011,radin_charge_2013,seo_calibrating_2015,skone_nonempirical_2016,das_band_2019}

Although most hybrid functionals have a preset HF exchange parameter,{\cite{becke_density-functional_1993,stephens_ab_1994}} they are less sensitive to material types or specific reactions and properties.\cite{illas_magnetic_1998,de_p_r_moreira_effect_2002,kaltsoyannis_performance_2004,alfredsson_electronic_2004,alfredsson_structural_2005,illas_spin_2006,marsman_hybrid_2008,da_silva_hybrid_2007,janesko_screened_2009,paier_dielectric_2008,chevrier_hybrid_2010,canepa_comparison_2011,yang_range-separated_2023,tran_hybrid_2006,graciani_comparative_2011,radin_charge_2013,seo_calibrating_2015,skone_nonempirical_2016,das_band_2019,franchini_ground-state_2007} A few non-empirical methods exist to estimate HF exchange.
\cite{gorling_hybrid_1997,perdew_rationale_1996,adamo_toward_1999,heyd_hybrid_2003,jaramillo_local_2003,janesko_local_2007,henderson_range_2008,janesko_self-consistent_2008,skone_nonempirical_2016} This paper explores
several global r\textsuperscript{2}SCAN-based hybrid functionals with
X\% of exact HF exchange, termed r\textsuperscript{2}SCANX.

Building on the idea that HF provides self-interaction-free,
uncorrelated densities, an elegant solution is to use HF electronic
charge densities to compute r\textsuperscript{2}SCAN total
energies.\cite{clementi_comparative_1990,gill_investigation_1992,scuseria_comparison_1992,oliphant_systematic_1994,janesko_hartreefock_2008,verma_increasing_2012,kanungo_unconventional_2024} This approach is called the
Hartree-Fock density functional theory or DFA@HF,\cite{kanungo_unconventional_2024} and has been successfully applied to
molecules.\cite{clementi_comparative_1990,gill_investigation_1992,scuseria_comparison_1992,oliphant_systematic_1994,janesko_hartreefock_2008,verma_increasing_2012,kanungo_unconventional_2024} In almost all cases,
r\textsuperscript{2}SCAN@HF is significantly more accurate or slightly
less accurate than SCAN/r\textsuperscript{2}SCAN.\cite{Santra2021a,kanungo_unconventional_2024} There
are instances where HF densities are more accurate than those of specific XC functionals when
describing molecular systems, especially charge transfer reactions. In
other cases, an "unconventional error cancellation" discussed in Refs.~\onlinecite{kanungo_unconventional_2024,kaplan_how_2024,pangeni_hartree-fock_2025} occurs, leading to surprisingly good numerical results, even when
HF electronic densities are less accurate than r\textsuperscript{2}SCAN
densities. This paper proposes an r{\textsuperscript{2}}SCAN@HF-like approach to predict the electronic and magnetic properties of open-\emph{d} 1{\textsuperscript{st}} row transition metal oxides and their oxidation reactions, avoiding the unconventional error cancellation which we show does not occur in these transition metal oxides.

Unconventional error cancellations observed in
r\textsuperscript{2}SCAN@HF is demonstrated in {\bf Eq.~\ref{eq:1}},\cite{kim_understanding_2013,wasserman_importance_2017,sim_quantifying_2018,vuckovic_density_2019} expressing the error of the DFT total energy
\(E\) as a sum of a functional-driven error, FE --intrinsic to the
DFA-- and a density-driven error --DE caused by inaccuracies carried
by the DFA electron density \(n\).
\begin{align}\label{eq:1}
\Delta E_\mathrm{DFA} &= E_\mathrm{DFA}\left\lbrack n_\mathrm{DFA} \right\rbrack - E_{exact}\left\lbrack n_{exact} \right\rbrack = \mathrm{FE} + \mathrm{DE}, \\ 
\mathrm{FE} &= E_\mathrm{DFA}\left\lbrack n_{exact} \right\rbrack - E_{exact}\lbrack n_{exact}\rbrack, \nonumber\\ 
\mathrm{DE} &= E_\mathrm{DFA}\left\lbrack n_\mathrm{DFA} \right\rbrack - E_\mathrm{DFA}\lbrack n_{exact}\rbrack \nonumber
\end{align}

where \(n_\mathrm{DFA}\) is the ``inaccurate'' charge density provided by the
DFA, and \(n_{exact}\ \)is the unknown exact density. To evaluate FE
and DE of {\bf Eq.~\ref{eq:1}}, an exact (or nearly exact) electron
density is needed. Refs. \onlinecite{kanungo_unconventional_2024,kaplan_how_2024,pangeni_hartree-fock_2025} used the coupled-cluster CCSD(T) density, which is nearly exact in typical \emph{sp}-bonded systems, and found that, for those systems, it produced FE and DE values very close to using the r\textsuperscript{2}SCAN50 (the global hybrid of r\textsuperscript{2}SCAN with 50\% of HF exchange) density as the exact density. Then r\textsuperscript{2}SCAN50 was used as the proxy for the
exact density in all such systems.\cite{kanungo_unconventional_2024,kaplan_how_2024,pangeni_hartree-fock_2025} The
r\textsuperscript{2}SCAN50 density was understood to provide correct
electron transfers\cite{perdew_density-functional_1982} from one atomic site to another,
a feature of the electron density to which the total energy appears
especially sensitive. Increasing the exact-exchange fraction X is
expected\cite{mori-sanchez_many-electron_2006} to increase the tendency to put an integer
number of electrons on each species. Our experience with self-consistent DFAs shows that DE is generally much smaller than FE for transition-metal oxidation energies, main-group barrier heights, and water binding energies in clusters.{\cite{kanungo_unconventional_2024,kaplan_how_2024,kaplan_understanding_2023}} DE is dominated by electron transfer errors, and insensitive to other density errors.\cite{kanungo_unconventional_2024}

In	\textit{sp-}bonded systems with minimal or no electron transfer, the r\textsuperscript{2}SCAN densities were more accurate than HF densities.\cite{mezei_electron_2017,gubler_accuracy_2025} The
improvement in density from HF to LSDA, to PBE, and to
r\textsuperscript{2}SCAN shows the predictive power of including
systematically more exact constraints in the
DFA.\cite{kaplan_predictive_2023,kaplan_how_2024,pangeni_hartree-fock_2025} Refs.~\onlinecite{kaplan_predictive_2023,kaplan_how_2024,pangeni_hartree-fock_2025} concluded that DFA@HF, in
which a DFA is evaluated on the HF orbitals and densities, often works
through an unconventional error cancellation between the FE of the
DFA and the DE of the HF density,
$E_{DFA}\lbrack n_{HF}\){]}-\(E_{DFA}\left\lbrack n_{exact} \right\rbrack$ (called the non-variational density overlocalization in  Ref.~\cite{pangeni_hartree-fock_2025}).
For example, Ref.~\cite{kanungo_unconventional_2024} found that unconventional error cancellation improved barrier heights in molecular reactions, mainly affected by SIE. This observation has been linked to error cancellation between large negative values of FE ({\bf Eq.~\ref{eq:1}}) counterbalanced by correspondingly large positive values of \(DE\); the latter caused by uncorrelated and overlocalized HF charge densities.\cite{kanungo_unconventional_2024}

We show that DFA@HF fails for transition-metal oxides due to insufficient error cancellation. Therefore, more accurate methodologies for the SIE in transition metal oxides are introduced. By independently correcting FE and DE ({\bf Eq.~\ref{eq:1}}), we address the SIE of the XC functional in predicting electronic, magnetic, and thermodynamic properties. This is achieved by mixing different fractions of exact
HF exchange into the DFA is used to evaluate the total energy and compute
the electronic charge density. This strategy is applied with the
previously proposed r\textsuperscript{2}SCAN, resulting in a generalized
r\textsuperscript{2}SCANY@r\textsuperscript{2}SCANX approach, where Y is
the percentage of exact HF exchange mixed with the r\textsuperscript{2}SCAN
functional used to evaluate energy. X represents the \% of exact HF
exchange mixed with r\textsuperscript{2}SCAN used to compute the
orbitals and, hence, the electronic charge density.

We will show that when
r\textsuperscript{2}SCANY@r\textsuperscript{2}SCANX is applied to
transition metal oxides' oxidation energies, as values of Y and/or
X increase, the error in oxidation energies drops noticeably minimizing at Y$\sim$10\% and X$\sim$50\%.  Importantly, these values for the transition-metal oxides are those we expected, based on experience with \emph{s-p}-bonded systems: Since r$^2$SCAN has less self-interaction error than PBE, its non-negligible functional-driven error of the energy can be reduced by using 10\% of HF exchange, less than the 25\% that PBE needs. Since the smaller density-driven error of the energy is dominated by electron-transfer error, the density and orbitals need 50\% of HF exchange to reduce these delocalization errors.

We will further propose a computationally efficient, non-self-consistent,
r\textsuperscript{2}SCAN10@r\textsuperscript{2}SCAN approach that
significantly improves the accuracy of oxidation energies compared to
r\textsuperscript{2}SCAN and is on par with highly parametrized
r\textsuperscript{2}SCAN+\emph{U} approaches. We will also demonstrate that r\textsuperscript{2}SCAN10@r\textsuperscript{2}SCAN outperforms r\textsuperscript{2}SCAN and r\textsuperscript{2}SCAN+\emph{U} in accuracy for band gaps.

\section{Results}\label{sec:results}
\subsection{Improving \emph{ab initio} Predictions of meta-GGA-type r\textsuperscript{2}SCAN Functionals}

All XC functionals suffer from functional-driven and
density-driven errors ({\bf Eq.~\ref{eq:1}}). SIE, a significant part of both
errors, leads to inaccurate predictions of energetics in molecules and
materials. LSDA, GGA, and meta-GGA XC functionals struggle to accurately
describe properties, such as reaction energies, inter-atomic charge
transfer, and electronic structure in systems with
strongly localized open-shell \emph{d} (or \emph{f)} electrons, as in
M\textsubscript{i}O\textsubscript{j}s.\cite{stevanovic_correcting_2012,sai_gautam_evaluating_2018,long_evaluating_2020,wang_oxidation_2006,jain_formation_2011,dudarev_electron-energy-loss_1998,tekliye_accuracy_2024,swathilakshmi_performance_2023,swathilakshmi_performance_2023,sawatzky_magnitude_1984,zaanen_band_1985,anisimov_band_1991,imada_metal-insulator_1998,lany_semiconducting_2015}

To address the SIE, we present the generalized
r\textsuperscript{2}SCANY@r\textsuperscript{2}SCANX approach, which
builds upon r\textsuperscript{2}SCAN. In
r\textsuperscript{2}SCANY@r\textsuperscript{2}SCANX, the SIE is
addressed by independently tuning the fraction of exact HF exchange
directly in the r\textsuperscript{2}SCAN {hybrid} functional definition used to
evaluate the total energy of a given set of orbitals and the
r\textsuperscript{2}SCAN {hybrid} functional used to compute these orbitals.
In r\textsuperscript{2}SCANY@r\textsuperscript{2}SCANX, Y
is the percent of HF exact exchange appearing in the XC energy of
r\textsuperscript{2}SCAN to correct the functional-driven error. In
contrast, in r\textsuperscript{2}SCANY@r\textsuperscript{2}SCANX, X is
the percentage of HF exchange used in the
r\textsuperscript{2}SCAN functional to determine the electronic orbitals
and potentially correct for the density-driven error. By independently varying the X\% and Y\% parameters in
r\textsuperscript{2}SCANY@r\textsuperscript{2}SCANX, the proposed method
enables a more systematic correction of the functional- and
density-driven errors inherent in r\textsuperscript{2}SCAN's
formulation, with the aim of improving the accuracy in describing
transition metal oxides and potentially other correlated systems.

In r\textsuperscript{2}SCANY@r\textsuperscript{2}SCANX, the XC energy
functional (\(E_{xc} = E_{xc}^{Y}\lbrack n^{X}\rbrack\)) is defined:
%

\begin{align}\label{eq:2}	
E_{xc}^{Y}\lbrack n^{X}\rbrack & = \frac{Y}{100} E_{x}^{\mathrm{HF}}\lbrack n^{X}\rbrack + \left(1-\frac{Y}{100}\right)E_{x}^{\mathrm{r^2SCAN}}  \lbrack n^{X}\rbrack \, + \\ & +  E_{c}^{\mathrm{r^2SCAN}}\lbrack n^{X}\rbrack \nonumber
\end{align}

where, in Eq.~\ref{eq:2} \(n^{X}\) is the self-consistent electron density from the r\textsuperscript{2}SCAN energy functional, including X\% of HF exchange. Therefore, the
r\textsuperscript{2}SCANY@r\textsuperscript{2}SCANX definition
incorporates different fractions of exact exchange in the energy and 
self-consistent electron density, ensuring a more systematic
correction of SIE. Replacing a small fraction of r\textsuperscript{2}SCAN exchange with the same fraction of HF exchange should have minimal or no effect on satisfying the exact DFT constraints of r\textsuperscript{2}SCAN.

\begin{figure}
\centering
 \includegraphics[width=1.\columnwidth]{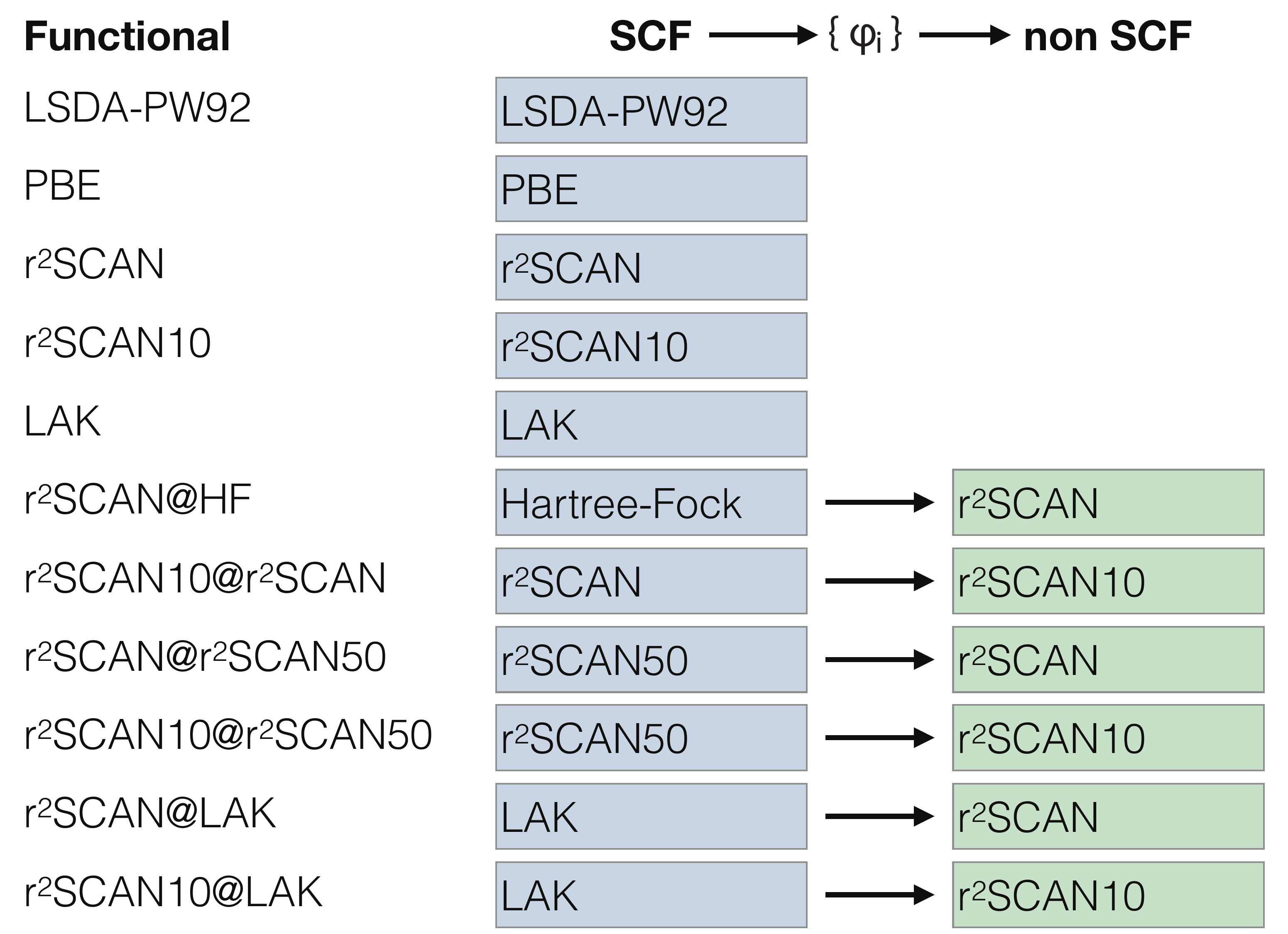}
\caption{Schematically connecting the self-consistent
and non-self-consistent (non-SCF) approaches required in the
r\textsuperscript{2}SCANY@r\textsuperscript{2}SCANX methods.
Self-consistent functionals used in this work are LSDA, PBE,
r\textsuperscript{2}SCAN, r\textsuperscript{2}SCAN10, and LAK.
Non-self-consistent hybrid functionals, including a fraction of exact HF
exchange, such as r\textsuperscript{2}SCAN@HF,
r\textsuperscript{2}SCAN10@r\textsuperscript{2}SCAN,
r\textsuperscript{2}SCAN@r\textsuperscript{2}SCAN50,
r\textsuperscript{2}SCAN10@r\textsuperscript{2}SCAN50,
r\textsuperscript{2}SCAN@LAK, r\textsuperscript{2}SCAN10@LAK, require a
self-consistent step to generate orbitals on 
which the energy is evaluated non-self-consistently.}\label{fig:1}
\end{figure}

Using {\bf Eq.~\ref{eq:2}} and specific X and Y values, we propose different r\textsuperscript{2}SCANY@r\textsuperscript{2}SCANX XC functionals ({\bf Fig.~\ref{fig:1}}). The approach of {\bf Eq.~\ref{eq:2}} is not self-consistent unless $\mathrm{Y=X}$. {\bf Fig.~\ref{fig:1}} lists in its second column (SCF) DFAs of interest here that can be implemented self-consistently to find orbitals, a density, and a total energy. The third column (non-scf), in {\bf Fig.~\ref{fig:1}} lists functionals that could be evaluated on those orbitals and that density to find a possibly-better total energy.  In the SCF part, functionals:
LDSA-PW92,\cite{perdew_accurate_1992} the GGA PBE,\cite{perdew_generalized_1996} the
meta-GGAs r\textsuperscript{2}SCAN\cite{sun_strongly_2015,furness_accurate_2020} (plain or
hybrid), or LAK were used.\cite{lebeda_balancing_2024}

\subsection{r\textsuperscript{2}SCAN Predictions of Transition-metal Oxides Properties}

 Here, we consider a diverse group of binary transition metal oxides of
the type M\textsubscript{i}O\textsubscript{j}. M is a
1\textsuperscript{st} row transition metal Ti, V, Cr, Mn, Fe, Co, Ni 
Cu, and Zn with closed or open-shell electronic structure. The predicted
energetics of these compounds are then used to compute the oxidation
energies of M\textsubscript{i}O\textsubscript{j}s' oxidation reactions.
The M\textsubscript{i}O\textsubscript{j}s comprise:
TiO\textsubscript{2} (\(P4_{2}/mnm\)),\cite{sugiyama_crystal_1991}
Ti\textsubscript{2}O\textsubscript{3}
(\(R\overline{3}c\)),\cite{abrahams_magnetic_1963} VO
(\(Fm\overline{3}m\)),\cite{kumarakrishnan_cation_1985}
V\textsubscript{2}O\textsubscript{3} (\(I2/a\)),\cite{dernier_crystal_1970}
{VO}\textsubscript{2} (\(P2_1/c\)),\cite{Rogers_1993}
V\textsubscript{2}O\textsubscript{5}\(\ (Pmmn\)),\cite{enjalbert_refinement_1986}
Cr\textsubscript{2}O\textsubscript{3}
(\(R\overline{3}c\)),\cite{hill_crystallographic_2010} CrO\textsubscript{3}
(\(C2cm\)),\cite{stephens_crystal_1970} CrO\textsubscript{2}
(\(P4_{2}/mnm\)),\cite{porta_chromium_1972} MnO
(\(Fm\overline{3}m\)),\cite{sasaki_x-ray_1979}
MnO\textsubscript{2}(\(P4_{2}/mnm\)),\cite{curetti_high-temperature_2021}
Mn\textsubscript{3}O\textsubscript{4}
(\(I4_{1}/amd\)),\cite{jarosch_crystal_1987}
Fe\textsubscript{2}O\textsubscript{3}
(\(R\overline{3}c\)),\cite{gokhan_unlu_structure_2019} FeO
(\(Fm\overline{3}m\)),\cite{yamamoto_modulated_1982}
Fe\textsubscript{3}O\textsubscript{4}
(\(Fd\overline{3}m\)),\cite{wright_charge_2002} CoO
(\(Fm\overline{3}m\)),\cite{sasaki_x-ray_1979}
Co\textsubscript{3}O\textsubscript{4}
(\(Fd\overline{3}m\)),\cite{picard_croissance_1980} 
NiO {(\(Fm\overline{3}m\)),\cite{Deng2024}}
CuO (\(C2/c\)),\cite{soejima_structural_2000} 
Cu\textsubscript{2}O (\(Pn\overline{3}m\)),\cite{restori_charge_1986} and 
ZnO {(\(P6_{3}mc\)).\cite{Abrahams1969}} These
M\textsubscript{i}O\textsubscript{j}s were selected based on the
availability of reliable experimental formation enthalpies, good
structural reports sourced from the
Inorganic Crystal Structure Database (ICSD),\cite{zagorac_recent_2019} and
M\textsubscript{i}O\textsubscript{j} ground-state magnetic orderings,
whenever required.

{\bf Table~\ref{tab:1}} summarizes the experimental and predicted lattice
parameters, magnetic moments, and band gaps of all materials using
r\textsuperscript{2}SCAN and r\textsuperscript{2}SCAN+\emph{U} (with
\emph{U} values from Ref.~\onlinecite{swathilakshmi_performance_2023}). Data reported in
{\bf Table~\ref{tab:1}} used r\textsuperscript{2}SCAN geometry-optimized
structures, including the relaxation of atomic positions, volumes, and
cell shapes. For the r\textsuperscript{2}SCAN+\emph{U} data
set, this functional was used for geometry optimization (atomic
positions, volumes, and shapes). Further, we tested M{\textsubscript{i}}O{\textsubscript{j}}s' oxidation energies and their electronic properties for a fixed {\emph{U}}=~2~eV on transition metals.

\begin{table*}[!ht]
\footnotesize
\caption{\scriptsize r\textsuperscript{2}SCAN and
r\textsuperscript{2}SCAN+\emph{U} (with
\emph{U} values from Ref.~\onlinecite{swathilakshmi_performance_2023}) predicted lattice parameters, on-site
magnetic moments (in \(\mu\)\textsubscript{B}), and band gaps (in eV) of
M\textsubscript{i}O\textsubscript{j}s. Experimental quantities are
marked as Exp. {Unless mentioned,} all calculations use experimentally determined magnetic orderings (M.O.), referenced in the magnetic moment (M.M.) column.
Antiferromagnetic (AFM), ferrimagnetic (FEM), ferromagnetic (FM), {diamagnetic-via-dimerization (DM$-$d)}, and
non-magnetic orderings (NM) are indicated. Spin-polarized calculations were used for all FM, AFM, and FEM systems, while NM materials were treated with spin-restricted calculations.
See {\bf Supplementary Table~1} for $U$ values used.
}\label{tab:1}
\begin{ruledtabular}
\begin{tabular}{llccccccccl}
{\bf System}  &  {\bf Method}  & {\bf \emph{a}} & {\bf \emph{b}} & {\bf \emph{c}} & $\pmb{\alpha}$ & $\pmb{\beta}$ & $\pmb{\gamma}$& {\bf M.M.} & {\bf M.O.} & {\bf Band Gap} \\ 
\hline

\multirow{3}{*}{\textbf{TiO\textsubscript{2}} (\(P4_{2}/mnm\))\cite{sugiyama_crystal_1991}}
& Exp. & 4.59 & 4.59 & 2.96 & 90 & 90 & 90 & 0.00 & NM &
3.0\cite{serpone_is_2006} \\
& r\textsuperscript{2}SCAN & 4.60 & 4.60 & 2.96 & 90 & 90 & 90 & 0.00 &
NM & 2.24 \\
& r\textsuperscript{2}SCAN+\emph{U} & 4.62 & 4.62 & 2.99 & 90 & 90 & 90
& 0.00 & NM & 2.51 \\

\hline 

\multirow{3}{*}{\textbf{Ti\textsubscript{2}O\textsubscript{3}} (\(R\overline3c\))\cite{abrahams_magnetic_1963}}
& Exp. & 5.43 & 5.43 & 5.43 & 56.57 & 56.57 & 56.57 &
{ $\le$ 0.03}\cite{moon_absence_1969} & {DM$-$d} & 0.20\cite{uchida_charge_2008} \\
& r\textsuperscript{2}SCAN & 5.46 & 5.46 & 5.46 & 57.76 & 57.76 & 55.76
& 0.00 & NM & 0.00 \\
& r\textsuperscript{2}SCAN+\emph{U} & 5.42 & 5.42 & 5.42 & 57.53 & 57.53
& 57.53 & 0.0 & NM & 0.59 \\

\hline 

\multirow{3}{*}{\textbf{VO} (\(Fm\overline{3}m\))\cite{kumarakrishnan_cation_1985}}
& Exp. & 2.88 & 2.88 & 4.99 & 73.22 & 90 & 120 & N/A & AFM & N/A \\
& r\textsuperscript{2}SCAN & 3.16 & 3.16 & 4.89 & 71.12 & 90 & 120 &
2.45 & AFM & 1.66 \\
& r\textsuperscript{2}SCAN+\emph{U} & 3.16 & 3.16 & 5.00 & 71.54 & 90 &
119.99 & 2.55 & AFM & 2.35 \\

\hline 

\multirow{3}{*}{\textbf{V\textsubscript{2}O\textsubscript{3}} (\(I2/a\))\cite{dernier_crystal_1970}} & Exp. &
7.25 & 5.00 & 5.55 & 90 & 96.75 & 90 & 1.2/2.37\cite{moon_antiferromagnetism_1970,shin_observation_1992}
& AFM & 0.20\cite{shin_vacuum-ultraviolet_1990} \\
& r\textsuperscript{2}SCAN & 7.28 & 5.00 & 5.51 & 90 & 97.50 & 90 & 1.70
& AFM & 0.00 \\
& r\textsuperscript{2}SCAN+\emph{U} & 7.28 & 5.08 & 5.56 & 90 & 96.42 &
90 & 1.80 & AFM & 0.68 \\

\hline 

\multirow{3}{*}{\textbf{VO\textsubscript{2}} (\(P2_1/c\))\cite{Rogers_1993}} & Exp. &
5.75 & 4.53 & 5.38 & 90 & 122.69 & 90 & 0.00 \cite{goodenough_VO2}
& {DM$-$d} & 0.70\cite{shin_vacuum-ultraviolet_1990} \\
& r\textsuperscript{2}SCAN & {5.88} & 4.49 & 5.35 & 90 & 123.21 & 90 & 0.93
& AFM & 0.17 \\
& r\textsuperscript{2}SCAN+\emph{U} & 5.94 & 4.46 & 5.37 & 90 & 123.65 &
90 & 0.98 & AFM & 0.69 \\

\hline 

\multirow{3}{*}{\textbf{V\textsubscript{2}O\textsubscript{5}} (\(Pmmn\))\cite{enjalbert_refinement_1986}} & Exp. &
11.51 & 3.56 & 4.37 & 90 & 90 & 90 & 0.00 & NM &
2.5\cite{kumar_structural_2008} \\
& r\textsuperscript{2}SCAN & 11.59 & 3.55 & 4.25 & 90 & 90 & 90 & 0.00 &
NM & 2.04 \\
& r\textsuperscript{2}SCAN+\emph{U} & 11.59 & 3.56 & 4.25 & 90 & 90 & 90
& 0.00 & NM & 2.14 \\

\hline

\multirow{2}{*}{\textbf{Cr\textsubscript{2}O\textsubscript{3}} (\(R\overline{3}c\))\cite{hill_crystallographic_2010}}
& Exp. & 4.95 & 4.95 & 13.60 & 90 & 90 & 120 & 2.76\cite{corliss_magnetic_1965}
& AFM & 3.2\cite{abdullah_structural_2014} \\
& r\textsuperscript{2}SCAN & 4.94 & 4.94 & 13.62 & 90 & 90 & 120 & 2.58
& AFM & 2.58 \\

\hline 

\multirow{2}{*}{\textbf{CrO\textsubscript{3}} (\(C2cm\))\cite{stephens_crystal_1970}} & Exp. &
4.79 & 8.56 & 5.74 & 90 & 90 & 90 & 0.00 & NM &
3.8\cite{misho_preparation_1989} \\
& r\textsuperscript{2}SCAN & 4.86 & 8.25 & 5.70 & 90 & 90 & 87.91 & 0.00
& NM & 2.30 \\

\hline 

\multirow{2}{*}{\textbf{CrO\textsubscript{2}} (\(P4_2/mnm\))\cite{porta_chromium_1972}}
& Exp. & 4.42 & 4.42 & 2.92 & 90 & 90 & 90 & 2.00\cite{coey_half-metallic_2002} &
FM & 0.00\cite{coey_half-metallic_2002} \\
& r\textsuperscript{2}SCAN & 4.40 & 4.40 & 2.91 & 90 & 90 & 90 & 2.06 &
FM & 0.00 \\

\hline 

\multirow{3}{*}{\textbf{MnO} (\(Fm\overline{3}m\))\cite{sasaki_x-ray_1979}}
& Exp. & 3.14 & 3.14 & 6.29 & 60 & 60 & 60 & 4.58\cite{cheetham_magnetic_1983} &
AFM & 3.6/3.8\cite{messick_direct_1972} \\
& r\textsuperscript{2}SCAN & 3.14 & 3.14 & 6.17 & 59.36 & 59.36 & 59.98
& 4.30 & AFM & 1.74 \\
& r\textsuperscript{2}SCAN+\emph{U} & 3.15 & 3.15 & 6.21 & 59.50 & 59.50
& 59.99 & 4.42 & AFM & 2.13 \\

\hline 

\multirow{3}{*}{\textbf{MnO\textsubscript{2}} (\(P4_2/mnm\))\cite{curetti_high-temperature_2021}}
& Exp. & 4.40 & 4.40 & 2.87 & 90 & 90 & 90 & 2.35\cite{regulski_incommensurate_2003} &
AFM & 0.27/0.3\cite{druilhe_electron_1967,islam_studies_2005} \\
& r\textsuperscript{2}SCAN & 4.38 & 4.38 & 2.86 & 90 & 90 & 90 & 2.62 &
AFM & 0.39 \\
& r\textsuperscript{2}SCAN+\emph{U} & 4.39 & 4.39 & 2.88 & 90 & 90 & 90
& 2.77 & AFM & 0.74 \\

\hline 
\multirow{3}{*}{\textbf{Mn\textsubscript{3}O\textsubscript{4}} (\(I4_{1}/amd\))\cite{jarosch_crystal_1987}}
& Exp. & 5.76 & 5.76 & 6.24 & 117.52 & 117.52 & 90 &
4.34, 3.25/3.6\cite{jensen_magnetic_1974} & FEM &
2.3-2.5\cite{xu_chemical_2006} \\
& r\textsuperscript{2}SCAN & 5.72 & 5.72 & 6.22 & 117.39 & 117.39 & 90 &
4.19, 3.51 & FEM & 0.96 \\
& r\textsuperscript{2}SCAN+\emph{U} & 5.76 & 5.76 & 6.24 & 117.50 &
117.50 & 90 & 4.36, 3.64 & FEM & 1.39 \\

\hline 

\multirow{3}{*}{\textbf{FeO} (\(Fm\overline{3}m\))\cite{yamamoto_modulated_1982}}
& Exp. & 6.08 & 6.08 & 6.08 & 60 & 60 & 60 &
3.32/4.2\cite{roth_magnetic_1958,battle_magnetic_1979} & AFM & 2.20\cite{bowen_electrical_1975} \\
& r\textsuperscript{2}SCAN & 5.87 & 6.13 & 5.96 & 62.31 & 60.48 & 61.39
& 3.42 & AFM & 0.43 \\
& r\textsuperscript{2}SCAN+\emph{U} & 6.11 & 6.10 & 6.10 & 61.04 & 59.91
& 59.95 & 3.54 & AFM & 1.58 \\

\hline 

\multirow{3}{*}{\textbf{Fe\textsubscript{2}O\textsubscript{3}}

(\(R\overline{3}c\))\cite{gokhan_unlu_structure_2019}}
& Exp. & 5.03 & 5.03 & 13.76 & 90 & 90 & 120 & 4.9\cite{shull_neutron_1951}
& AFM & 2.20\cite{droubay_structure_2007} \\
& r\textsuperscript{2}SCAN & 5.00 & 5.00 & 13.74 & 90 & 90 & 120 & 3.86
& AFM & 1.52 \\
& r\textsuperscript{2}SCAN+\emph{U} & 5.04 & 5.04 & 13.75 & 90 & 90 &
120 & 4.12 & AFM & 1.50 \\

\hline 
\multirow{3}{*}{\textbf{Fe\textsubscript{3}O\textsubscript{4}}

(\(Fd\overline{3}m\))\cite{wright_charge_2002}}
& Exp. & 8.39 & 8.39 & 8.39 & 90 & 90 & 90 &
4.44, 4.1\cite{wright_charge_2002} & FEM & 0.14\cite{park_charge-gap_1998} \\
& r\textsuperscript{2}SCAN & 8.34 & 8.34 & 8.34 & 90 & 90 & 90 & 3.73;
3.70 & FEM & 0.00 \\
& r\textsuperscript{2}SCAN+\emph{U} & 8.44 & 8.47 & 8.37 & 90.01 & 90.28
& 90.03 & 4.12; 3.58 & FEM & 0.23 \\

\hline 

\multirow{3}{*}{\textbf{CoO}

(\(Fm\overline{3}m\))\cite{sasaki_x-ray_1979}}
& Exp. & 3.01 & 3.01 & 6.03 & 60 & 60 & 60 &
3.35/3.8\cite{roth_magnetic_1958,khan_magnetic_1970} & AFM &
2.40\cite{wei_insulating_1994,van_elp_electronic_1991} \\
& r\textsuperscript{2}SCAN & 2.99 & 2.99 & 5.96 & 59.92 & 59.92 & 60 &
2.54 & AFM & 0.85 \\
& r\textsuperscript{2}SCAN+\emph{U} & 3.01 & 3.01 & 5.97 & 59.78 & 59.78
& 60 & 2.62 & AFM & 2.12 \\

\hline 

\multirow{3}{*}{\textbf{Co\textsubscript{3}O\textsubscript{4}}

(\(Fd\overline{3}m\))\cite{picard_croissance_1980}}
& Exp. & 8.07 & 8.07 & 8.07 & 90 & 90 & 90 & 3.02\cite{roth_magnetic_1964} &
AFM & 1.60\cite{van_elp_electronic_1991} \\
& r\textsuperscript{2}SCAN & 8.03 & 8.03 & 8.03 & 90 & 90 & 90 & 2.45 &
AFM & 1.12 \\
& r\textsuperscript{2}SCAN+\emph{U} & 8.06 & 8.06 & 8.06 & 90 & 90 & 90
& 2.57 & AFM & 1.94 \\

\hline 

\multirow{3}{*}{\textbf{NiO}

(\(Fm\overline{3}m\))\cite{Deng2024}} & Exp. &
2.93 & 2.93 & 5.87 & 60 & 60 & 60 & 1.64/1.90\cite{cheetham_magnetic_1983,anisimov_band_1991} & AFM
& 4.3\cite{sawatzky_magnitude_1984} \\
& r\textsuperscript{2}SCAN & 2.94 & 2.94 & 5.88 & 59.91 & 59.91 & 60.07 &
1.59 & AFM & 2.42 \\
& r\textsuperscript{2}SCAN+\emph{U} & 2.95 & 2.95 & 5.90 & 59.97 & 59.97 & 60.03
& 1.69 & AFM & 3.52 \\

\hline 

\multirow{2}{*}{\textbf{CuO}

(\(C2/c\))\cite{soejima_structural_2000}} & Exp. &
6.32 & 3.42 & 7.50 & 90 & 95.23 & 90 & 0.68\cite{yang_neutron_1988} & AFM
& 1.40\cite{ghijsen_electronic_1988} \\
& r\textsuperscript{2}SCAN & 6.35 & 3.90 & 6.95 & 90 & 100.89 & 90 &
0.56 & AFM & 0.69 \\

\hline 

\multirow{2}{*}{\textbf{Cu\textsubscript{2}O}

(\(Pn\overline{3}m\))\cite{restori_charge_1986}}
& Exp. & 4.27 & 4.27 & 4.27 & 90 & 90 & 90 & 0.00 & NM &
2.17/2.4\cite{zhang_symmetry-breaking_2020,ghijsen_electronic_1988} \\
& r\textsuperscript{2}SCAN & 4.24 & 4.24 & 4.24 & 90 & 90 & 90 & 0.00 &
NM & 2.20 \\

\hline 

\multirow{2}{*}{\textbf{ZnO}

(\(P6_{3}mc\))\cite{Abrahams1969}} & Exp. &
3.25 & 3.25 & 5.21 & 90 & 90 & 120 & 0.0 & NM
& 3.4\cite{Reynolds1999} \\
& r\textsuperscript{2}SCAN & 3.24 & 3.24 & 5.20 & 90 & 90 & 119.99
& 0.00 & NM & 1.25 \\

\hline

\multirow{3}{*}{{\textbf{CeO$_{2}$}}

(\(Fm\overline{3}m\))\cite{CeO2_struc_WOLCYRZ1992409}}
& Exp. & 5.41 & 5.41 & 5.41 & 90 & 90 & 90 & 0.00 & NM &
3.32\cite{Kolodiazhnyi2018} \\
& r\textsuperscript{2}SCAN & 5.44 & 5.44 & 5.44 & 90 & 90 & 90 & 0.00 &
NM & 2.20 \\
& r\textsuperscript{2}SCAN+\emph{U} & 5.45 & 5.45 & 5.45 & 90 & 90 & 90 & 0.00 &
NM & 2.35 \\
\hline

\multirow{3}{*}{{\textbf{Ce$_2$O$_3$}}

(\(P\overline{3}m\))\cite{Ce2O3_struc_BARNIGHAUSEN1985385}}
& Exp. & 3.89 & 3.89 & 6.06 & 90 & 90 & 120 & 1.08\cite{Ce2O3_mag_PINTO198281} & AFM & 2.34\cite{Ce2O3_bg_PROKOFIEV199641,Kolodiazhnyi2018} \\
& r\textsuperscript{2}SCAN & 3.87 & 3.87 & 6.05 & 90 & 89.99 & 120 & 0.94 & AFM & 0.57 \\
& r\textsuperscript{2}SCAN+\emph{U} & 3.89 & 3.89 & 6.09 & 90 & 90 & 120 & 0.97 & AFM & 1.90 \\

\end{tabular}
\end{ruledtabular}
\end{table*}

Previous studies have shown that r\textsuperscript{2}SCAN (SCAN)
geometries agree well with experimental
data.\cite{kitchaev_energetics_2016,sai_gautam_evaluating_2018,long_evaluating_2020,tekliye_accuracy_2024,swathilakshmi_performance_2023}  Data in {\bf Table~\ref{tab:1}}
confirms the accuracy of r\textsuperscript{2}SCAN in predicting the
lattice parameters and magnetic configurations of
M\textsubscript{i}O\textsubscript{j}s and is consistent with Ref. \onlinecite{swathilakshmi_performance_2023}.
Exceptions to this trend are VO ($\sim$8.9\%) and CuO
($\sim$14\%), whose lattice
constants deviate from experimental reports.\cite{kumarakrishnan_cation_1985,soejima_structural_2000}  

Mn{\textsubscript{3}}O{\textsubscript{4} is known to exhibit complex
magnetism,{\cite{jensen_magnetic_1974}} which was approximated by setting a
collinear magnetic configuration that best approximates {the}
 experimentally observed magnetic
ground state.} Ti{\textsubscript{2}}O{\textsubscript{3}} is diamagnetic upon dimerization,{\cite{moon_absence_1969}} and modeled as non-magnetic here. VO{\textsubscript{2}} also dimerizes, which is described with an antiferromagnetic configuration.{\cite{10.1063/5.0180315}}
{These choices of magnetic configurations in Ti{\textsubscript{2}}O{\textsubscript{3}} and VO{\textsubscript{2}} are based on the best approach proposed here (r\textsuperscript{2}SCAN10@r2SCAN50), see {\bf Supplementary~Fig.~1}. For 
many of the non-magnetic materials in \textbf{Table~{\ref{tab:1}}} (TiO$_2$,
V$_2$O$_5$, CrO$_3$, Cu$_2$O, and ZnO), the assumption that each oxygen is a closed-shell O{\textsuperscript{2-}}
anion leaves no valence electron on the transition-metal cation.

Experimentally, Ti\textsubscript{2}O\textsubscript{3}
($\sim$0.2 eV), V\textsubscript{2}O\textsubscript{3}
($\sim$0.2 eV), and Fe\textsubscript{3}O\textsubscript{4} ($\sim$0.14
eV) have small experimental band gaps, respectively, but are predicted
to be metallic with r\textsuperscript{2}SCAN and consistent with
previous results in Ref.~\onlinecite{swathilakshmi_performance_2023}.

\subsection{Benchmarking Oxidation Enthalpies of
M\textsubscript{i}O\textsubscript{j}s with
r\textsuperscript{2}SCANY@r\textsuperscript{2}SCANX
Functionals}

 The XC functionals of {\bf Fig.~\ref{fig:1}} were used to assess the
oxidation enthalpies of several
M\textsubscript{i}O\textsubscript{j}s (M = Ti, V, Cr, Mn, Fe, Co, and
Cu) as defined for \(z > x\) in {\bf Eq.~\ref{eq:3}}.
\begin{equation}\label{eq:3}
\frac{2}{z-x}\mathrm{MO_x} + \mathrm{O_2} \rightarrow \frac{2}{z-x}\mathrm{MO_z}
\end{equation}
Following {\bf Eq.~\ref{eq:3}}, and using various
r\textsuperscript{2}SCANY@r\textsuperscript{2}SCANX DFAs, we computed all possible oxidation enthalpy reactions
using the M\textsubscript{i}O\textsubscript{j}s in {\bf Table~\ref{tab:1}} and
{\bf Eq.~\ref{eq:4}}, resulting in {18} distinct reactions summarized in {\bf Table~\ref{tab:2}}. The negative
chemical energy change in the reaction is:
\begin{widetext}
\begin{equation}\label{eq:4}
\mathrm{\Delta}H_0 = \frac{2}{z-x}E(\mathrm{MO_z})^\mathrm{r^2SCANY@r^2SCANX} - \frac{2}{z-x}E(\mathrm{MO_x})^\mathrm{r^2SCANY@r^2SCANX} - E(\mathrm{O_2})^\mathrm{r^2SCANY@r^2SCANX}
\end{equation}
\end{widetext}
where, \(E\left( \mathrm{MO_{z}} \right)^\mathrm{r^{2}SCANY@r2SCANX}\),
\(E\left( \mathrm{MO_{x}} \right)^\mathrm{r^{2}SCANY@r2SCANX}\),
\(E\left( \mathrm{O_{2}} \right)^\mathrm{r^{2}SCANY@r2SCANX}\), are the
r\textsuperscript{2}SCANY@r\textsuperscript{2}SCANX total energies of
the oxidized, reduced M\textsubscript{i}O\textsubscript{j}s, and $\mathrm{O_2}$ gas,
respectively. Our DFT predictions are compared to
experimental oxidation {reaction enthalpies shown in} {\bf Table \ref{tab:2}}, {obtained from formation energies} of \(\mathrm{MO_{x}}\) and \(\mathrm{MO_{z}}\)
({\bf Supplementary Table~2}) extracted from experimental
reports.\cite{Kubaschewski1979,thomas_c_allison_nist-janaf_2013}
Thermochemical tables provide oxidation energies with an uncertainty range that should include the true value 95\% of the time. Furthermore, the definition of the bounds of the mean absolute error used in theoretical work is narrower than the typical standard deviation set to a 95\% level of confidence imposed in thermochemical tables. In
addition, here we approximated the oxidation enthalpies by the DFT total
energies by imposing \(\Delta H_{0} \approx \Delta E\), thus neglecting
the pV contributions, which are expected to be minimal.


\begin{table}[!ht]
\scriptsize
\caption{{Oxidation reactions and their experimental $\Delta H_0$ (in eV/O$_2$) as defined in Eq.}~\ref{eq:3} and Eq.~\ref{eq:4}.}
\label{tab:2}
\begin{ruledtabular}
\begin{tabular}{llc}

 {\bf Reaction} & {\bf Change in oxidation state} & $\pmb{\Delta H_0}$  \\ 

\hline 
2Ti\textsubscript{2}O\textsubscript{3} + O$_2$ $\rightarrow$ 4TiO$_2$ & Ti$^{3+}$ $\rightarrow$ Fe$^{4+}$   & --7.6166\cite{thomas_c_allison_nist-janaf_2013} \\ 
\hline 
  4VO + O$_2$ $\rightarrow$ 2V$_2$O$_3$ &  V$^{2+}$ $\rightarrow$ V$^{3+}$ & --7.3632\cite{Kubaschewski1979}  \\
  $\frac{4}{3}$VO + O$_2$ $\rightarrow$ $\frac{2}{3}$V$_2$O$_5$ & V$^{2+}$ $\rightarrow$ V$^{5+}$ & --4.7469\cite{Kubaschewski1979}  \\
  V$_2$O$_3$ + O$_2$ $\rightarrow$ V$_2$O$_5$ & V$^{3+}$ $\rightarrow$ V$^{5+}$ & --3.4387\cite{Kubaschewski1979}  \\
  2VO + O$_2$ $\rightarrow$ 2VO$_2$ & V$^{2+}$ $\rightarrow$ V$^{4+}$ & --5.8368\cite{Kubaschewski1979}  \\
  2V$_2$O$_3$ + O$_2$ $\rightarrow$ 4VO$_2$ & V$^{3+}$ $\rightarrow$ V$^{4+}$ & --4.3104\cite{Kubaschewski1979}  \\
  2VO$_2$ + O$_2$ $\rightarrow$ V$_2$O$_5$ & V$^{4+}$ $\rightarrow$ V$^{5+}$ & --2.5670\cite{Kubaschewski1979}  \\
\hline 
 2Cr$_2$O$_3$ + O$_2$ $\rightarrow$ 4CrO$_2$ & Cr$^{6+}$ $\rightarrow$ Cr$^{4+}$ & --0.7286\cite{Kubaschewski1979} \\
 $\frac{2}{3}$Cr$_2$O$_3$ + O$_2$ $\rightarrow$  $\frac{4}{3}$CrO$_3$ & Cr$^{3+}$ $\rightarrow$ Cr$^{6+}$ & --0.2023\cite{Kubaschewski1979} \\
  2CrO$_2$ + O$_2$ $\rightarrow$ 2CrO$_3$ & Cr$^{4+}$ $\rightarrow$ Cr$^{6+}$ & 0.0608\cite{Kubaschewski1979}  \\
\hline 
  6MnO+O$_2$ $\rightarrow$  2Mn$_3$O$_4$ & Mn$^{2+}$ $\rightarrow$ Mn$^{2+}$, Mn$^{3+}$&  --4.7862\cite{Kubaschewski1979}  \\ 
  2MnO + O$_2$ $\rightarrow$ 2MnO$_2$  & Mn$^{2+}$ $\rightarrow$ Mn$^{4+}$ & --2.7942\cite{Kubaschewski1979}  \\
  Mn$_3$O$_4$ + O$_2$ $\rightarrow$ 3MnO$_2$ & Mn$^{2+}$, Mn$^{3+}$ $\rightarrow$ Mn$^{4+}$ & --1.7982\cite{Kubaschewski1979}  \\
\hline 
 6FeO + O$_2$ $\rightarrow$ 2Fe$_3$O$_4$ & Fe$^{2+}$ $\rightarrow$ Fe$^{2+}$, Fe$^{3+}$&  --6.7038\cite{Kubaschewski1979}  \\
 4FeO + O$_2$ $\rightarrow$ 2Fe$_2$O$_3$  & Fe$^{2+}$ $\rightarrow$ Fe$^{3+}$ &  --6.0620\cite{Kubaschewski1979}  \\
 4Fe$_3$O$_4$ + O$_2$ $\rightarrow$ 6Fe$_2$O$_3$ & Fe$^{2+}$, Fe$^{3+}$ $\rightarrow$  Fe$^{3+}$ & --4.7784\cite{Kubaschewski1979}  \\ 
\hline 
 6CoO + O$_2$ $\rightarrow$ 2Co$_3$O$_4$ & Co$^{2+}$ $\rightarrow$  Co$^{2+}$, Co$^{3+}$& --3.9014\cite{thomas_c_allison_nist-janaf_2013}  \\
\hline 
 2Cu$_2$O + O$_2$ $\rightarrow$ 4CuO & Cu$^{+}$ $\rightarrow$  Cu$^{2+}$ & --2.8744\cite{thomas_c_allison_nist-janaf_2013} \\
\hline 
 2Ce$_2$O$_3$ + O$_2$ $\rightarrow$ 4CeO$_2$ & Ce$^{3+}$ $\rightarrow$  Ce$^{4+}$ & --7.4238\cite{Kubaschewski1979} \\

\end{tabular}
\end{ruledtabular}
\end{table}

{\bf Fig.~\ref{fig:2}} displays the deviation of predicted oxidation
enthalpies including common XC functionals, such as
LSDA,\cite{perdew_accurate_1992} PBE,\cite{perdew_generalized_1996} and meta-GGA,
including r\textsuperscript{2}SCAN\cite{furness_accurate_2020} and
LAK,\cite{lebeda_balancing_2024} as well as
r\textsuperscript{2}SCANY@r\textsuperscript{2}SCANX
(r\textsuperscript{2}SCANY@LAK) derivatives. From the benchmarking of {18}
predicted oxidation energies with experimental values, we derived the
error distributions in {\bf Fig.~\ref{fig:2}b}, and the mean absolute errors
(MAEs) of {\bf Fig.~\ref{fig:2}c}. Note, errors in oxidation energies with
all methods were evaluated using r\textsuperscript{2}SCAN geometries.
For the r\textsuperscript{2}SCAN+\emph{U} dataset, the
M\textsubscript{i}O\textsubscript{j}s geometries were optimized using
r\textsuperscript{2}SCAN+\emph{U}.

\begin{figure}[!ht]
\centering
\includegraphics[width=0.95\columnwidth]{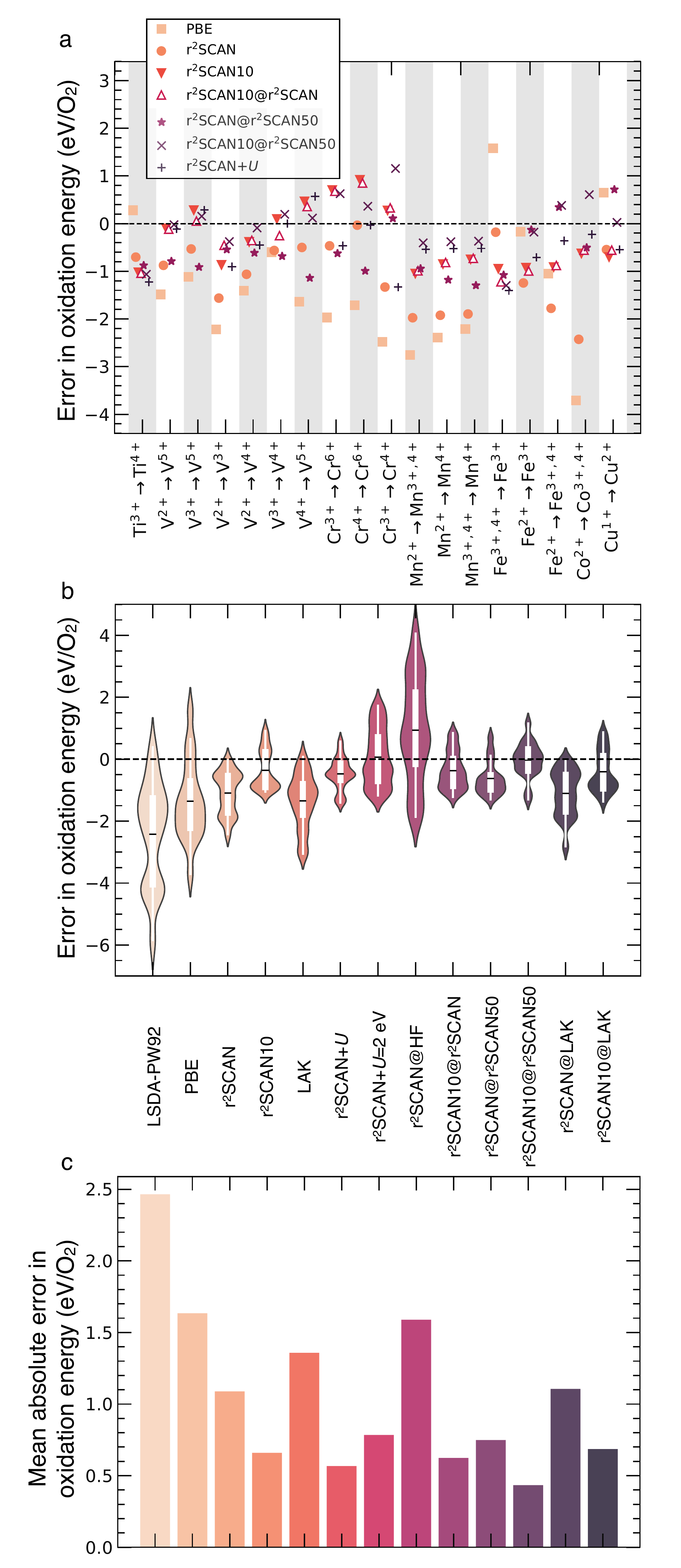}
\caption{Prediction error of 3{\emph{d}} M\textsubscript{i}O\textsubscript{j} oxidation energies {of reactions in {\bf Table~{\ref{tab:2}}}} (except for Ce{\textsubscript{i}}O{\textsubscript{j}} compounds)  with several XC
functionals, including
r\textsuperscript{2}SCANY@r\textsuperscript{2}SCANX and
r\textsuperscript{2}SCANY@LAK as defined in {\bf Fig.~\ref{fig:1}}. (a) Error
in oxidation energies of 18 reactions considered. (b) Violin
representation of error distributions. (c) The mean absolute errors.
Errors in predicting oxidation energies of all DFT functionals except
r\textsuperscript{2}SCAN+\emph{U} are evaluated at r\textsuperscript{2}SCAN
geometries. The mean experimental oxidation energy is --3.82~eV/O\textsubscript{2}.}\label{fig:2}
\end{figure}

As shown in {\bf Fig.~\ref{fig:2}b} and {\bf Fig.~\ref{fig:2}c}, the errors in
predicting oxidation energies for M\textsubscript{i}O\textsubscript{j}s
with LSDA, PBE, and r\textsuperscript{2}SCAN reduce systematically as
the quality of the exchange-correlation functional increases from LSDA
to PBE, and to meta-GGA. In this sequence, the number of exact
constraints on the exchange-correlation energy that can be satisfied
increases from 8 to 11 to 17. Predictions with these XC functionals are
affected to different extents by the 
SIE.\cite{koch_chemists_2001} Therefore, we will focus on
r\textsuperscript{2}SCAN and its hybrids
r\textsuperscript{2}SCANY@r\textsuperscript{2}SCANX ({\bf Fig.~\ref{fig:1}}).

In {\bf Fig.~\ref{fig:2}c}, r\textsuperscript{2}SCAN exhibits a mean absolute
error (MAE) of $\sim${1.09}~eV/O\textsubscript{2} and a root mean
squared error (RMSE) of $\sim${1.29}~eV/O\textsubscript{2} in oxidation enthalpies in our dataset. r\textsuperscript{2}SCAN
systematically makes oxidation energies too negative across all tested
cases. This is largely a functional-driven error. 

The
r\textsuperscript{2}SCAN+\emph{U} implemenented here used \emph{U} parameters from Ref.~\onlinecite{swathilakshmi_performance_2023}. We have also tested r\textsuperscript{2}SCAN+\emph{U} calculations, with \emph{U} fixed to 2~eV.  When applied to the 18 oxidation reactions (excluding the one for Ce),
r\textsuperscript{2}SCAN+\emph{U} yields milder deviation
from experimental data, with an MAE of $\sim${0.57}~eV/O\textsubscript{2} and an RMSE of $\sim${0.70}~eV/O\textsubscript{2}. As expected, setting a standard {\emph{U}}~=~2~eV for all transition metal worsens the MAE to $\sim${0.78}~eV/O{\textsubscript{2}} and an RMSE to $\sim${0.92}~eV/{O\textsubscript{2}}.

A self-consistent global hybrid, such as r\textsuperscript{2}SCAN10,
incorporating 10\% of exact HF exchange should partly mitigate the SIE
inaccuracies of the r\textsuperscript{2}SCAN meta-GGA, but at the
expense of slightly higher computational costs than for GGA-based global
hybrids (and considerably higher than the cost of pure meta-GGAs). As
expected, the r\textsuperscript{2}SCAN10 functional significantly
reduces the errors, yielding an MAE of $\sim${0.66}~eV/O\textsubscript{2} and an RMSE of $\sim${0.73}~eV/O\textsubscript{2} ({\bf Fig.~\ref{fig:2}}). This represents approximately
a {40}\% reduction in errors compared to r\textsuperscript{2}SCAN.

When applying r\textsuperscript{2}SCAN@HF and using the HF
electronic density for the r\textsuperscript{2}SCAN total energy
evaluation, the MAE for our dataset is {1.59}~eV/O\textsubscript{2} and an RMSE of {1.92}~eV/O\textsubscript{2}. This
indicates a decline of approximately {45}\% in accuracy compared to
standard r\textsuperscript{2}SCAN. Similarly to
r\textsuperscript{2}SCAN@HF, one can reduce the fraction of HF exact
exchange to 50\% as in
r\textsuperscript{2}SCAN@r\textsuperscript{2}SCAN50 while producing a
potentially lower density-driven error than
r\textsuperscript{2}SCAN.\cite{adamo_toward_1999,ernzerhof_assessment_1999} This method gives our
dataset an MAE of $\sim${0.75}~eV/O\textsubscript{2} ({\bf Fig.~\ref{fig:2}c}) and an
RMSE of $\sim${0.82}~eV/O\textsubscript{2}.

Here, we propose the 
r\textsuperscript{2}SCAN10@r\textsuperscript{2}SCAN, which iterates to
self-consistency with the less expensive r\textsuperscript{2}SCAN, and 
requires only a single total-energy evaluation (not requiring a complete
self-consistent field electronic relaxation) with the more costly
global hybrid r\textsuperscript{2}SCAN10. Supposedly,
r\textsuperscript{2}SCAN10@r\textsuperscript{2}SCAN corrects
functional-driven errors by introducing 10\% exact exchange in the
functional while using r\textsuperscript{2}SCAN orbitals.

In {\bf Fig.~\ref{fig:2}b} and {\bf Fig.~\ref{fig:2}c},
r\textsuperscript{2}SCAN10@r\textsuperscript{2}SCAN yields an MAE of
$\sim${0.62}~eV/O\textsubscript{2} and a RMSE of $\sim${0.71}
eV/O\textsubscript{2}, which are comparable in magnitude to the hybrid
functional r\textsuperscript{2}SCAN10 but can be obtained at a far lower
computational cost. Indeed,
r\textsuperscript{2}SCAN10@r\textsuperscript{2}SCAN appears sufficient
to correct the functional-driven error of r\textsuperscript{2}SCAN
({\bf Fig.~\ref{fig:2}b}). We propose a more general approach in the form of
r\textsuperscript{2}SCAN10@r\textsuperscript{2}SCAN50 that corrects both
functional-driven errors, including 10\% exact exchange in
total energy estimation, and density-driven errors with 50\% exact
exchange in the density (orbital) generation. Applying
r\textsuperscript{2}SCAN10@r\textsuperscript{2}SCAN50 to our dataset
results in an MAE of $\sim${0.43}~eV/O\textsubscript{2} and an RMSE of $\sim${0.57}~eV/O\textsubscript{2}. r\textsuperscript{2}SCAN10@r\textsuperscript{2}SCAN50 is the most accurate approach in  the r\textsuperscript{2}SCANY@r\textsuperscript{2}SCANX family.

The XC functional LAK is expected to predict more accurate band gaps,
which are closely related to charge transfer processes in materials. For
this reason, we expected that LAK might yield generally improved electron
densities compared to r\textsuperscript{2}SCAN,\cite{lebeda_balancing_2024,lebeda_meta-gga_2025}
and hence, speculatively, r\textsuperscript{2}SCAN@LAK and
r\textsuperscript{2}SCAN10@LAK should predict oxidation energies in
better agreement with experimental data. However, for our dataset, the
performance of r\textsuperscript{2}SCAN@LAK (MAE: $\sim${1.12}~eV/O\textsubscript{2}) appears comparable {to or slightly worse than that of}
r\textsuperscript{2}SCAN (MAE: $\sim${1.09}~eV/O\textsubscript{2}), and
similarly, r\textsuperscript{2}SCAN10@LAK (MAE:~$\sim${0.69}~eV/O\textsubscript{2}) shows no evident improvements over
r\textsuperscript{2}SCAN10@r\textsuperscript{2}SCAN (MAE:~$\sim${0.62}~eV/
O\textsubscript{2}).
Meanwhile, LAK gave a higher error compared to r{\textsuperscript{2}}SCAN, with a MAE of 1.36 eV/O{\textsubscript{2}} (RMSE $\sim$1.60 eV/O{\textsubscript{2}}).

It is essential to analyze the type of distribution of oxidation
energies in {\bf Fig.~\ref{fig:2}b}. In {\bf Fig.~\ref{fig:2}b}, oxidation
energies with PBE form a largely unimodal distribution with a long tail
towards positive error, comprising reactions such as
Fe\textsubscript{3}O\textsubscript{4} \(\rightarrow\)
Fe\textsubscript{2}O\textsubscript{3}, Cu\textsubscript{2}O
\(\rightarrow\) CuO, Ti\textsubscript{2}O\textsubscript{3}
\(\rightarrow\) TiO\textsubscript{2}, and FeO \(\rightarrow\)
Fe\textsubscript{2}O\textsubscript{3}. The negative tail of this distribution is
set by the reaction CoO \(\rightarrow\)
Co\textsubscript{3}O\textsubscript{4}. In contrast,
r\textsuperscript{2}SCAN oxidation energies follow a bimodal
distribution ({\bf Fig.~\ref{fig:2}b}), with low and high error peaks. The
low (\emph{i.e.}, close to 0) error peak primarily comprises early
transition metal, such as Ti, V, Cr, and Cu reactions. The high error
peak includes almost all the Mn, Fe, and Co reactions ({\bf Fig.~\ref{fig:1}a}). 
r\textsuperscript{2}SCAN10 and r\textsuperscript{2}SCAN10@r\textsuperscript{2}SCAN data also
follows a bimodal distribution with {low and high error peaks}.
{Here, the low error peak is centered on the zero error line and consists of V and Cr reactions.
The high-error peak is attributed to the reactions involving Ti, Mn, Fe, Co, and Cu species.} {Origins of bimodal distributions are discussed in detail in Sec.~{\ref{sec:eroors}}.}

The distribution for r\textsuperscript{2}SCAN@r\textsuperscript{2}SCAN50
appears largely unimodal with a broad tail in the positive error
comprising
Cr\textsubscript{2}O\textsubscript{3}~\(\rightarrow\)~CrO\textsubscript{2},
and Cu\textsubscript{2}O~\(\rightarrow\)~CuO reactions.
r\textsuperscript{2}SCAN10@r\textsuperscript{2}SCAN50 also displays a
unimodal distribution with the positive part encompassing the
Cr\textsubscript{2}O\textsubscript{3}~\(\rightarrow\)~CrO\textsubscript{2}
reaction, whereas the negative end of the tail gathers
Ti\textsubscript{2}O\textsubscript{3}~\(\rightarrow\)~TiO\textsubscript{2},
and
Fe\textsubscript{3}O\textsubscript{4}~\(\rightarrow\)~Fe\textsubscript{2}O\textsubscript{3}.

{To test the applicability of r$^2$SCANY@r$^2$SCANX on $f$-electron systems, we have applied it to the Ce$_2$O$_3$~$\rightarrow$~CeO$_2$ reaction (with prediction errors for this reaction in \textbf{Supplementary Fig. 2}). 
Experimentally, CeO{\textsubscript{2}} is NM (with no valence electron on Ce),{\cite{Kolodiazhnyi2018}}  while Ce{\textsubscript{2}}O{\textsubscript{3}} is AFM (with one \emph{f} electron on Ce).{\cite{Ce2O3_mag_PINTO198281}} \textbf{Table~{\ref{tab:1}}} shows good agreement with the experiment for the r{\textsuperscript{2}}SCAN
and r{\textsuperscript{2}}SCAN\emph{+U} ($U$ from Ref.~{\cite{sai_gautam_evaluating_2018}}) lattice constants
and magnetic moments.

{At r{\textsuperscript{2}}SCAN, the Ce$_2$O$_3$~$\rightarrow$~CeO$_2$ reaction has an error of --0.95 eV/O{\textsubscript{2}}. 
The error drops to 0.15 and 0.18 eV/O{\textsubscript{2}} for r{\textsuperscript{2}}SCAN10 and r{\textsuperscript{2}}SCAN10@r{\textsuperscript{2}}SCAN.}
{However, using global hybrid densities in the DFA, the error surprisingly increases to $\sim$1.40 (r{\textsuperscript{2}}SCAN@r{\textsuperscript{2}}SCAN50) and $\sim$1.66 
eV/O{\textsubscript{2}} (r{\textsuperscript{2}}SCAN10@r{\textsuperscript{2}}SCAN50), respectively.} 
{With r{\textsuperscript{2}}SCAN+$U$, the predicted oxidation energy is overestimated by $\sim$1.1~eV/O{\textsubscript{2}}.}
{For this reaction r{\textsuperscript{2}}SCAN@HF yields a large error of $\sim$5.76~eV/O{\textsubscript{2}}, a very substantial underbinding of the extra O{\textsubscript{2}}, aligning with observations of transition metal oxides.}


\subsection{Achieving Optimal Fractions of Exact
Exchange in the r\textsuperscript{2}SCANY@r\textsuperscript{2}SCANX
Formulations}

 We now focus on potential improvements to the r\textsuperscript{2}SCAN
functional, as elucidated in {\bf Fig.~\ref{fig:1}}. We will explain how
optimal X and Y percentages of exact HF exchange are
incorporated in the r\textsuperscript{2}SCAN functionals during the
non-self-consistent and self-consistent steps. By utilizing all the
oxidation energies predicted ({\bf Fig.~\ref{fig:2}}) with various
r\textsuperscript{2}SCANY@r\textsuperscript{2}SCANX functionals, we
identify the optimal combination of X and Y percentages of HF exchange
that minimizes their prediction errors in {\bf Fig.~\ref{fig:3}}. In
{\bf Supplementary~Fig.~3}, {\bf Supplementary~Fig.~4}, and {\bf Supplementary~Fig.~5} (which also shows the individual
reactions), the behavior of errors for individual reactions appears
non-trivial; no correlation could be identified for any of the
reactions. Analysis of the MAE and RMSE trends in {\bf Fig.~\ref{fig:3}}a
and {\bf Fig.~\ref{fig:3}b} for the self-consistent r\textsuperscript{2}SCANX and
non-self-consistent r\textsuperscript{2}SCANY@r\textsuperscript{2}SCAN
methods shows that the optimal exact-exchange fraction lies in the
{9--14}\% range.

\begin{figure}[!t]
\centering
 \includegraphics[width=1.0\columnwidth]{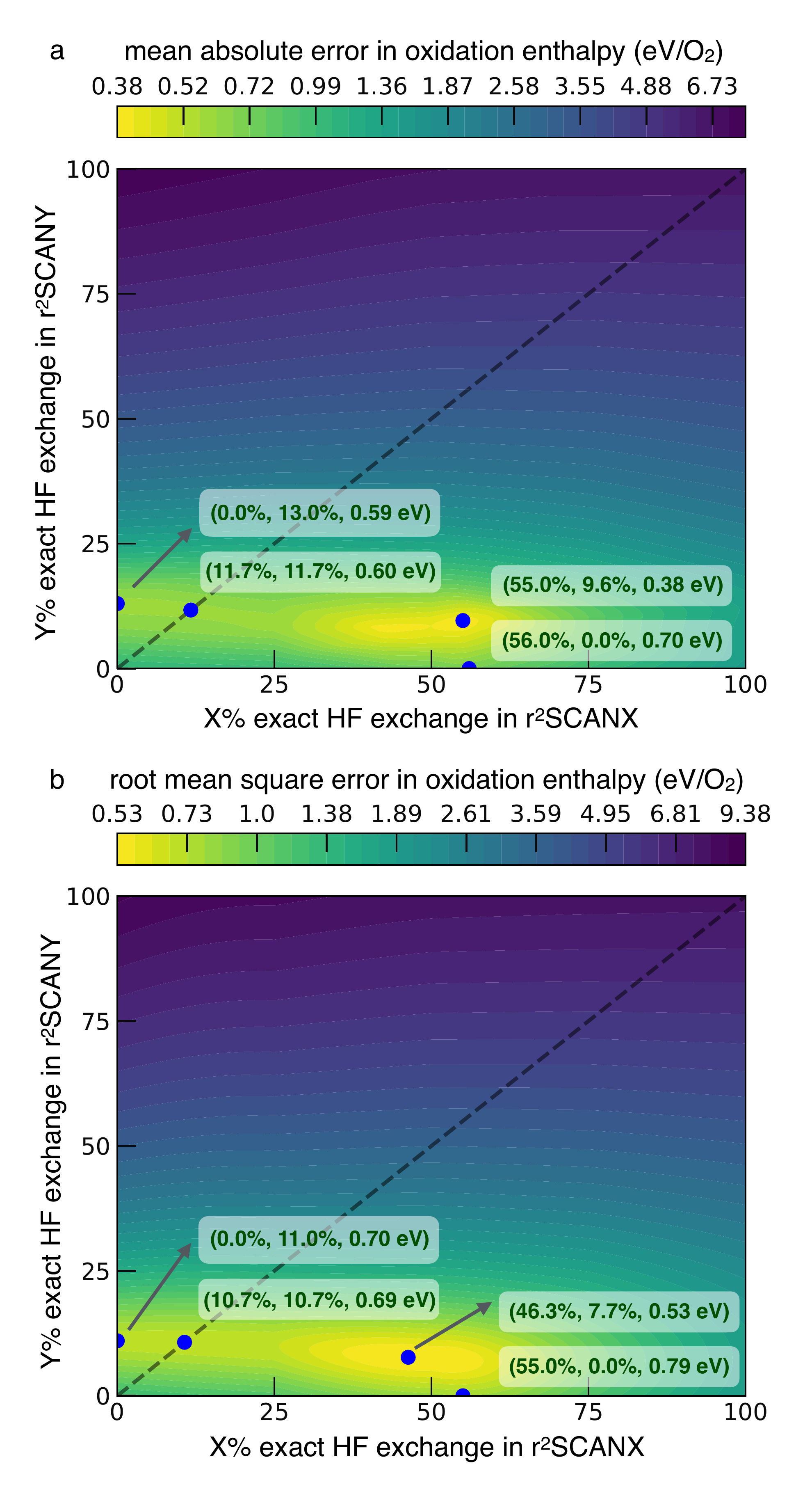}
\caption{Identification of optimal values of X and Y
in r\textsuperscript{2}SCANY@r\textsuperscript{2}SCANX functionals, by
locating minima in mean absolute and root mean square errors of
oxidation energies of all possible oxidation reactions of
M\textsubscript{i}O\textsubscript{j}s. Panel \textbf{a} shows that the
mean absolute error minimizes at {55.0}\% exact exchange in orbitals and
{9.6}\% exact exchange in the functional, forming
{r\textsuperscript{2}SCAN10@r\textsuperscript{2}SCAN55}, which produces an
error of $\sim${0.38}~ eV/O\textsubscript{2}. Panel \textbf{b}
shows that the root mean square error minimizes, with an error value of
$\sim${0.53} eV at {46.3}\% exact exchange in orbital and {7.7}\%
exact exchange in functional, thus
{r\textsuperscript{2}SCAN8@r\textsuperscript{2}SCAN46}. See {\bf Sec.~{\ref{sec:DFTmethod}}} for details on the interpolation scheme to coarse-grain values
of X and Y.}\label{fig:3}
\end{figure}

Looking at the \emph{x}-axis of {\bf Fig.~\ref{fig:3}a} and {\bf Fig.~\ref{fig:3}b}, for
r\textsuperscript{2}SCAN@r\textsuperscript{2}SCANX, the minimum error
(MAE and RMSE) is observed in the range $\sim${54--58}\%
exact HF exchange, indicating a significantly higher
requirement for exact exchange in the Hamiltonian used for the orbital
generation when the underlying functional is r\textsuperscript{2}SCAN.
From the global minima in {\bf Fig.~\ref{fig:3}a} and {\bf Fig.~\ref{fig:3}b}, the
r\textsuperscript{2}SCANY@r\textsuperscript{2}SCANX approach achieves
its lowest MAE and RMSE when Y --the HF fraction in the functional
definition-- is in the range of $\sim${7--10}\%, and X (in the HF
fraction in orbitals) is between $\sim${45--56}\%.

While the optimal exact exchange fraction varies across methods, a general trend shows errors are minimized with about $\sim$10\% HF exchange in the functional and around $\sim$50\% in orbital generation. These values balance the correction of density-functional-approximation SIEs.

\subsection{Effects of Exact-exchange Fractions X
and Y on the Binding Energy of the Oxygen
Molecule}

\begin{figure}[!h]
\centering
 \includegraphics[width=0.85\columnwidth]{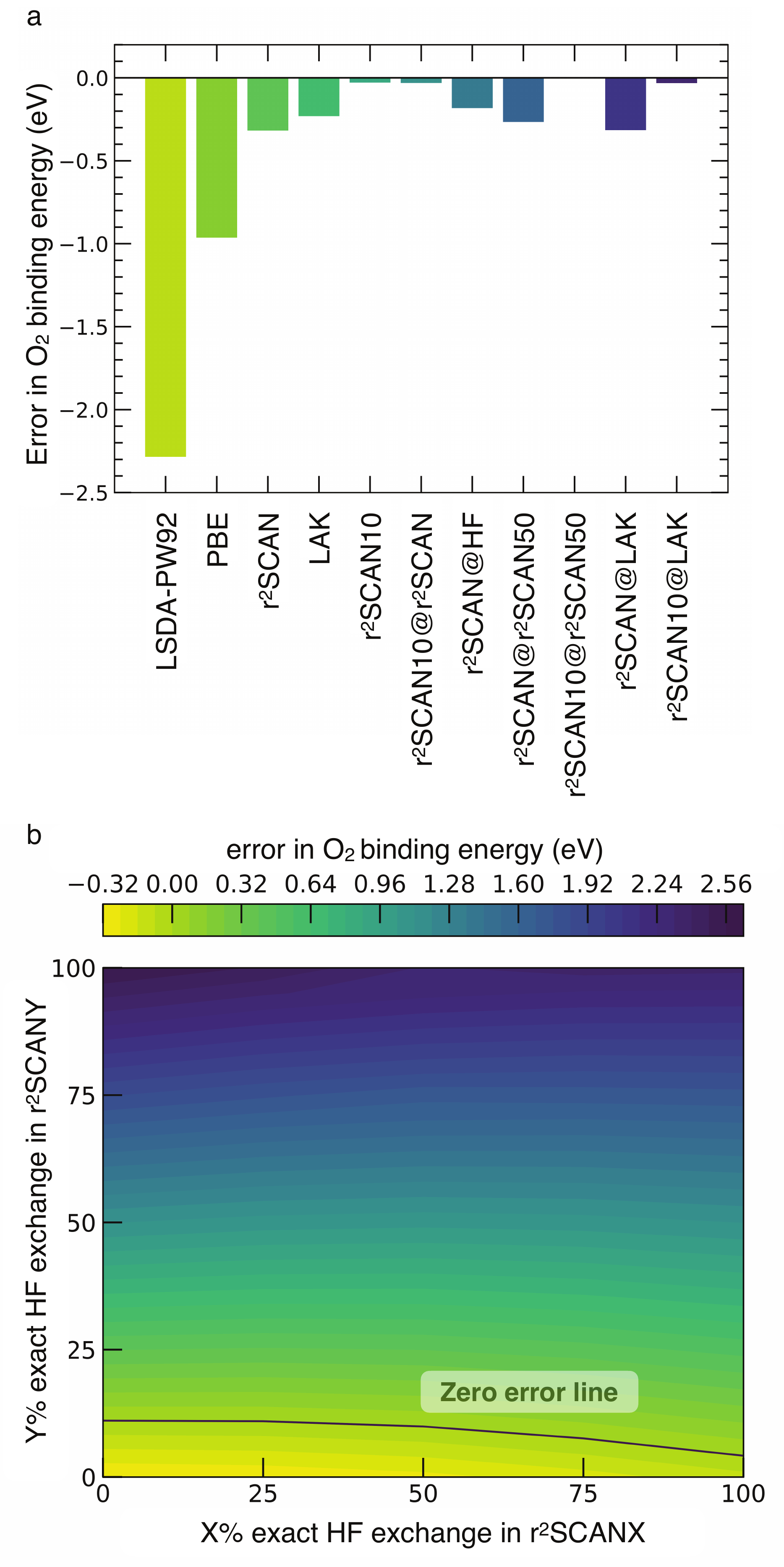}
\caption{Error in the negative binding energy of the
O\textsubscript{2} molecule using
r\textsuperscript{2}SCANY@r\textsuperscript{2}SCANX.
Panel {\textbf{a}} shows the error for DFAs introduced in {\textbf{Fig.~\ref{fig:1}}}. All calculations are
performed using r\textsuperscript{2}SCAN geometries. In panel
\textbf{b}, optimal values of X and Y, in various
r\textsuperscript{2}SCANY@r\textsuperscript{2}SCANX functionals for
O\textsubscript{2} binding energy, are demonstrated using the
generalized r\textsuperscript{2}SCANY@r\textsuperscript{2}SCANX method.
See Sec.~{\ref{sec:DFTmethod}} for details on the interpolation scheme to
coarse-grain values of X and Y in panel \textbf{b}.}\label{fig:4}
\end{figure}

 {\bf Fig.~\ref{fig:4}} displays the error of O\textsubscript{2} binding
energies introduced by several XC functionals, including the new
r\textsuperscript{2}SCANY@r\textsuperscript{2}SCANX proposed here.
{\bf Fig.~\ref{fig:4}a} shows that all XC functionals considered here tend to
overbind the O\textsubscript{2} molecule, with errors
diminishing progressively from LSDA-PW92 (--2.2~eV/O\textsubscript{2}) $\gg$
GGA PBE ($\sim$--1.0~eV/O\textsubscript{2}) $\gg$ meta-GGA
r\textsuperscript{2}SCAN and LAK (--0.3 --- --0.2~eV/O\textsubscript{2}) $\gtrsim$
r\textsuperscript{2}SCANY@r\textsuperscript{2}SCANX.

Note that zero-point energy corrections are not included in this
analysis and are a constant energy shift of
$\sim$0.1~ eV/O\textsubscript{2}.\cite{pople_gaussian-1_1989,irikura_experimental_2007} The inadequacy of LSDA XC functionals, which overbinds O\textsubscript{2} molecule had already been noted by Perdew and
Zunger\cite{perdew_self-interaction_1981} in agreement with LSDA-PW92 in
{\bf Fig.~\ref{fig:4}a}. Similar inaccuracies were also identified for GGA functionals in Refs.~\cite{blochl_first-principles_2000,wang_oxidation_2006}

Both LSDA and GGA strongly overestimate O\textsubscript{2} binding energy, causing systematic errors in predicted oxidation energies of M\textsubscript{i}O\textsubscript{j}s, which could be corrected \emph{ad hoc} in {\bf Eq.~\ref{eq:3}}. Both SCAN\cite{sun_strongly_2015} and
r\textsuperscript{2}SCAN\cite{furness_accurate_2020} meta-GGA significantly
alleviate the O\textsubscript{2} overbinding of PBE. In
{\bf Fig.~\ref{fig:4}a}, the error in binding energy drops from
$\sim$--1~eV/O\textsubscript{2} with PBE to $\sim$--0.3~eV/O\textsubscript{2} with r\textsuperscript{2}SCAN. However, a systematic error remains in r\textsuperscript{2}SCAN-predicted oxidation energies due to r\textsuperscript{2}SCAN overbinding O\textsubscript{2}.

 {\bf Fig.~\ref{fig:4}}
suggests that O\textsubscript{2} binding energies can be accurately
predicted by r\textsuperscript{2}SCANY@r\textsuperscript{2}SCANX to eliminate systematic errors and achieve accurate oxidation
energies of M\textsubscript{i}O\textsubscript{j}s. As indicated by the
contour line in {\bf Fig.~\ref{fig:4}b}, setting Y at approximately 10\% of
the exact HF exchange in the non-self-consistent part, along with any
value of X\% for exact exchange in the orbital definition, results in a
small error in O\textsubscript{2} binding energy ($\sim$--0.03~eV/O\textsubscript{2}).

In {\bf Fig.~\ref{fig:4}a} r\textsuperscript{2}SCAN10,
r\textsuperscript{2}SCAN10@r\textsuperscript{2}SCAN, and
r\textsuperscript{2}SCAN10@r\textsuperscript{2}SCAN50 further reduce the
error in O\textsubscript{2} binding energy to --0.031, --0.030, and 0.002~eV/O\textsubscript{2}, respectively, thereby minimizing inaccuracies in predicting
oxidation energies of M\textsubscript{i}O\textsubscript{j}s.
Additionally, r\textsuperscript{2}SCAN@r\textsuperscript{2}SCAN50, which
only deals with the density-driven error, does not seem to improve the
O\textsubscript{2} binding energy error compared to
r\textsuperscript{2}SCAN, indicating that the overbinding in
O\textsubscript{2} binding energy is almost entirely due to the
functional-driven error.

Using a harder PAW potential for oxygen (\texttt{O\_h 06Feb2004} in VASP) overbinds the O$_2$ molecule, resulting in an increased error in O$_2$ binding energy of  $\sim$0.15 eV/O$_2$ (see {\bf Supplementary~Fig.~6}) across all  methods in {\textbf{Fig.~\ref{fig:4}}.  This increased error can be empirically mitigated by including the zero-point energy correction of  $\sim$0.1 eV/O$_2$.{\cite{pople_gaussian-1_1989,irikura_experimental_2007}

\subsection{Effects of Exact-Exchange Fractions X and Y on On-site Magnetic Moments}

 The strong correlation in M\textsubscript{i}O\textsubscript{j}s seems to
be captured at least in part by the DFA for
normal correlation through spin symmetry
breaking.\cite{zhang_symmetry-breaking_2020,trimarchi_polymorphous_2018,varignon_origin_2019,xiong_symmetry_2025,perdew_symmetry_2023} Here, we have used experimental or nearly experimental magnetic orders ({\bf Table~\ref{tab:1}}).

The artificial delocalization of \emph{d} electrons due to SIE of XC
functionals affects the predicted on-site magnetic moments in transition
metals.\cite{illas_magnetic_1998,de_p_r_moreira_effect_2002,illas_spin_2006,franchini_ground-state_2007,horton_high-throughput_2019,arale_brannvall_predicting_2024,houchins_quantifying_2017}  We investigate the
variation in the error of predicted magnetic moments in our
r\textsuperscript{2}SCANY@r\textsuperscript{2}SCANX. The magnetic
moments are calculated by integrating the net spin density over the
projector augmented wave (PAW) potential spheres of the transition metal
atoms.

The calculated on-site magnetic moments of transition metals depend only
on the accuracy of electronic orbitals, particularly the fraction of
exact exchange used in generating those orbitals. The on-site magnetic
moments can be considered independent of the functional or the
percentage of exact HF exchange employed in the non-self-consistent step
(\emph{i.e.}, Y in r\textsuperscript{2}SCANY@r\textsuperscript{2}SCANX),
which only affects the energy evaluation and not the electronic charge
density. Therefore, only the X percent of the exact exchange in 
r\textsuperscript{2}SCANY@r\textsuperscript{2}SCANX is
relevant.

{\bf Fig.~\ref{fig:5}a} and {\bf Fig.~\ref{fig:5}b} show the mean percent and absolute
errors in magnetic moments of M\textsubscript{i}O\textsubscript{j}s.
Increasing the percentage X of Hartree-Fock exchange increases the
on-site magnetic moment associated with \emph{d} electrons. For example,
in Fe\textsubscript{2}O\textsubscript{3}, the average magnetic moment on
Fe atoms increases from 3.86 \(\mu_{B}\) with r\textsuperscript{2}SCAN
to 4.01~\(\mu_{B}\) with r\textsuperscript{2}SCAN10, 4.31~\(\mu_{B}\) with
r\textsuperscript{2}SCAN50, 4.48~\(\mu_{B}\) with
r\textsuperscript{2}SCAN100, and 4.51~\(\mu_{B}\) with Hartree-Fock,
\emph{i.e.}, X = 100\% and no correlation in the DFA.
This trend aligns with the expected behavior of hybrid functionals,
enhancing the localization of magnetic moments. {\bf Supplementary Table~3} delineates this trend for all  M{\textsubscript{i}}O{\textsubscript{j}}s.

\begin{figure*}[!ht]
\centering
 \includegraphics[width=1.0\textwidth]{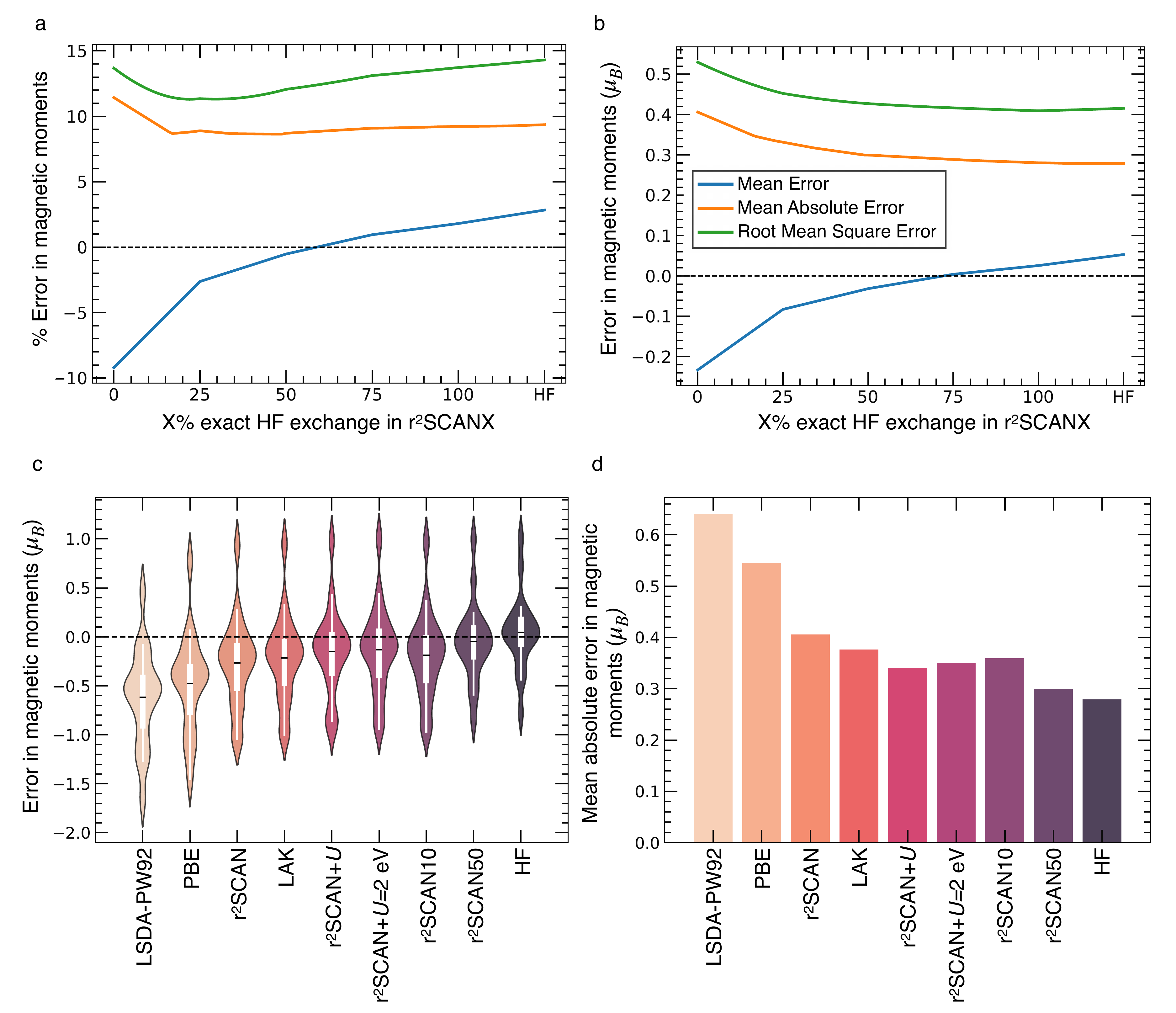}
\caption{Prediction errors of magnetic moments of
M\textsubscript{i}O\textsubscript{j}s. Panels \textbf{a} and \textbf{b}
show percent and non-relative errors in magnetic moments vs. X \% exact
HF exchange in r\textsuperscript{2}SCANX. Panels \textbf{c} and
\textbf{d} show distributions and mean absolute errors of magnetic
moments for selected DFAs introduced in {\textbf{Fig.~{\ref{fig:1}}}}. The mean
experimental magnetic moment is $\sim$3.03~\(\mu\)\textsubscript{B}. See {\bf Supplementary Table~3} for
numerical values of computed magnetic moments.}\label{fig:5}
\end{figure*}

From {\bf Fig.~\ref{fig:5}a}, the percent MAE and RMSE of magnetic moments
are minimized in an interval of HF exact exchange in the
r\textsuperscript{2}SCANX functional definition. There is no significant
improvement in percent MAE for HF fractions larger than {30}\%, whereas
percent RMSE increases progressively. Conversely, as shown in
{\bf Fig.~\ref{fig:5}b}, the absolute magnitude of errors in magnetic moments
decreases gradually as the fraction of exact HF exchange increases in
defining the orbitals.

{\bf Fig.~\ref{fig:5}c} and {\bf Fig.~\ref{fig:5}d} demonstrate that, starting from
r\textsuperscript{2}SCAN magnetic moments, r\textsuperscript{2}SCAN50
reduces the MAE by 29\%, and HF reduces it by 34\%. In contrast, 
r\textsuperscript{2}SCAN+{\emph{U}} (r\textsuperscript{2}SCAN+{\emph{U}}~=~2~eV) reduces error only by about 17\% (15\%).
The MAE is
especially sensitive to M\textsubscript{i}O\textsubscript{j}s with larger magnetic
moments (\emph{e.g.}, MnO, FeO, Fe\textsubscript{2}O\textsubscript{3},
and Fe\textsubscript{3}O\textsubscript{4} in {\bf Table~\ref{tab:1}}), where
relatively small fractional moment changes lead to larger absolute
differences. However, the fact that the distribution center in
{\bf Fig.~\ref{fig:5}c} falls near the zero-error line indicates that
r\textsuperscript{2}SCAN50 yields a better electron density in this
sense. {\bf Fig.~\ref{fig:5}c} and {\bf Fig.~\ref{fig:5}d}  show an improvement 
from LAK over r\textsuperscript{2}SCAN  by only about 7\% for the magnetic
moments.

\subsection{Effects of Exact-exchange Fractions X and Y on the Prediction of Fundamental Band Gaps}

 Band gaps of materials are typically affected by SIE, with band gaps
often underestimated by LSDA, GGA, and meta-GGA
functionals.\cite{sai_gautam_evaluating_2018,long_evaluating_2020,tekliye_accuracy_2024,swathilakshmi_performance_2023,lany_semiconducting_2015,chan_efficient_2010,perdew_understanding_2017,Peng2012} We investigate the
error in predicted band gaps using our
r\textsuperscript{2}SCANY@r\textsuperscript{2}SCANX method. The
fundamental band gap is the difference between the lowest unoccupied and
the highest occupied orbital energies. Our orbital energies are
expectation values of the r\textsuperscript{2}SCANY one-electron
generalized Kohn-Sham Hamiltonian using the
r\textsuperscript{2}SCANX orbitals. Y=100\% includes the full Fock
operator and strongly overestimates the band gaps.

\begin{figure*}[!ht]
\centering
 \includegraphics[width=1.0\textwidth]{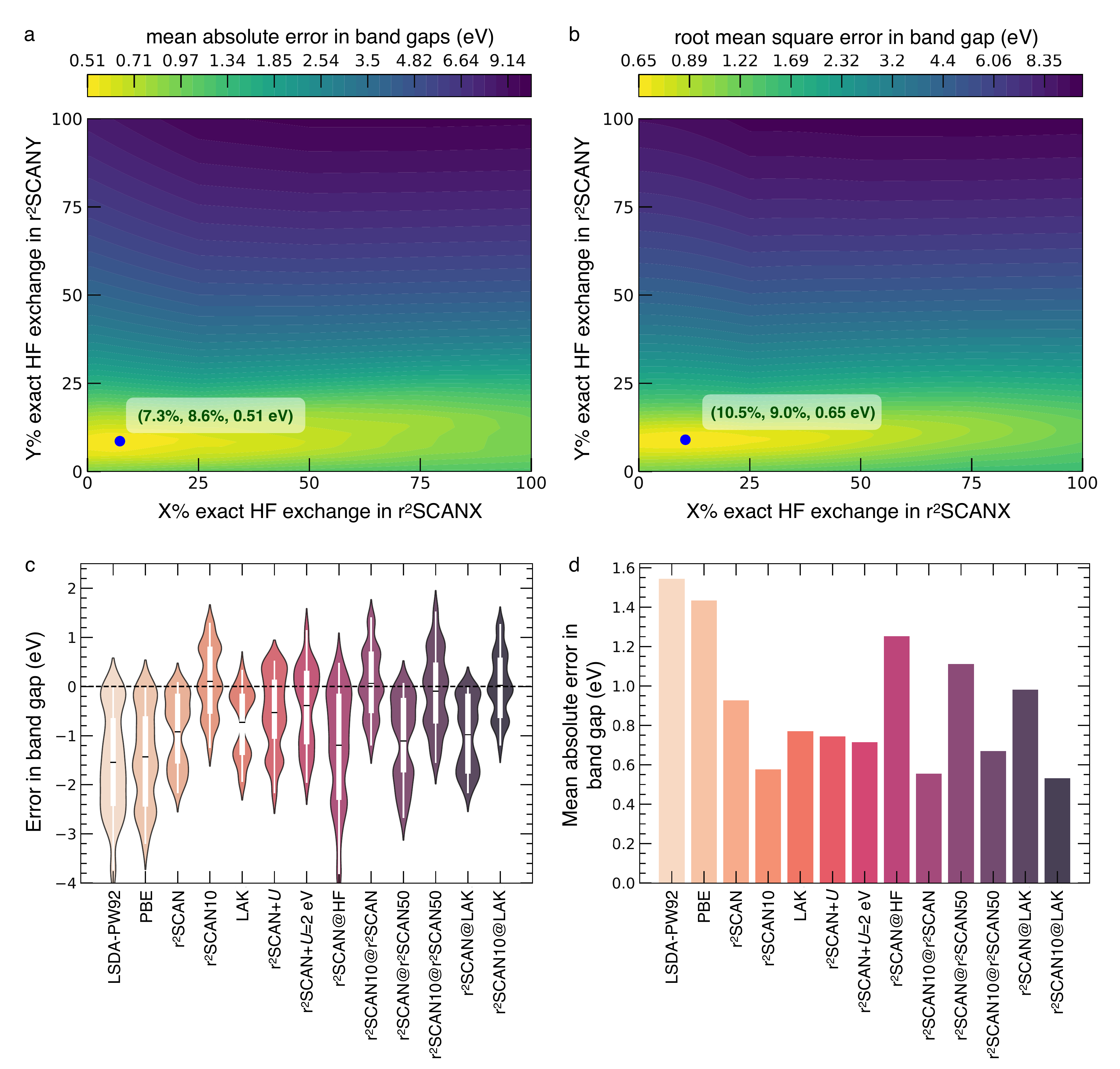}
\caption{Error in the fundamental band gap of
M\textsubscript{i}O\textsubscript{j}s. Panels \textbf{a} and \textbf{b}
display the mean absolute and root mean square errors in band gaps for
the generalized r\textsuperscript{2}SCANY@r\textsuperscript{2}SCANX
functionals. Likewise, panels \textbf{c} and \textbf{d} show the
distribution and mean absolute errors in band gap for various
r\textsuperscript{2}SCANY@r\textsuperscript{2}SCANX. See Sec.~{\ref{sec:DFTmethod}} for details on the interpolation scheme adopted to coarse-grain
the values of X and Y of panels \textbf{a} and \textbf{b}. The mean band gap
of M\textsubscript{i}O\textsubscript{j} is $\sim$2.11~eV. See
{\bf Supplementary Tables~4} and {\bf 5} for
numerical values of computed band gaps.}\label{fig:6}
\end{figure*}

{\bf Fig.~\ref{fig:6}a} and {\bf Fig.~\ref{fig:6}b} show the MAE and the RMSE in
predicted band gaps using the
r\textsuperscript{2}SCANY@r\textsuperscript{2}SCANX method. Predicted
band gaps with  r\textsuperscript{2}SCANY@r\textsuperscript{2}SCANX
 depend on both the X\% of exact HF exchange in the functional for
the orbitals and the Y\% of exact HF exchange in the functional used in
the non-self-consistent step to evaluate the orbital energies. For
M\textsubscript{i}O\textsubscript{j}s ({\bf Table~\ref{tab:1}}),
{\bf Fig.~\ref{fig:6}a} and {\bf Fig.~\ref{fig:6}b} show that the MAE minimizes at
$\sim$7.3\% exact HF exchange is used in the orbital-generation
(\emph{i.e.}, r\textsuperscript{2}SCANX), and around $\sim$8.6\% exact HF
exchange in the functional used in the non-self-consistent step
(r\textsuperscript{2}SCANY). Similarly, the RMSE is minimized at
$\sim$10.5\% exact HF exchange is used in the orbital-generation
(\emph{i.e.}, r\textsuperscript{2}SCANX), and around $\sim$9.0\% exact HF
exchange in the functional used in the non-self-consistent step
(r\textsuperscript{2}SCANY). 
{Indeed, the predicted band gaps attain an optimal value for Y ($\sim$9\%), and the error increases rapidly beyond this. 
In contrast, the error increases slowly as X increases, reaching an optimal value.}

{\bf Fig.~\ref{fig:6}c} is the error distribution, and {\bf Fig.~\ref{fig:6}d} is the mean absolute error in predicted band gaps with XC
functionals of {\bf Fig.~\ref{fig:1}}. In {\bf Fig.~\ref{fig:6}b} the MAE in band
gaps systematically decreases from 1.54~eV for LSDA
(PW91) \textgreater{} 1.43~eV for GGA(PBE) $\gg$
0.93~eV in r\textsuperscript{2}SCAN, and \textgreater{}
0.77~eV for LAK, following the number of exact
constraints satisfied by the DFAs in that sequence.

Turning the attention to
r\textsuperscript{2}SCANY@r\textsuperscript{2}SCANX-type functionals,
r\textsuperscript{2}SCAN10,
r\textsuperscript{2}SCAN10@r\textsuperscript{2}SCAN, and
r\textsuperscript{2}SCAN10@LAK have similar MAE errors of
$\sim$0.53---0.58~eV, with noticeable improvement over
r\textsuperscript{2}SCAN. However,
r\textsuperscript{2}SCAN@r\textsuperscript{2}SCAN50, which performed
well for oxidation energies, performs poorly in predicting band gaps,
with a substantial MAE of $\sim$1.11~eV. This error increase mainly comes from three closed \textit{d}-shell oxides: TiO\textsubscript{2}, V\textsubscript{2}O\textsubscript{5}, and CrO\textsubscript{3} ({\bf Supplementary~Fig.~7}). 
Notably, r\textsuperscript{2}SCAN+\emph{U}  (r{\textsuperscript{2}}SCAN+{\emph{U}}~=~2~eV)  with a MAE error of 0.74~eV (0.71~eV) is
outperformed by r\textsuperscript{2}SCAN10 (0.58~eV), r\textsuperscript{2}SCAN10@r\textsuperscript{2}SCAN (0.55~eV) and matched by r\textsuperscript{2}SCAN10@r\textsuperscript{2}SCAN50 (0.67~eV). NiO is often considered to be a prototype Mott insulator,{\cite{sawatzky_magnitude_1984,anisimov_band_1991}} but its band gap (\textbf{Supplementary~Table~4}) is remarkably accurate with r{\textsuperscript{2}}SCAN10@r{\textsuperscript{2}}SCANX for X~=~0 or 10.

\subsection{Relative Stability of MnO, NiO and ZnO Polymorphs with r\textsuperscript{2}SCANY@r\textsuperscript{2}SCANX   Approaches}
\begin{figure}[!ht]
\centering
 \includegraphics[width=1.0\columnwidth]{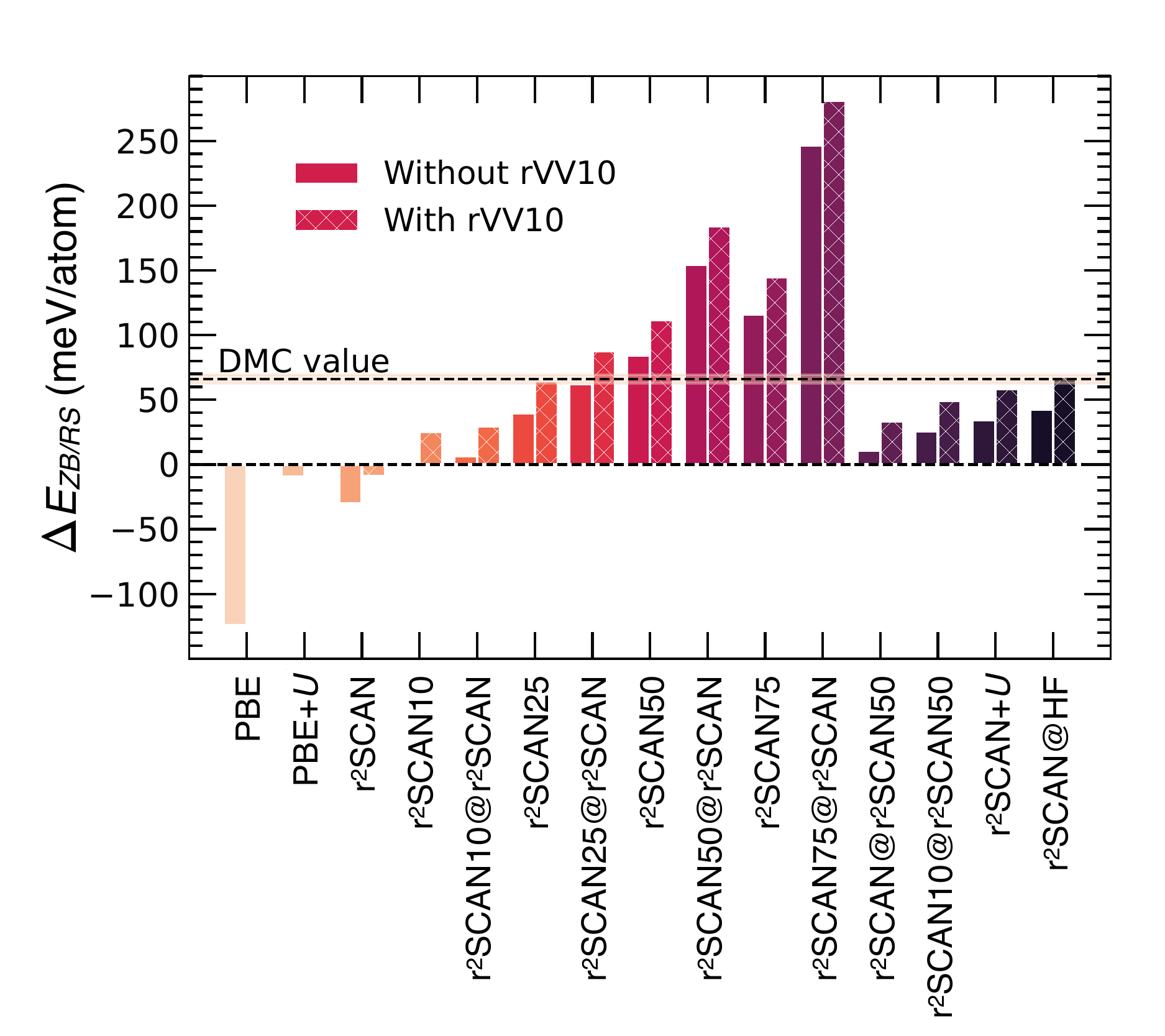}
\caption{ Predicted relative energy difference between zinc blende (ZB) and rocksalt (RS) phases of MnO ($\Delta E_{ZB/RS}$) with a variety of r{\textsuperscript{2}}SCANY@r{\textsuperscript{2}}SCANX DFAs. Results for rVV10 van der Waals-corrected  r{\textsuperscript{2}}SCANY@r{\textsuperscript{2}}SCANX analogs are also shown.   Values of $\Delta E_{ZB/RS}$  were computed using total energies from PAW potentials of GW type (as in Ref.~\cite{lany_rpa_2013}), where electrons from only {\emph{3d}} and {\emph{4s}} orbitals are explicitly considered. See {\bf Supplementary~Fig.~8} for dependence of $\Delta E_{ZB/RS}$ on PAW potentials.
}
\label{fig:7}
\end{figure}

A key quality of XC functionals is their ability to predict the correct polymorphism of transition metal oxides accurately.  A known issue with traditional DFAs concerns transition-metal monoxides MnO and CoO, which are predicted to be stable in the zincblende (ZB) phase rather than the experimentally observed rocksalt (RS) structure.{\cite{lany_rpa_2013,mno_dmc_2015,u_vdw_mno_haowei}}  This failure is common among GGA (PBE,  see {\textbf{Fig.}}~{\ref{fig:7}}), GGA+$U$ (or PBE+$U$),  the range-separated hybrid HSE06, SCAN, and r$^2$SCAN DFA.{\cite{lany_rpa_2013,mno_dmc_2015,u_vdw_mno_haowei}} We will show that the r\textsuperscript{2}SCANY@r\textsuperscript{2}SCANX DFA  reproduces the correct polymorphism of MnO.

In low-temperature experiments, MnO adopts a slightly distorted antiferromagnetic RS structure, while the ZB phase is not the bulk ground state.} {As shown in {\bf Fig.~\ref{fig:7}}, PBE predicts the ZB structure as more stable than the RS polymorph, with a negative $\Delta E_{ZB/RS}$$\approx$--244~meV/f.u.{\cite{u_vdw_mno_haowei}} Albeit with less negative values, r{\textsuperscript{2}}SCAN, and van der Waals corrected r{\textsuperscript{2}}SCAN+rVV10 cannot reproduce the correct MnO polymorphism.} {Adding a Hubbard $U$ (PBE+$U$) or separately a pairwise vdW correction (TS-vdW) to the DFA reduces this error but still does not recover the RS ground state for MnO.} {Peng and Perdew showed that the correct polymorphism ($\Delta E_{ZB/RS}$$\approx$88~meV/f.u.) is recovered by PBE+$U$+vdW.{\cite{u_vdw_mno_haowei}}} {Peng and Lany showed that the random phase approximation (RPA) can recover the RS phase as the ground state for MnO, with an energy difference $\Delta E_{ZB/RS}$$\approx$+131~meV/f.u.\ (+65.5~meV/atom) as shown in {\bf Fig.~\ref{fig:7}}.{\cite{lany_rpa_2013}}} {This was confirmed by diffusion Monte Carlo (DMC) simulations, with $\Delta E_{ZB/RS}$=132$\pm$6.5~meV/f.u.\ (66~meV/atom).{\cite{mno_dmc_2015}}} {Ref.~ \cite{u_vdw_mno_haowei} demonstrated that SCAN+rVV10+$U${\ produces a $\Delta E_{ZB/RS} \sim$135~meV/f.u., reproducing the DMC and RPA data.}

{{\bf Fig.~\ref{fig:7}}  shows the predicted $\Delta E_{ZB/RS}$ with the r{\textsuperscript{2}}SCANY@r{\textsuperscript{2}}SCANX DFA. } {All r{\textsuperscript{2}}SCANY@r{\textsuperscript{2}}SCANX variants shown, predict a positive $\Delta E_{ZB/RS}$ (in meV/atom), correctly stabilizing RS MnO over the ZB polymorph, and in agreement with existing RPA/DMC benchmarks.{\cite{lany_rpa_2013,mno_dmc_2015}}} {The inclusion of the nonlocal rVV10 correlation systematically increases $\Delta E_{ZB/RS}$, pushing the r{\textsuperscript{2}}SCANY@r{\textsuperscript{2}}SCANX and r{\textsuperscript{2}}SCAN+{\emph{U}} predictions toward the DMC (RPA) value. Here we
have used a version of rVV10 adapted for 
r{\textsuperscript{2}}SCAN.{\cite{Ning2022}}  {\bf Fig.~\ref{fig:7}} shows that r{\textsuperscript{2}}SCAN25@r{\textsuperscript{2}}SCAN, r{\textsuperscript{2}}SCAN10@r{\textsuperscript{2}}SCAN50, and r{\textsuperscript{2}}SCAN@HF (and their rVV10 van der Waals analogs)  are in excellent quantitative agreement with DMC and RPA references.}

Previous work has demonstrated that GGA+{\emph{U}}, meta-GGA (SCAN), and range-separated hybrids (HSE06) all predict the correct polymorphs for FeO and NiO.{\cite{u_vdw_mno_haowei}}  We further confirm these predictions for the antiferromagnetic NiO configuration with r{\textsuperscript{2}}SCAN and the selected r{\textsuperscript{2}}SCANY@r{\textsuperscript{2}}SCANX  (see  \textbf{Supplementary~Fig.~9}). The relative stability of the nonmagnetic wurtzite ZnO phase to its RS and ZB polymorphs appears correctly described by r{\textsuperscript{2}}SCAN and r{\textsuperscript{2}}SCANY@r{\textsuperscript{2}}SCANX and in better quantitative agreement with RPA{\cite{lany_rpa_2013}} than GGA results (\textbf{Supplementary~Fig.~10}).


\section{Discussion}

 Standard DFAs suffer from significant SIEs, especially when applied to
strongly correlated open-shell transition metal oxides
(M\textsubscript{i}O\textsubscript{j}s). SIE leads to inaccuracies in
predicting many essential properties of
M\textsubscript{i}O\textsubscript{j}s, including their structural
parameters, electronic and magnetic structures, and thermochemical data.

Hybrid XC functionals mix a fraction of exact HF exchange with DFA to
improve the accuracy of the electronic structure, thereby correcting
functional-driven errors, particularly the SIE. The amount of HF
exchange remains a tunable parameter. A number of strategies have been proposed to tune the
amount of HF exchange in the DFA.\cite{stephens_ab_1994,becke_new_1993, perdew_rationale_1996,adamo_toward_1999,heyd_hybrid_2003,skone_nonempirical_2016,jaramillo_local_2003,janesko_local_2007,henderson_range_2008,janesko_self-consistent_2008} 
We
have shown that M\textsubscript{i}O\textsubscript{j}s' oxidation
energies generally improve when climbing the ladder from LSDA to PBE to
r\textsuperscript{2}SCAN, but further improvement is needed.

Unsurprisingly, we have shown that admixing exact HF exchange with
r\textsuperscript{2}SCAN achieves better electronic, magnetic, and
oxidation energies of M\textsubscript{i}O\textsubscript{j}s. For
example, r\textsuperscript{2}SCAN10 (with 10\% HF exchange) reduced the
error in oxidation energies by $\sim$40\% and by
$\sim$38\% for band gaps of
M\textsubscript{i}O\textsubscript{j}s compared to
r\textsuperscript{2}SCAN. r\textsuperscript{2}SCAN10 also reduces
overbinding in O\textsubscript{2,} which is important for the accurate
prediction of oxidation energies of
M\textsubscript{i}O\textsubscript{j}s.

\subsection{ Electron Densities from Hartree-Fock, Hybrids, and 
DFT+\emph{U} Approaches} 
HF also provides self-interaction-free, albeit uncorrelated, electronic
charge densities. Electronic charge densities from HF or hybrid
functionals can {overcorrect or} correct density-driven errors. We have demonstrated that
HF (or hybrid functional) electronic charge densities can correct
r\textsuperscript{2}SCAN (or its hybrids) total energies for
M\textsubscript{i}O\textsubscript{j}s by proposing a generalized
r\textsuperscript{2}SCANY@r\textsuperscript{2}SCANX method. We have found optimal
combinations of HF exchange in the charge density and functional
definitions, such as r\textsuperscript{2}SCAN10,
r\textsuperscript{2}SCAN10@r\textsuperscript{2}SCAN,
r\textsuperscript{2}SCAN@r\textsuperscript{2}SCAN50, and
r\textsuperscript{2}SCAN10@r\textsuperscript{2}SCAN50, which reduce the
mean absolute (and relative) errors of predicted electronic and magnetic
properties, and oxidation energies of
M\textsubscript{i}O\textsubscript{j}s.

We have demonstrated that
r\textsuperscript{2}SCANY@r\textsuperscript{2}SCANX can match or
outperform  r\textsuperscript{2}SCAN+\emph{U}.  In predicting the oxidation energies of
M\textsubscript{i}O\textsubscript{j}s, the
r\textsuperscript{2}SCAN10@r\textsuperscript{2}SCAN50 performed best,
with a mean absolute error of $\sim${0.43}~ eV/O\textsubscript{2},
lower than the r\textsuperscript{2}SCAN+\emph{U} {0.57}
eV/O\textsubscript{2}. This can be rationalized as independently
correcting the functional-driven error with 10\% exact HF exchange and
the density-driven error with 50\% HF exchange.

Kulik and collaborators demonstrated that the ability of DFT+\emph{U} to
localize electrons on transition metals can vary significantly from that
of hybrid functionals; these latter tend to localize the minority spin
density (of the transition metal) away from the metal, and towards
oxygen atoms in transition metal complexes.\cite{gani_where_2016,zhao_where_2018} We
observe the same behavior in M\textsubscript{i}O\textsubscript{j}s. In
agreement with Kulik \emph{et al}.,\cite{gani_where_2016,zhao_where_2018} we also
show that a higher fraction of HF exchange in the XC functional
increases the on-site magnetic moments on transition metals in
M\textsubscript{i}O\textsubscript{j}s, which is rationalized by the rise	
of majority spin in the \emph{d}-manifolds. Considering these
observations, we can qualitatively state that an increasing amount of HF
in the DFA increases the ionic character of
M\textsubscript{i}O\textsubscript{j}s. We quantify these behaviors for
Fe\textsubscript{2}O\textsubscript{3} {\bf Supplementary~Fig.~11} by
change in minority- and majority-spin electron numbers. In addition,
{\bf Supplementary~Fig.~12} shows a progressive increase in the
electron number on the oxygen sites with increasing X in
r\textsuperscript{2}SCANX. This is preceded by a similar increase from
LSDA to PBE to r\textsuperscript{2}SCAN. The electron number at the
transition-atom site decreases with increasing values of X. Exceptions to this trend are the M\textsubscript{i}O\textsubscript{j}s where strong charge
disproportionation of the transition metal is favored, for example,
Fe\textsubscript{3}O\textsubscript{4} (Fe\textsuperscript{3+} and
Fe\textsuperscript{2+}), Mn\textsubscript{3}O\textsubscript{4}
(Mn\textsuperscript{3+} and Mn\textsuperscript{4+}). 

In addition, in transition metal complexes, Ref.~\onlinecite{zhao_where_2018}
concluded that DFT+\emph{U} charge densities might differ substantially
from hybrid functionals. To test this observation, we compare
r\textsuperscript{2}SCAN@r\textsuperscript{2}SCAN+\emph{U}
M\textsubscript{i}O\textsubscript{j}s band gaps and oxidation energies
({\bf Supplementary~Fig.~13}) to r\textsuperscript{2}SCAN and
r\textsuperscript{2}SCAN10@r\textsuperscript{2}SCAN50 results. We
demonstrate that
r\textsuperscript{2}SCAN@r\textsuperscript{2}SCAN+\emph{U} oxidation
energies and band gaps are comparable to r\textsuperscript{2}SCAN.

In contrast, when using HF densities with r\textsuperscript{2}SCAN, as
in r\textsuperscript{2}SCAN@HF, we have shown that
M\textsubscript{i}O\textsubscript{j} oxidation energies worsen compared
to r\textsuperscript{2}SCAN. MAEs increased from {1.09}~eV/O\textsubscript{2} at r\textsuperscript{2}SCAN to {1.59}~eV/O\textsubscript{2} at r\textsuperscript{2}SCAN@HF. But, we
demonstrated that the MAE in M\textsubscript{i}O\textsubscript{j}
oxidation energies can be halved (w.r.t r\textsuperscript{2}SCAN@HF) to
$\sim${0.75}~eV/O\textsubscript{2} when implementing hybrid
electronic charge densities of r\textsuperscript{2}SCAN admixed with
50\% HF exchange as in
r\textsuperscript{2}SCAN@r\textsuperscript{2}SCAN50.

\subsection{Sources of Uncertainty and Error \label{sec:eroors}}
The widely accepted standard for representing uncertainties in
experimentally obtained oxidation enthalpies and other
thermochemical data requires estimates of the 95\% confidence intervals,
which all thermochemical tables universally
follow.\cite{wagman_selected_1971,Kubaschewski1979,thomas_c_allison_nist-janaf_2013} Ruscic  argued that the MAE 
of calculated formation energies relative to experimental ones underestimates the error of the 
calculated values.\cite{ruscic_uncertainty_2014} For this reason,
when benchmarking the accuracy of
r\textsuperscript{2}SCANY@r\textsuperscript{2}SCANX (Fig.~\ref{fig:3}), we used
 the root mean square error.

Reproducing errorless O\textsubscript{2} binding energies is important
for accurate predictions of oxidation energies of
M\textsubscript{i}O\textsubscript{j}s. LSDA and PBE drastically overbind
the O\textsubscript{2} molecule with large negative binding
energies,\cite{wang_oxidation_2006,perdew_self-interaction_1981,blochl_first-principles_2000} $\sim$--2.2~eV/O\textsubscript{2} and $\sim$--1~eV/O\textsubscript{2,}
respectively, resulting in a systematic shift in predicted oxidation
energies. With r\textsuperscript{2}SCAN, this negative error drops in
magnitude to a sizeable value of $\sim$--0.3~eV/O\textsubscript{2}. Applying a constant shift (of --0.3~eV/O\textsubscript{2}) to correct the spurious O\textsubscript{2}
overbinding in r\textsuperscript{2}SCAN would further underestimate
M\textsubscript{i}O\textsubscript{j} oxidation energies (see {\bf Fig.~\ref{fig:2}a}), further exacerbating the error in
r\textsuperscript{2}SCAN predictions. Previously, similar corrections
have been applied to PBE O\textsubscript{2} and
M\textsubscript{i}O\textsubscript{j} {formation}
energies.\cite{wang_oxidation_2006,jain_formation_2011} Notably, we have demonstrated that
r\textsuperscript{2}SCAN10@r\textsuperscript{2}SCAN,
r\textsuperscript{2}SCAN10, and
r\textsuperscript{2}SCAN10@r\textsuperscript{2}SCAN50 decrease the error
in the O\textsubscript{2} binding energy to values \(\leq\) 0.03~eV/O\textsubscript{2} in magnitude, and without requiring any fitting
procedures.\cite{wang_oxidation_2006,jain_formation_2011}
 
Since the errors of O\textsubscript{2} binding energy are minimal with
r\textsuperscript{2}SCANY@r\textsuperscript{2}SCANX methods, errors in
the oxidation energies of M\textsubscript{i}O\textsubscript{j}s must
reside in the prediction of the reduced and oxidized oxides. This is a
bimodal distribution of r\textsuperscript{2}SCAN oxidation energies
({\bf Fig.~\ref{fig:2}b}), consisting of a low (near-zero) peak for early
transition metals (Ti, V, Cr) reactions, and higher errors
for late transition metals (Mn, Fe, Co) reactions.
Early transition metals (Ti, V, and Cr) with a small number of
\emph{d-}electrons in their electronic configurations probably reduce
correlation effects, which are more prominent for the late transition
metals (Mn, Fe, and Co) with more {\emph{d}} electrons. Furthermore, redox reactions involving transition metals' \emph{d}-shell changes, such as
Ti\textsuperscript{4+}(\emph{d}\textsuperscript{0}) $\rightarrow$
Ti\textsuperscript{3+} (\emph{d}\textsuperscript{1}), or
Cu\textsuperscript{+} (\emph{d}\textsuperscript{10}) $\rightarrow$
Cu\textsuperscript{2+} (\emph{d}\textsuperscript{9}), appear to be in
better agreement with experimental data. Milder correlation
effects in closed-shell transition metals make r\textsuperscript{2}SCAN
sufficient to describe the electronic structure of their oxides (TiO\textsubscript{2}, V\textsubscript{2}O\textsubscript{5},
\emph{etc}.).

From these observations, we expect that the electron density by
r\textsuperscript{2}SCAN may be too delocalized to capture localization
effects in late transition metals, \emph{i.e.}, Mn, Fe, Co, with a
larger number of \emph{d} electrons. Better electronic charge densities
as provided by r\textsuperscript{2}SCAN@r\textsuperscript{2}SCAN50,
correct the electron number in the \emph{d} manifolds, recovering a
unimodal distribution ({\bf Fig.~\ref{fig:2}b}), with a substantial decrease
of oxidation energies errors for M\textsubscript{i}O\textsubscript{j}s
with M~=~Mn, Fe, and Co. In addition, due to the functional-driven error,
the average error of r\textsuperscript{2}SCAN@r\textsuperscript{2}SCAN50 is
non-zero, resulting in an offset distribution around the zero-error
line. In r\textsuperscript{2}SCAN10@r\textsuperscript{2}SCAN50
the functional-driven error reduces, and the unimodal distribution
centers near the zero-error line.

Magnetic moments are also typically impacted by SIE and are
underestimated due to the overdelocalization of \emph{d}(\emph{f})
electrons on transition metals by GGA and meta-GGA functionals.
As we climb the Jacob's ladder\cite{perdew_jacobs_2001} of the XC
functionals from LSDA (PW91) to GGA, and to meta-GGAs, the error in
magnetic moments decreases accordingly, as shown by the functional
MAE distributions of {\bf Fig.~\ref{fig:5}a} and {\bf Fig.~\ref{fig:5}b}. We
observed that the error in magnetic moments drops slowly as the
percentage of HF exchange in the orbital increases. Compared to
r\textsuperscript{2}SCAN, for r\textsuperscript{2}SCAN50, the error in
magnetic moment reduces by $\sim$29\%, and for HF, the error
reduces by $\sim$34\%. In contrast,
r\textsuperscript{2}SCAN+\emph{U} minimizes the error in magnetic
moments {only by} $\sim$17\%. Note that the variance in
experimentally reported magnetic moments can be as high as +1.00
\(\mu_{B}\). For example, magnetic moments in V\textsuperscript{3+} in
V\textsubscript{2}O\textsubscript{3} are reported to vary between 1.2 to
2.37~\(\mu_{B}\) ({\bf Table~\ref{tab:1}}),\cite{moon_antiferromagnetism_1970,shin_observation_1992} which
corresponds to variations larger than 100\%. Note, we have neglected the orbital component of the magnetic moment and considered only the spin component. It has been shown that the orbital component can be substantial in specific cases. For example,  0.6--1~{\(\mu_{B}\)} for FeO,{\cite{Svane1990,tran_hybrid_2006,Radwanski2008,Schroen2013}}
 1--1.6~{\(\mu_{B}\)} for CoO,{\cite{Svane1990,tran_hybrid_2006,Radwanski2008,Schroen2013,Solovyev1998,Shishidou1998,Neubeck2001,Jauch2002,Ghiringhelli2002,Radwanski2004,Boussendel2010}} 0.3--0.45~{\(\mu_{B}\)} for NiO,{\cite{Svane1990,Radwanski2008,Neubeck2001,Radwanski2004,Fernandez1998}}  but can have negligible contributions in other oxides.{\cite{Tran2020}} Despite this considerable uncertainty in experimentally reported magnetic moments and neglected orbital components, we observed a noticeable decrease in MAE of M\textsubscript{i}O\textsubscript{j}s
predicted magnetic moments from r\textsuperscript{2}SCAN to
r\textsuperscript{2}SCAN50.

Band gaps of M\textsubscript{i}O\textsubscript{j}s are typically
underestimated by standard DFT predictions, with absolute errors rapidly
decreasing as LSDA (PW92) \textgreater{} GGA (PBE) $\gg$ meta-GGAs
(r\textsuperscript{2}SCAN and LAK) \textgreater{} hybrid meta-GGA
(r\textsuperscript{2}SCANX), see {\bf Fig.~\ref{fig:6}c} and {\bf Fig.~\ref{fig:6}d}. Using r\textsuperscript{2}SCANY@r\textsuperscript{2}SCANX, we
demonstrated that errors in band gaps are minimized with approximately
7--11\% exact exchange in the functional and the orbital definitions,
which differs from the optimal HF fractions minimizing oxidation energy
errors of M\textsubscript{i}O\textsubscript{j}s. We have shown that band
gaps predicted by r\textsuperscript{2}SCAN10@r\textsuperscript{2}SCAN
are more accurate than r\textsuperscript{2}SCAN+\emph{U}, with MAEs of
0.55~eV/O\textsubscript{2} and 0.74~eV/O\textsubscript{2}, respectively.

\subsection{r\textsuperscript{2}SCANY@r\textsuperscript{2}SCANX as a Generalization of Single-Shot and Density-Corrected Approaches}

The r\textsuperscript{2}SCANY@r\textsuperscript{2}SCANX approach resembles single-shot hybrid schemes, which serve as cost-effective proxies for fully self-consistent hybrid calculations in large-scale solid-state simulations with supercells containing hundreds or thousands of atoms. Single-shot hybrid approaches start with a self-consistent calculation using an inexpensive (semi)local DFA to get initial charge density and orbitals, then perform a one-shot non-self-consistent evaluation with a superior DFA, often a hybrid functional, based on those initial orbitals.\cite{Alkauskas2007,Tran2012,Lany2013,Lany2017,Bischoff2019,Bischoff2019a,Bauers2019,Yang2022,Gant2022,Cordell2022,Gorai2021,Sharan2022,Xiong2023}

{Single-shot approaches have been proven to produce band-gap-corrected single-particle and quasiparticle energies approaching the quality of hybrid DFAs or even GW quality for semiconductor alloys,{\cite{Alkauskas2007,Tran2012,Lany2017,Bischoff2019,Bischoff2019a,Bauers2019,Yang2022,Gant2022}} transition metal oxides,{\cite{Tran2012,Lany2013}} disordered phases,{\cite{Cordell2022}} point defects,{\cite{Gorai2021}} and interfaces.{\cite{Sharan2022,Xiong2023}}}

Single-shot approaches can significantly improve prediction accuracy for errors in the DFA, like inaccuracies in predicting orbital energies. Using a single non-self-consistent hybrid (or more advanced) evaluation based on a semi-local density functional can produce better orbital energies at much lower computational cost than a fully self-consistent hybrid calculation. This resembles the “single-shot” G$_0$W$_0$ variant of GW, employing a more accurate self-energy operator non-iteratively on orbitals from inexpensive DFT calculations.

We generalized these approaches by replacing the single-shot hybrid-type with parameterized global hybrid r\textsuperscript{2}SCANY and r\textsuperscript{2}SCANX for density, yielding r\textsuperscript{2}SCANY@r\textsuperscript{2}SCANX that addresses both functional and density-driven inaccuracies.

It is worth noting that r{\textsuperscript{2}}SCANY@r{\textsuperscript{2}}SCANX is also a more general form of the density--corrected DFT, also known as DFA@HF, previously proposed in the literature.{\cite{clementi_comparative_1990,gill_investigation_1992,scuseria_comparison_1992,oliphant_systematic_1994,janesko_hartreefock_2008,verma_increasing_2012,kanungo_unconventional_2024}}

\subsection{False Metals and Structure Symmetry Breaking}
Materials exhibit many degrees of freedom in their crystal structures. Deformation of cation polyhedra, polyhedra
tilting, and Jahn-Teller distortions all decrease the crystal symmetry
in materials.\cite{zhang_symmetry-breaking_2020,xiong_symmetry_2025} Specific magnetic
orderings of transition metals in M\textsubscript{i}O\textsubscript{j}s,
accessible by neutron scattering and magnetic spectroscopies,
further reduce crystal symmetries. On the one hand, standard structural
techniques, such as X-ray and neutron diffraction, generally show less
sensitivity to such distortions and defects. As such, the reported crystal
structures may be inherently more symmetric than the actual structures.
On the other hand, band-gap measurements are sensitive to
symmetry-breaking motifs, defects in materials, and charge and magnetic
orderings. Zunger and collaborators demonstrated that DFT of overly
symmetrized M\textsubscript{i}O\textsubscript{j}s and other materials
tends to close band gaps, predicting ``false
metals''.\cite{zhang_symmetry-breaking_2020,trimarchi_polymorphous_2018,varignon_origin_2019,xiong_symmetry_2025,Zunger2025}

Starting from the highly symmetrized experimental structures, further
optimized with r\textsuperscript{2}SCAN, our  r\textsuperscript{2}SCAN band-gap predictions of
Ti\textsubscript{2}O\textsubscript{3},\cite{abrahams_magnetic_1963}
V\textsubscript{2}O\textsubscript{3},\cite{dernier_crystal_1970} and
Fe\textsubscript{3}O\textsubscript{4}\cite{wright_charge_2002} is $\sim$0.0~eV ({\bf Table~\ref{tab:1}}). Upon removing all possible symmetry layers
(perturbing atomic positions from high-symmetry sites of large supercell
models, imposing ground state magnetic orderings, removing intrinsic
symmetry of wavefunctions, orbitals, and time reversal),
Ti\textsubscript{2}O\textsubscript{3},
V\textsubscript{2}O\textsubscript{3}, and
Fe\textsubscript{3}O\textsubscript{4} remain metallic. However, the
experimentally reported band gaps for
Ti\textsubscript{2}O\textsubscript{3},
V\textsubscript{2}O\textsubscript{3}, and
Fe\textsubscript{3}O\textsubscript{4} are smaller than or equal to
$\sim$0.2~eV ({\bf Table~\ref{tab:1}}), which makes band-gap opening
unlikely, even with substantial symmetry breaking as demonstrated here. We found no structure symmetry breaking for Ti\textsubscript{2}O\textsubscript{3} and V\textsubscript{2}O\textsubscript{3}, but observed it for Fe\textsubscript{3}O\textsubscript{4}. 

There can
be multiple symmetry breakings,\cite{zhang_symmetry-breaking_2020,trimarchi_polymorphous_2018} and we may not
have found the lowest-energy one. However,  {\bf Supplementary~Table~4}
shows that (without structure symmetry breaking) non-zero gaps in these materials arise in the self-consistent hybrid r\textsuperscript{2}SCAN10.

\subsection{Computational Efficiency Considerations of r\textsuperscript{2}SCANY@r\textsuperscript{2}SCANX Approaches}  

We comment on the computational costs of
r\textsuperscript{2}SCANY@r\textsuperscript{2}SCANX.
With VASP, the self-consistent (SCF) global hybrids, such as r\textsuperscript{2}SCAN10, are $\sim$10 to 100 times more expensive than typical SCF r\textsuperscript{2}SCAN calculations.
However, using the r\textsuperscript{2}SCAN10@r\textsuperscript{2}SCAN, which gives similar accuracy as SCF r\textsuperscript{2}SCAN10 the cost can be as low as $\sim$2-5 times that of SCF r\textsuperscript{2}SCAN, since it only requires one non-self-consistent evaluation of r\textsuperscript{2}SCAN10.
Methods such as r\textsuperscript{2}SCAN@r\textsuperscript{2}SCAN50 and r\textsuperscript{2}SCAN10@r\textsuperscript{2}SCAN50 have a similar computational cost as global hybrids, as they require SCF
r\textsuperscript{2}SCAN50 orbitals but provide improved accuracy over a
 SCF r\textsuperscript{2}SCAN10. The qualitative trend in the computational cost required for  r\textsuperscript{2}SCANY@r\textsuperscript{2}SCANX is as follows: 
r\textsuperscript{2}SCAN $<$ r\textsuperscript{2}SCANY@r\textsuperscript{2}SCAN $\ll$ r\textsuperscript{2}SCANX $\approx$ r\textsuperscript{2}SCAN@r\textsuperscript{2}SCANX $\approx$ r\textsuperscript{2}SCANY@r\textsuperscript{2}SCANX.

\section{Conclusions}

We have found that prediction accuracy for M{\textsubscript{i}}O{\textsubscript{j}} improves as more exact constraints and suitable non-bonded norms are met by approximating the density functional for exchange-correlation energy. 
We have developed 
r\textsuperscript{2}SCANY@r\textsuperscript{2}SCANX to mitigate the
 self-interaction error of exchange and correlation
functionals for the accurate simulations of electronic, magnetic, and
thermochemical properties of transition metal oxides.
r\textsuperscript{2}SCANY@r\textsuperscript{2}SCANX uses different fractions of exact exchange to define energy and density simultaneously:  X affects the electronic density, and Y changes the density functional approximation of the total energy, addressing density- and functional-driven errors of the energy.  While Y=10 and X=50 are fitted or optimized for the oxidation energies of the transition-metal oxides, they are also the values expected from experience with \emph{s-p}-bonded systems, as in the next-to-last paragraph of Sec.~{\ref{sec:introduction}}.

In r\textsuperscript{2}SCANY@r\textsuperscript{2}SCANX, we found a dependence of the X and Y optimal percentages of exact
Hartree--Fock exchange justified by their
performance on the  M\textsubscript{i}O\textsubscript{j} oxidation energies, their magnetic moments, and band
gaps. Predicted uncertainties are minimized for:
M\textsubscript{i}O\textsubscript{j}s' oxidation energies by
r\textsuperscript{2}SCAN10@r\textsuperscript{2}SCAN50, and band gaps
with r\textsuperscript{2}SCAN10@r\textsuperscript{2}SCAN. These properties improve from LSDA to PBE to r{\textsuperscript{2}}SCAN to r{\textsuperscript{2}}SCAN10@r{\textsuperscript{2}}SCAN. Replacing a small part of r$^2$SCAN with the same amount of Hartree-Fock exchange is likely to have minimal impact on satisfying DFA constraints.

r\textsuperscript{2}SCANY@r\textsuperscript{2}SCANX improves predictions, outperforming the DFT(r\textsuperscript{2}SCAN)+\emph{U}  commonly used for strongly correlated materials. 
r\textsuperscript{2}SCAN10@r\textsuperscript{2}SCAN is computationally more affordable than hybrid
functionals while maintaining comparable accuracy for oxidation energies and band gaps. Further studies should investigate the transferability of r\textsuperscript{2}SCANY@r\textsuperscript{2}SCANX to other correlated systems, such as ionically bound transition-metal fluorides or polyanion systems (\emph{e.g.}, phosphates and silicates) with strong covalent character.

Self-consistent r\textsuperscript{2}SCAN10,
already reduces the MAE of the oxidation energies to $\sim${0.66}
eV~ eV/O\textsubscript{2}. r\textsuperscript{2}SCAN10 improves upon
r\textsuperscript{2}SCAN, for transition-metal oxides and
for \emph{sp}-bonded systems.\cite{maniar_atomic_2025} r\textsuperscript{2}SCAN10
even gives oxide band gaps slightly better than
r\textsuperscript{2}SCAN+\emph{U}. For the prototype Mott insulator NiO, r{\textsuperscript{2}}SCAN10@r{\textsuperscript{2}}SCANX gives a remarkably accurate band gap for X=0 or 10.  To reduce the MAE of oxidation energies in M{\textsubscript{i}}O{\textsubscript{j}}s to $\sim$0.43 eV/O{\textsubscript{2}}, r{\textsuperscript{2}}SCAN10@r{\textsuperscript{2}}SCAN50 is needed, which lowers the density-driven error of r{\textsuperscript{2}}SCAN and r{\textsuperscript{2}}SCAN10.
The r\textsuperscript{2}SCAN10@r\textsuperscript{2}SCAN is nearly as accurate as the self-consistent r\textsuperscript{2}SCAN10 hybrid for oxidation energies. 

%


This and recent works{\cite{kanungo_unconventional_2024,kaplan_how_2024,pangeni_hartree-fock_2025,kaplan_understanding_2023,gubler_accuracy_2025,maniar_atomic_2025}} suggest that, in many cases with main-group and transition-metal elements, the functional-driven errors of r\textsuperscript{2}SCAN energy differences can far exceed small density-driven errors. This allows significant improvements in r\textsuperscript{2}SCAN energy differences by performing the costly self-consistent iteration and geometry optimization with the efficient r\textsuperscript{2}SCAN (with van der Waals correction), then applying a more expensive nonlocal functional (\emph{e.g.}, a hybrid) in a single post-self-consistent calculation, as in r\textsuperscript{2}SCAN10@r\textsuperscript{2}SCAN.

 Standard GGA global hybrids require approximately 25{\%} of exact exchange. SCAN and r$^2$SCAN meta-GGAs have less self-interaction error than GGAs, with the most accurate self-consistent SCAN hybrid for main-group molecules using 10{\%} exact exchange,{\cite{Santra2021}}  aligning with r$^2$SCAN10 to reduce functional-driven errors in transition-metal oxides. The functional-driven error dominates the density-driven{\cite{kim_understanding_2013}} error of the energy in main-group 
molecules.{\cite{kanungo_unconventional_2024,kaplan_how_2024,pangeni_hartree-fock_2025}} 
The density-driven error is smaller and less sensitive to many density features, as expected from the variational principle for DFA energy, but more sensitive to long-range charge-transfer errors than other density errors.\cite{pangeni_hartree-fock_2025} 50\% of exact exchange reduces the density-driven errors of the energy in transition-metal oxides, 
 as in main-group molecules.{\cite{kanungo_unconventional_2024,kaplan_how_2024,pangeni_hartree-fock_2025}} }

{For structural energy differences in M\textsubscript{i}O\textsubscript{j}s, RPA is the 
accepted standard,{\cite{lany_rpa_2013,Pines_2016}} capturing the correct long-range physics. RPA yields realistic van der Waals interactions and accurate long-range electron transfer in ionic materials, including transition-metal oxides, and corrects energy differences between polymorphs.  
RPA includes exact exchange and nearly exact long-range nonlocal correlation, but it is inaccurate at short distances.
It is known that RPA provides very accurate electron densities for molecules,{\cite{qhr1-788v,9wpv-qlwk}}  surpassing r$^2$SCAN and Hartree-Fock when compared to CCSD(T). The reason for that is unclear, but the RPA error for short-range correlation does not affect the density.
RPA is similar to coupled-cluster double excitations, which keep the short-range or exchange-like correction to the correlation energy that RPA ignores. RPA, CCD, CCSD, and CCSD(T) are all accurate for the density, with accuracy likely improving in that order. The same correct long-range physics should be found in CCSD and CCSD(T).
Our r$^2$SCAN10@r$^2$SCAN50+rVV10 appears to capture the same physics in a different way.  While r{\textsuperscript{2}}SCAN50 strongly improves electron transfer relative to r{\textsuperscript{2}}SCAN, it is expected to be less accurate than r{\textsuperscript{2}}SCAN for other features of the electron density to which the density-driven error of the energy is insensitive. 

%
We believe that r$^2$SCAN10@r$^2$SCAN50  may accurately predict total energies and band gaps for many non-metallic solids with \emph{s}, \emph{p}, or \emph{d} electrons, and may be similarly accurate for molecules. r$^2$SCAN is already reasonably accurate for many molecular and material geometries. It exhibits intermediate-range van der Waals interactions but requires a long-range correction, particularly for layered geometries. r$^2$SCAN10@r$^2$SCANX+rVV10 could be accurate for total and single-electron energies in many nonmetallic or weakly metallic systems with \emph{s}, \emph{p}, and \emph{d} valence electrons.

\section{First-Principles Calculation Details}\label{sec:DFTmethod}

 All calculations presented here are performed using the DFT
formalism, as implemented in the Vienna Ab initio Simulation Package
(VASP).\cite{kresse_ab_1993,kresse_efficiency_1996,kresse_efficient_1996} The PAW potentials describe the core
electrons.\cite{blochl_projector_1994,kresse_ultrasoft_1999} The electrons from 3\emph{s},
3\emph{p}, 3\emph{d}, and 4\emph{s} orbitals are explicitly considered
for the transition metal atoms. Using a PAW set that treats fewer valence electrons explicitly  (3\emph{p}, 3\emph{d}, and 4\emph{s})
significantly impacts M\textsubscript{i}O\textsubscript{j} oxidation
energies and band gaps ({\bf Supplementary~Fig.~14}). The kinetic energy cutoff for the plane waves
was set to 700 eV, and the total energy was converged to 10$^{-6}$~eV per cell. Various DFT exchange and correlation
functionals were used in their collinear spin-polarized implementation.
These are the LSDA-PW92,\cite{perdew_accurate_1992} the PBE,\cite{perdew_generalized_1996} the 
r\textsuperscript{2}SCAN,\cite{furness_accurate_2020} and
LAK.\cite{lebeda_balancing_2024}  Global hybrid
r\textsuperscript{2}SCANX\cite{gill_investigation_1992} calculations were 
performed with percentages X of Hartree-Fock exchange as discussed in
the results. Ground-state magnetic orderings reported
experimentally ({\bf Table~\ref{tab:1}}), often derived from neutron
diffraction experiments, were imposed in all simulations of
M\textsubscript{i}O\textsubscript{j}s.

Geometries (coordinates, volumes, and cell shapes) converged when the
forces on all atoms were lower than 0.01~eV/\AA. All properties were
calculated with r\textsuperscript{2}SCAN geometries unless explicitly
mentioned. 
All calculations in this work use a $\Gamma$-centered Monkhorst-Pack\cite{monkhorst_special_1976} grid. For structure relaxation, a $k$-grid with a density of 48 $k$-points per \AA\textsuperscript{$-$1} is used for all systems. All other calculations used a \emph{k}-grid with a density of approximately 700 $k$-points per reciprocal atom.

The global hybrid r\textsuperscript{2}SCANX orbitals were evaluated at
the following percentages of exact Hartree-Fock exchange, with X~=~0\%,
10\%, 25\%, 50\%, 75\%, and 100\%, respectively. The non-self-consistent
field r\textsuperscript{2}SCANY energies were assessed with a fine grid
of Y values, from 0\% to 100\%, with 5\% intervals. In
r\textsuperscript{2}SCANY@r\textsuperscript{2}SCANX calculations, we
started from self-consistently converged r\textsuperscript{2}SCANX
orbitals (\texttt{WAVECAR} in VASP), which are used for a
non-self-consistent single electronic step (\texttt{ALGO = EIGENVAL})
with the r\textsuperscript{2}SCANY functional. A linear interpolation
scheme was employed to obtain the energies, band gaps, and magnetic
moments  at intermediate fractions of exact HF exchange, not
explicitly calculated.

Due to the high computational costs of global hybrids used in this
work, all DFT calculations are performed using unit cells that can
accommodate the expected experimental magnetic ordering for each
transition metal oxide. On-site magnetic moments of each transition
metal atom were obtained by integrating the spin density within the
projection radius of the PAW potentials. Changes of such radii have a
negligible influence on the numerical values of the magnetic moment.

For estimating band gaps with
r\textsuperscript{2}SCANY@r\textsuperscript{2}SCANX, we
utilize the orbitals obtained self-consistently with
r\textsuperscript{2}SCANX, subsequently, and apply the
r\textsuperscript{2}SCANY functional non-self-consistently on the
same orbitals to obtain the one-electron eigenenergies. These
one-electron eigenenergies are then used to calculate the band gap.

\section*{Author contributions}
P.C.\ and J.P.P.\ designed and supervised the project. H.R.G. prepared and performed the simulations, data collection, and data analysis. P.C., J.P.P., J.S., R.Z., Y.W., A.R., and A.P.\ contributed to the data analysis. H.R.G.\ and P.C.\ wrote the first draft of the manuscript. All authors contributed to the discussion and the final version of this manuscript.

\section*{Competing Interests}
 The authors declare that they have no competing interests.

\section*{Additional Information}
\noindent {\bf Supplementary Information.} The online version contains supplementary material available at

\section*{Data Availability}
\noindent All the computational data associated with this study, including the input and output files of the simulations, are available on Zenodo at \url{https://doi.org/10.5281/zenodo.15741824}.

\begin{acknowledgements}
\noindent The Welch Foundation is acknowledged for providing P.C. a Robert A.
Welch Professorship at the Texas Center for Superconductivity (grant No. L--E--001--19921203) and the
Welch Foundation grant No.\ E--2227--20250403. We are grateful for the
support of the Research Computing Data Core at the University of Houston
for assistance with the calculations carried out in this work. J.P.P.\
acknowledges support from the National Science Foundation under grant
DMR--2426275 and the {Department of Energy, Office of Science, Basic Energy Sciences, under grant
DE--SC--0018331. A.R. acknowledges support from the Department of Energy. Office of Science, Basic Energy Sciences, under grant DE-SC0026293.}
J.S. acknowledges support from the National Science
Foundation under grant DMR--2042618. We acknowledge Dr. Hong Tang for pointing out that Ti$_2$O$_3$
is dimerized and is a diamagnetic system.
We thank the referees for suggesting calculations and clarifications.
\end{acknowledgements}

\bibliography{biblio}

\pagebreak
\widetext
\begin{center}
\textbf{\large ---Supplemental Materials--- \\
Reducing Self-Interaction Error in Transition-Metal Oxides with
Different Exact-Exchange Fractions for Energy and Density}

\vspace{1.5em}  

Harshan Reddy Gopidi$^{1,2}$,
Ruiqi Zhang$^{3}$,
Yanyong Wang$^{3}$,
Abhirup Patra$^{4}$,
Jianwei Sun$^{3}$,
Adrienn Ruzsinszky$^{3}$,
John P. Perdew$^{3,*}$,
Pieremanuele Canepa$^{1,2,\dagger}$\\[0.7em]

$^{1}$\it{Department of Electrical and Computer Engineering, \\University of Houston, Houston, TX 77204, USA}\\
$^{2}$\it{Texas Center for Superconductivity, \\University of Houston, Houston, TX 77204, USA}\\
$^{3}$\it{Department of Physics and Engineering Physics, \\Tulane University, New Orleans, LA 70118, USA}\\
$^{4}$\it{Shell International Exploration and Production Inc., Houston, TX 77082, USA}\\

\end{center}

\begin{center}
$^{*}${perdew@tulane.edu}\\
$^{\dagger}${pcanepa@uh.edu}\\
\end{center}
\setcounter{equation}{0}
\setcounter{figure}{0}
\setcounter{table}{0}
\setcounter{page}{1}
\makeatletter
\renewcommand{\figurename}{{\bf Supplementary Figure}}
\renewcommand{\tablename}{{\bf Supplementary Table}}
\renewcommand{\theequation}{\bf{arabic{equation}}}
\renewcommand{\theHequation}{S{\arabic{equation}}}
\renewcommand{\thefigure}{{\bf \arabic{figure}}}
\renewcommand{\theHfigure}{S{\arabic{figure}}}
\renewcommand{\thetable}{{\bf \arabic{table}}}
\renewcommand{\theHtable}{S{\arabic{table}}}

\author{Harshan Reddy Gopidi}
\affiliation{%
 Department of Electrical and Computer Engineering, University of Houston, Houston, TX 77204, USA
}%
\affiliation{
Texas Center for Superconductivity, University of Houston, Houston, TX 77204, USA
}%

\author{Ruiqi Zhang}
\affiliation{
Department of Physics and Engineering Physics, Tulane University, New Orleans, LA 70118, USA
}%

\author{Yanyong Wang}
\affiliation{
Department of Physics and Engineering Physics, Tulane University, New Orleans, LA 70118, USA
}%

\author{Abhirup Patra}
\affiliation{
Shell International Exploration and Production Inc., Houston, TX 77082, USA
}%

\author{Jianwei Sun}
\affiliation{
Department of Physics and Engineering Physics, Tulane University, New Orleans, LA 70118, USA
}%

\author{Adrienn
Ruzsinszky}
\affiliation{
Department of Physics and Engineering Physics, Tulane University, New Orleans, LA 70118, USA
}%

\author{John P. Perdew}
\email{perdew@tulane.edu}
\affiliation{
Department of Physics and Engineering Physics, Tulane University, New Orleans, LA 70118, USA
}%

\author{Pieremanuele Canepa}
\email{pcanepa@uh.edu}
\affiliation{%
 Department of Electrical and Computer Engineering, University of Houston, Houston, TX 77204, USA
}%
\affiliation{
Texas Center for Superconductivity, University of Houston, Houston, TX 77204, USA
}%

\patchcmd{\frontmatter@abstract@produce}
  {\vskip200\p@\@plus1fil
   \penalty-200\relax
   \vskip-200\p@\@plus-1fil}
  {}
  {}
  {}
\makeatother


\newpage
\begin{figure*}[ht]
\centering
 \includegraphics[width=1.0\textwidth]{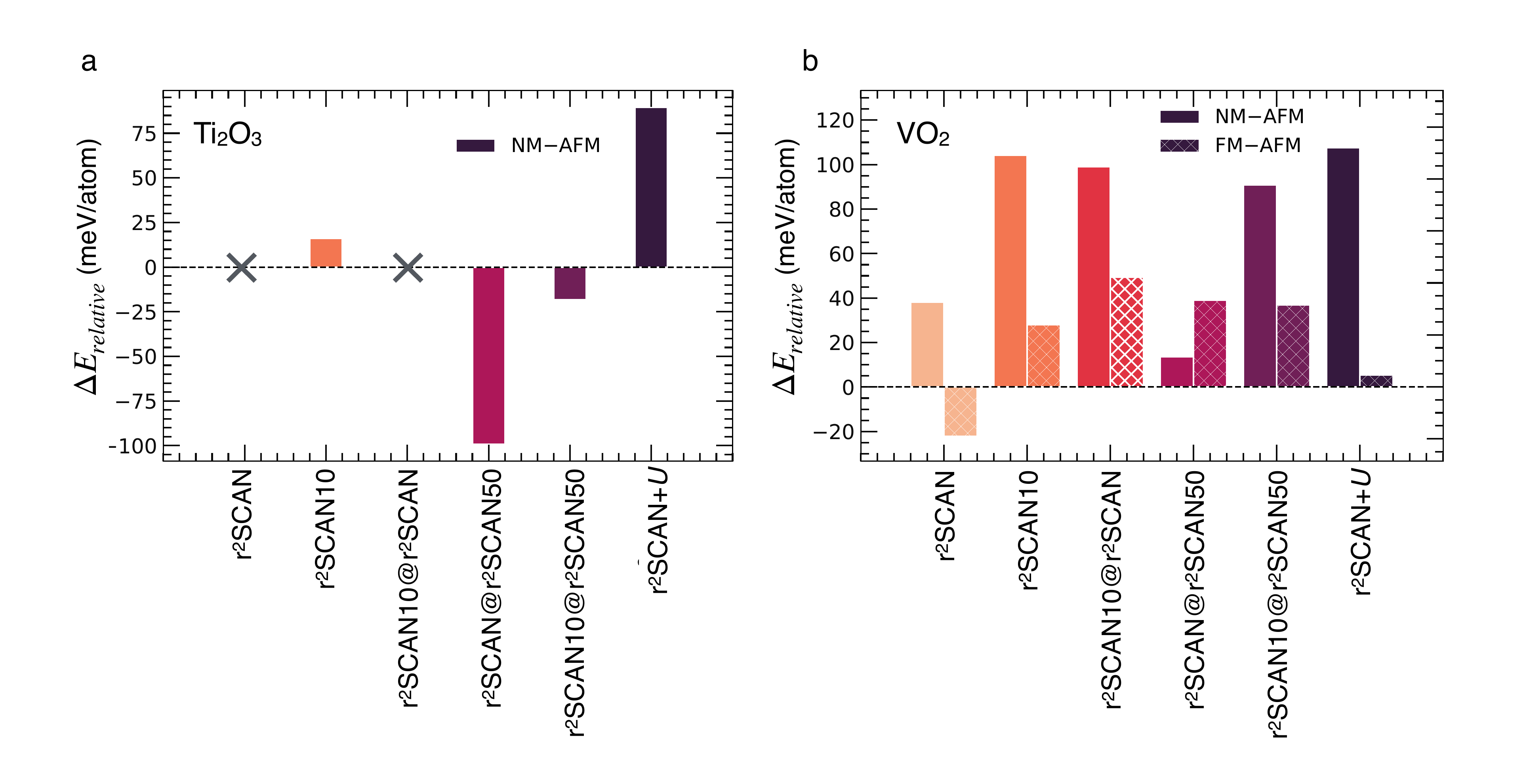}
\caption{Stabiltiy of various magnetic configurations in dimerized (a) Ti$_2$O$_3$ and (b) VO$_2$ using r$^{2}$SCANY@r$^{2}$SCANX approaches.  Here as elsewhere, we
use the r$^{2}$SCAN geometries. In {\bf Table 1} of the 
manuscript, we have used the magnetic state 
that achieves the lowest energy in our best 
hybrid, r$^{2}$SCAN10@r$^{2}$SCAN50: NM for Ti$_2$O$_3$
and AFM for VO$_2$.
}
\label{fig:s_afm_vs_nm}
\end{figure*}

\newpage
\begin{figure*}[t]
\centering
 \includegraphics[width=0.6\textwidth]{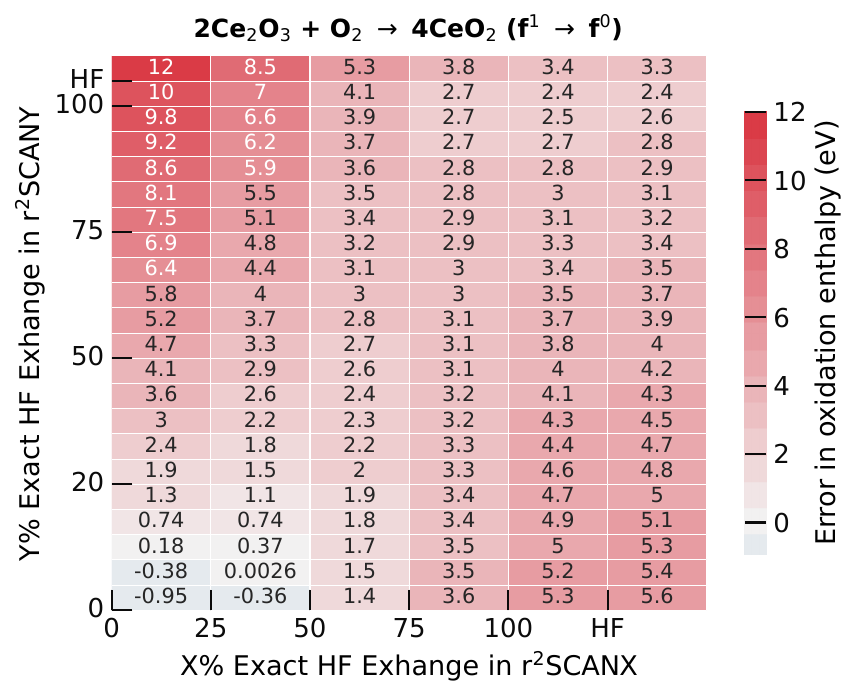}
\caption{Error in oxidation enthalpy of oxidation reaction of Cerium oxides with r$^2$SCANY@r$^2$SCANX method. Oxidation reaction is indicated. }\label{fig:s1_Ce}
\end{figure*}

\newpage
\begin{figure*}[t]
\centering
 \includegraphics[width=1.0\textwidth]{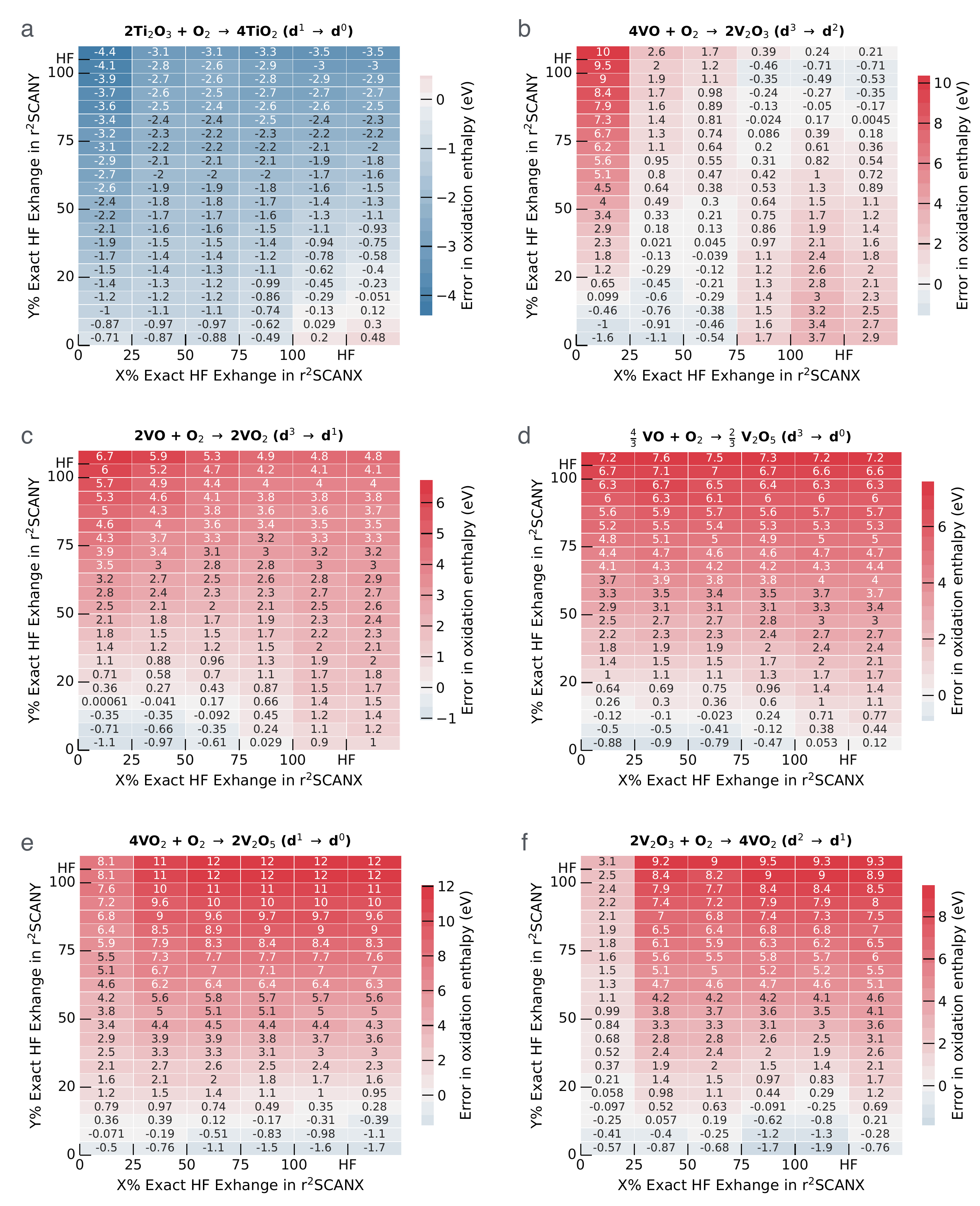}
\caption{{\bf(a-f)} Error in oxidation enthalpies of oxidation reactions between the selected M$_\mathrm{i}$O$_\mathrm{j}$s with r$^2$SCANY@r$^2$SCANX method. Oxidation reactions are indicated. }\label{fig:s1af}
\end{figure*}

\newpage
\begin{figure*}[t]
\centering
 \includegraphics[width=1.0\textwidth]{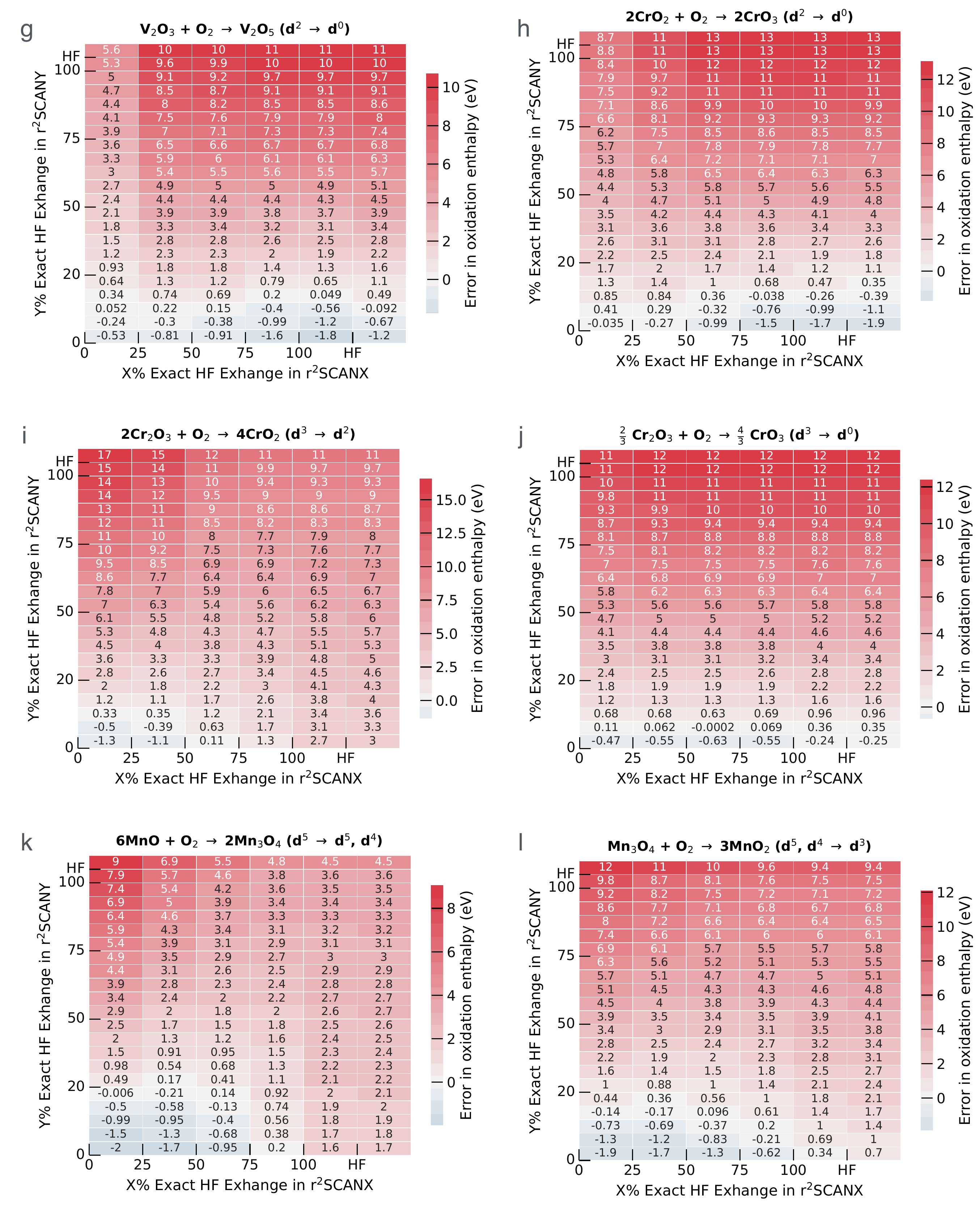}
\caption{{\bf(g-l)} Error in oxidation enthalpies of oxidation reactions between the selected M$_\mathrm{i}$O$_\mathrm{j}$s with r$^2$SCANY@r$^2$SCANX method. Oxidation reactions are indicated. }\label{fig:s1gl} 
\end{figure*}

\begin{figure*}[t]
\centering
 \includegraphics[width=1.0\textwidth]{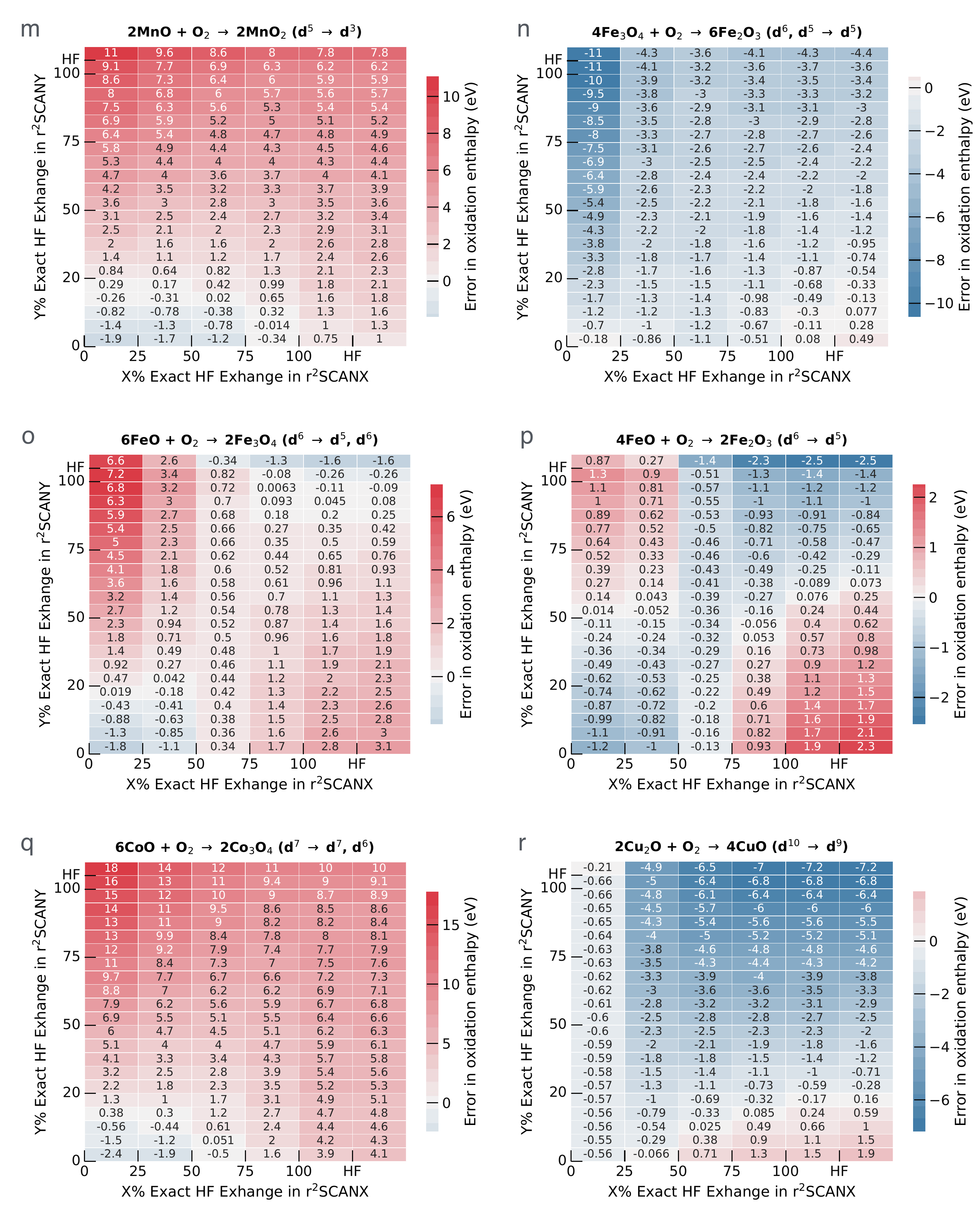}
\caption{{\bf(m-r)} Error in oxidation enthalpies of oxidation reactions between the selected M$_\mathrm{i}$O$_\mathrm{j}$s with r$^2$SCANY@r$^2$SCANX method. Oxidation reactions are indicated. }\label{fig:s1mo}
\end{figure*}

\newpage
\begin{figure*}[ht]
\centering
 \includegraphics[width=1.0\textwidth]{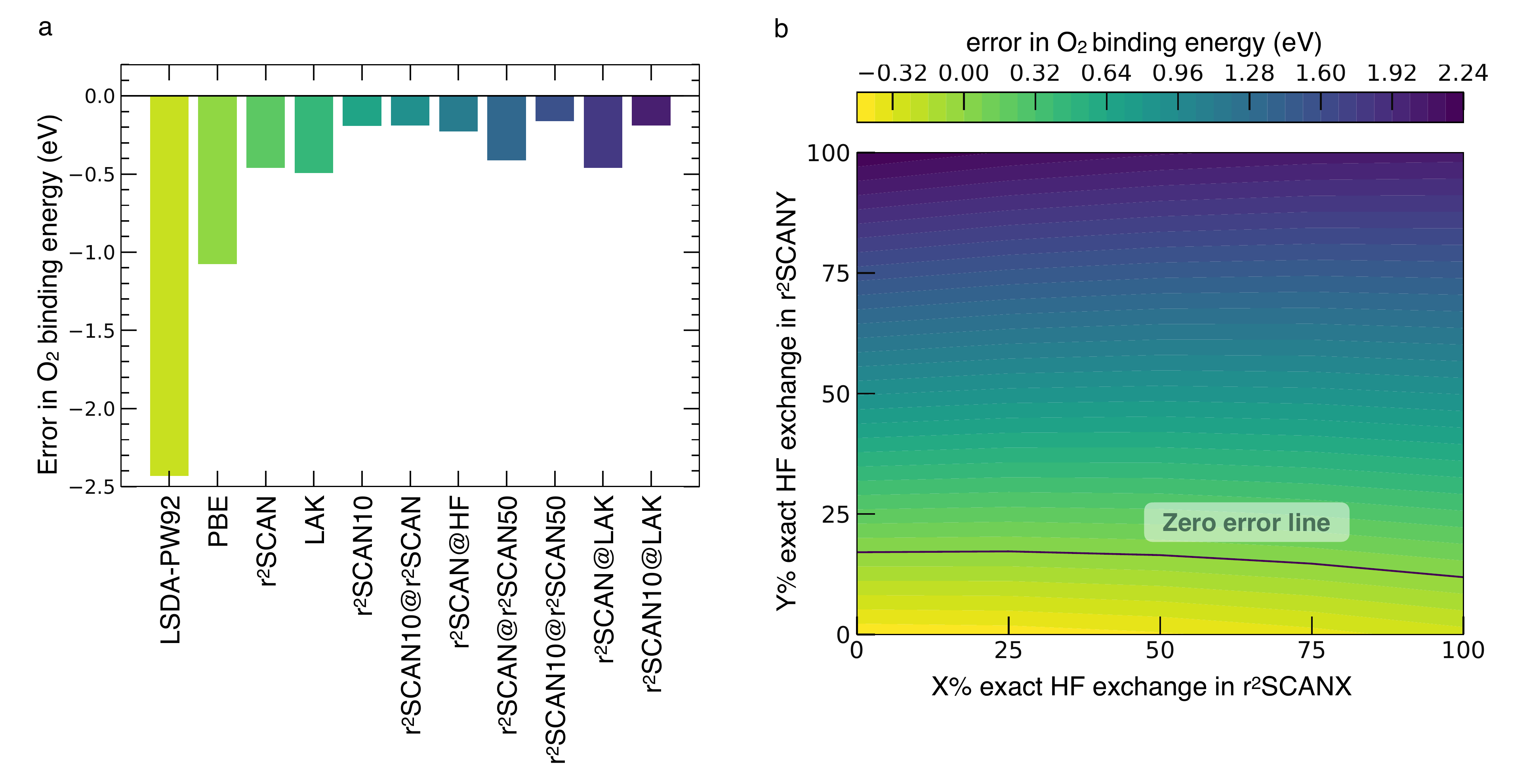}
\caption{r$^{2}$SCANY@r$^{2}$SCANX error in O$_2$ DFT binding energies calculated using hard Oxygen (O\_h) PAW potential. Panels {\bf a} and {\bf b} are obtained similarly to \textbf{Fig.~4} in the main text.}
\label{fig:s_o2_hard_paw}
\end{figure*}

\newpage
\begin{figure*}[t]
\centering
 \includegraphics[width=1.0\textwidth]{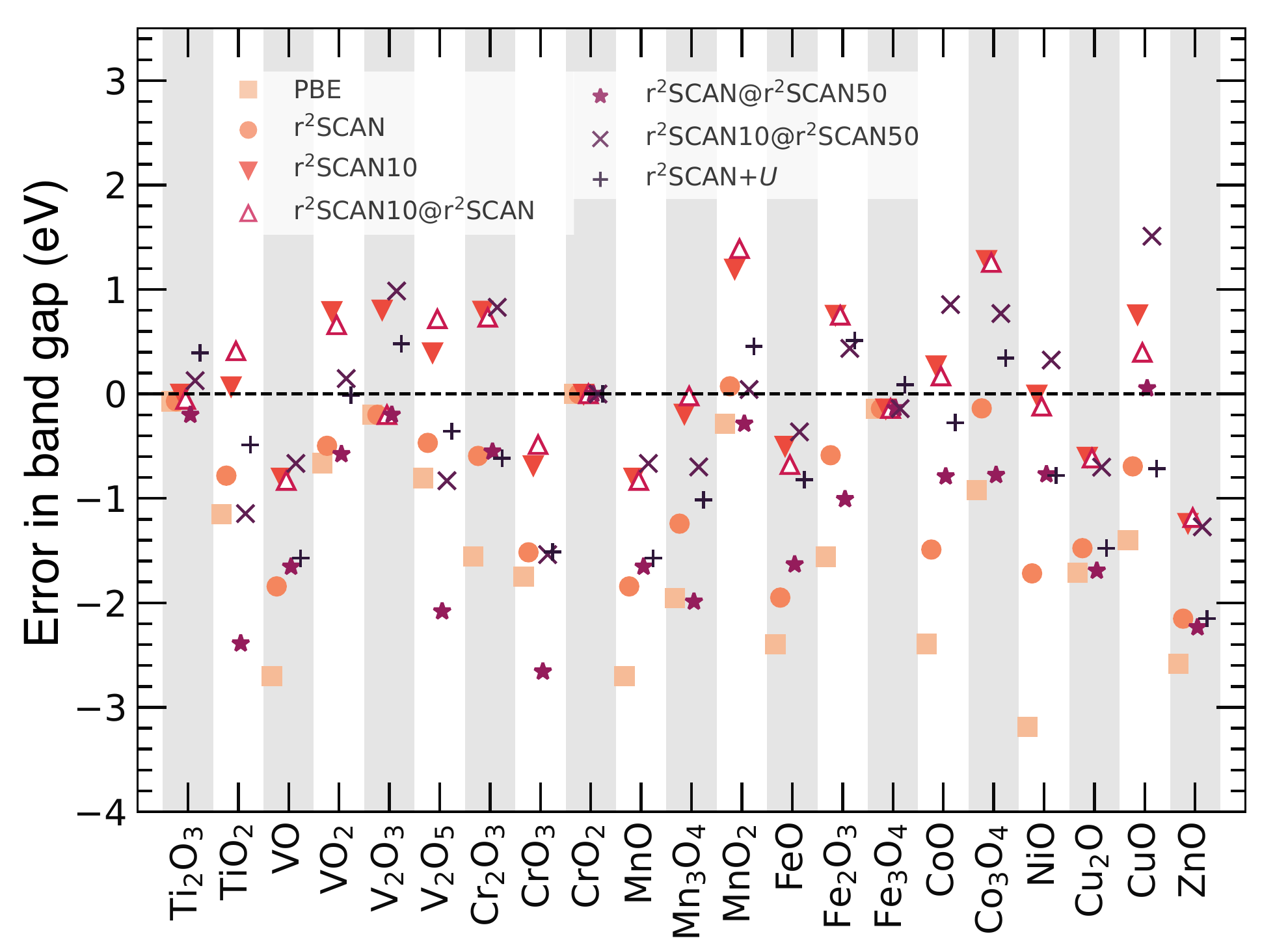}
\caption{Absolute error in band gaps for  M$_\mathrm{i}$O$_\mathrm{j}$s.}\label{fig:s2}
\end{figure*}

\newpage
\begin{figure*}[ht]
\centering
 \includegraphics[width=1.0\textwidth]{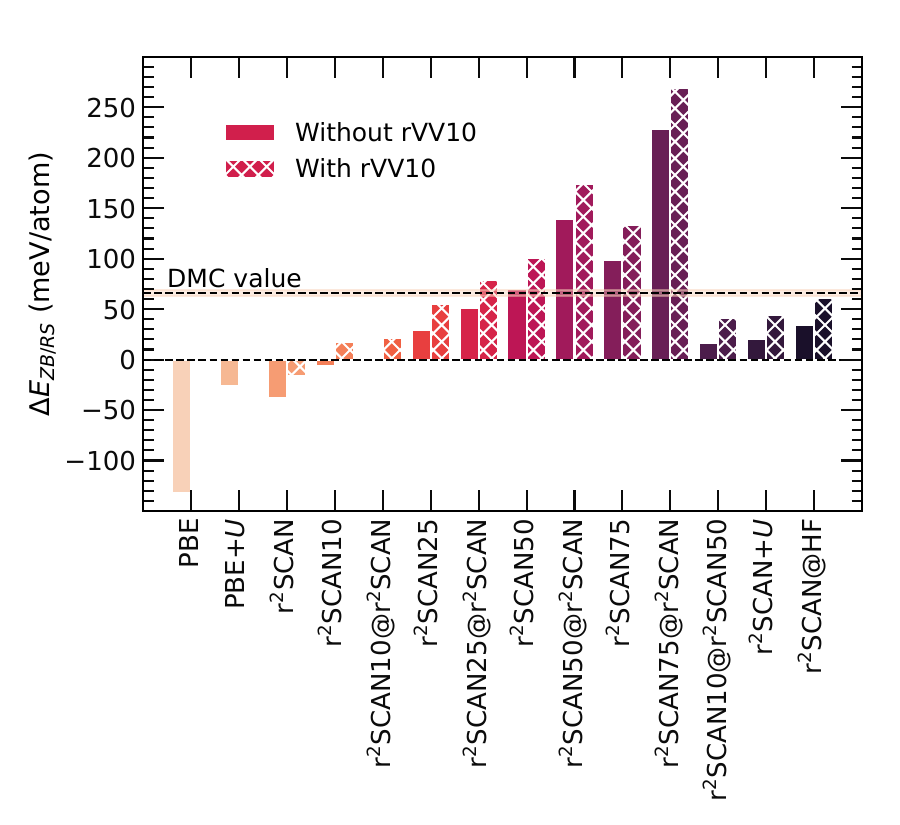}
\caption{ Predicted relative energy difference between zinc blende (ZB) and rocksalt (RS) phases of MnO ($\Delta E_{ZB/RS}$) with a variety of r{\textsuperscript{2}}SCANY@r{\textsuperscript{2}}SCANX DFAs. Results for rVV10 van der Waals-corrected  r{\textsuperscript{2}}SCANY@r{\textsuperscript{2}}SCANX analogs are also shown.  $\Delta E_{ZB/RS}$ were computed using total energies from PAW potentials where electrons from 3\emph{s} 3\emph{p} 3\emph{d} and 4\emph{s} orbitals are explicitly considered.}
\label{fig:s_MnO_RS_vs_ZB_sv}
\end{figure*}

\newpage
\begin{figure*}[ht]
\centering
 \includegraphics[width=1.0\textwidth]{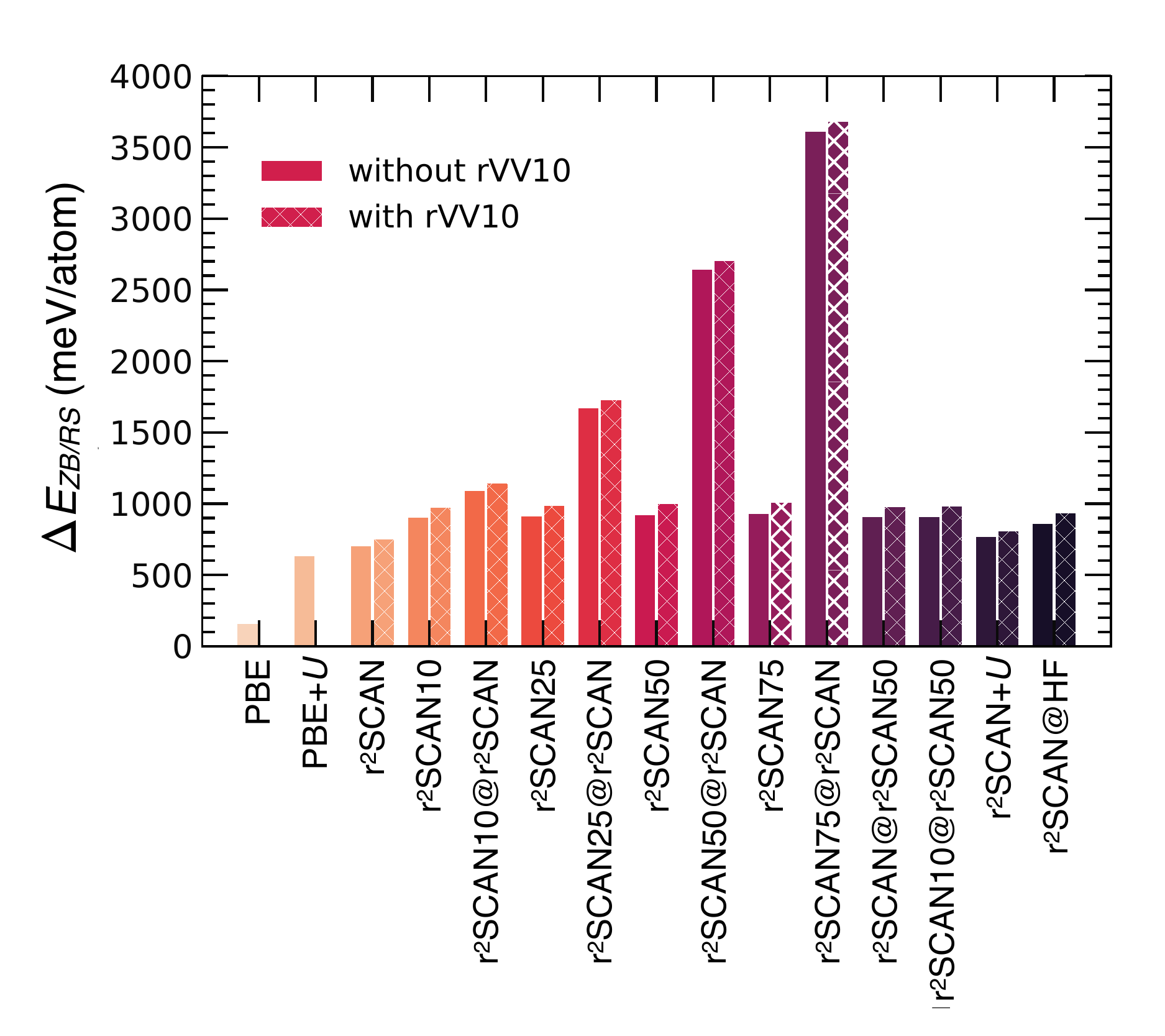}
\caption{ Predicted relative energy difference between zinc blende (ZB) and rocksalt (RS) phases of NiO ($\Delta E_{ZB/RS}$) with a variety of r{\textsuperscript{2}}SCANY@r{\textsuperscript{2}}SCANX DFAs. Results for rVV10 van der Waals-corrected  r{\textsuperscript{2}}SCANY@r{\textsuperscript{2}}SCANX analogs are also shown.  $\Delta E_{ZB/RS}$ were computed using total energies from GW potentials where electrons from 3\emph{d} and 4\emph{s} orbitals are explicitly considered.}
\label{fig:s_NiO_RS_vs_ZB}
\end{figure*}

\newpage
\begin{figure*}[ht]
\centering
 \includegraphics[width=1.0\textwidth]{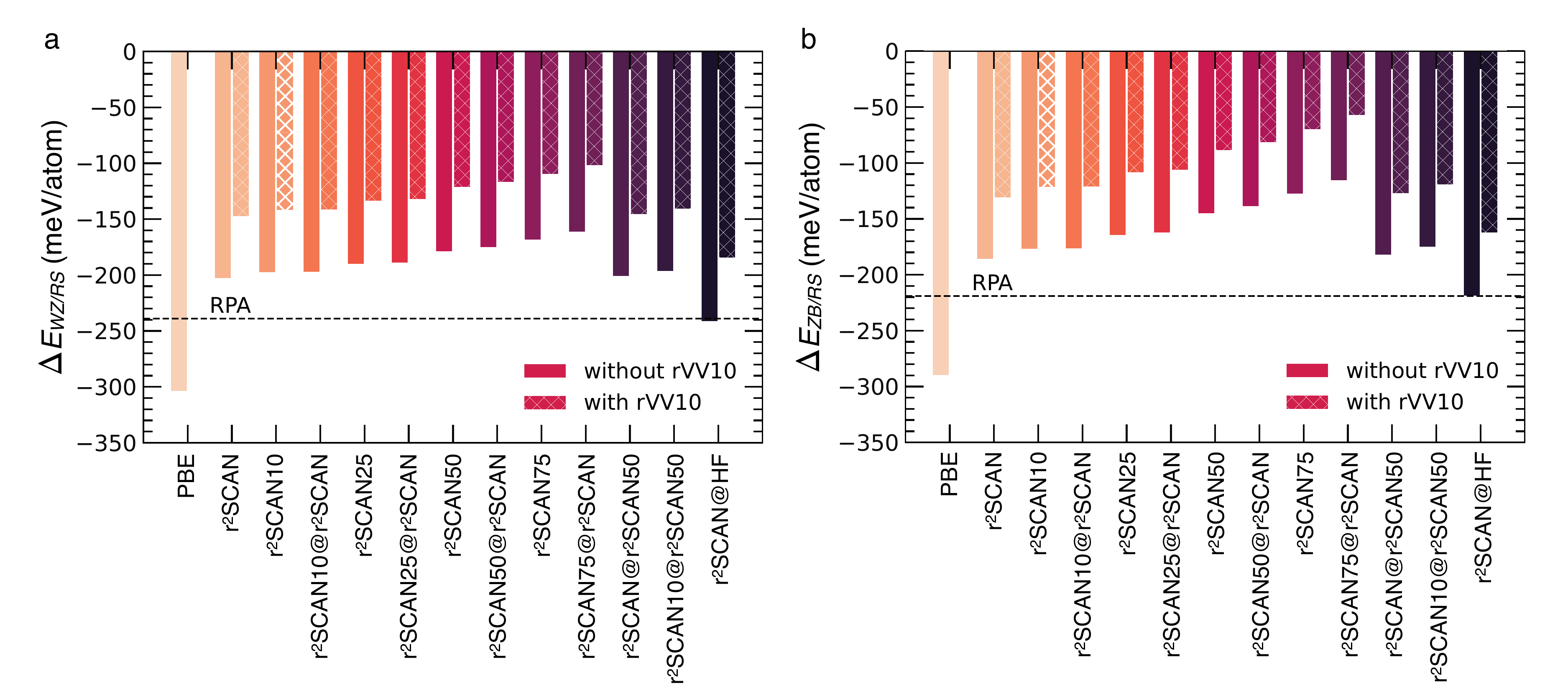}
\caption{ Predicted relative energy difference between (a) Wurtzite (WZ) and rocksalt (RS) phases (b) zinc blende (ZB) and rocksalt (RS) phases of ZnO ($\Delta E$) with a variety of r{\textsuperscript{2}}SCANY@r{\textsuperscript{2}}SCANX DFAs. Results for rVV10 van der Waals-corrected  r{\textsuperscript{2}}SCANY@r{\textsuperscript{2}}SCANX analogs are also shown.  $\Delta E_{ZB/RS}$ were computed using total energies from GW potentials where electrons from 3\emph{d} and 4\emph{s} orbitals are explicitly considered. RPA values are taken from Ref.{\cite{Peng2013}}}
\label{fig:s_ZnO_RS_ZB_WZ}
\end{figure*}

\newpage
\begin{figure*}[t]
\centering
 \includegraphics[width=1.0\textwidth]{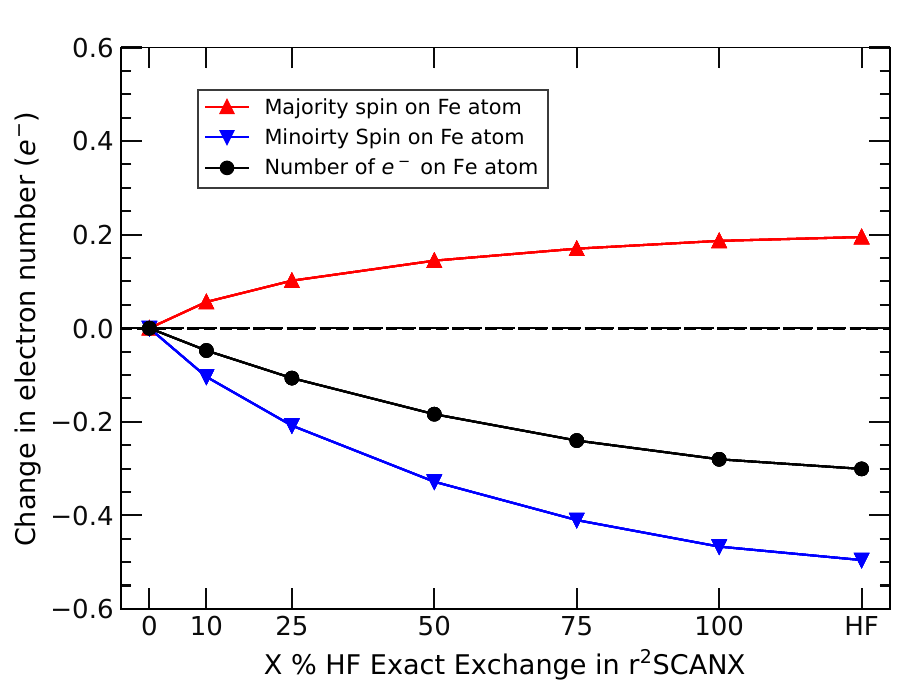}
\caption{Changes in electron numbers for minority spin and majority spin on the Fe atom in Fe$_2$O$_3$ (\(R\overline{3}c\)), as functions of the percentage X of exact exchange in r$^2$SCANX.}\label{fig:s3}
\end{figure*}

\newpage
\begin{figure*}[ht]
\centering
 \includegraphics[width=1.\textwidth]{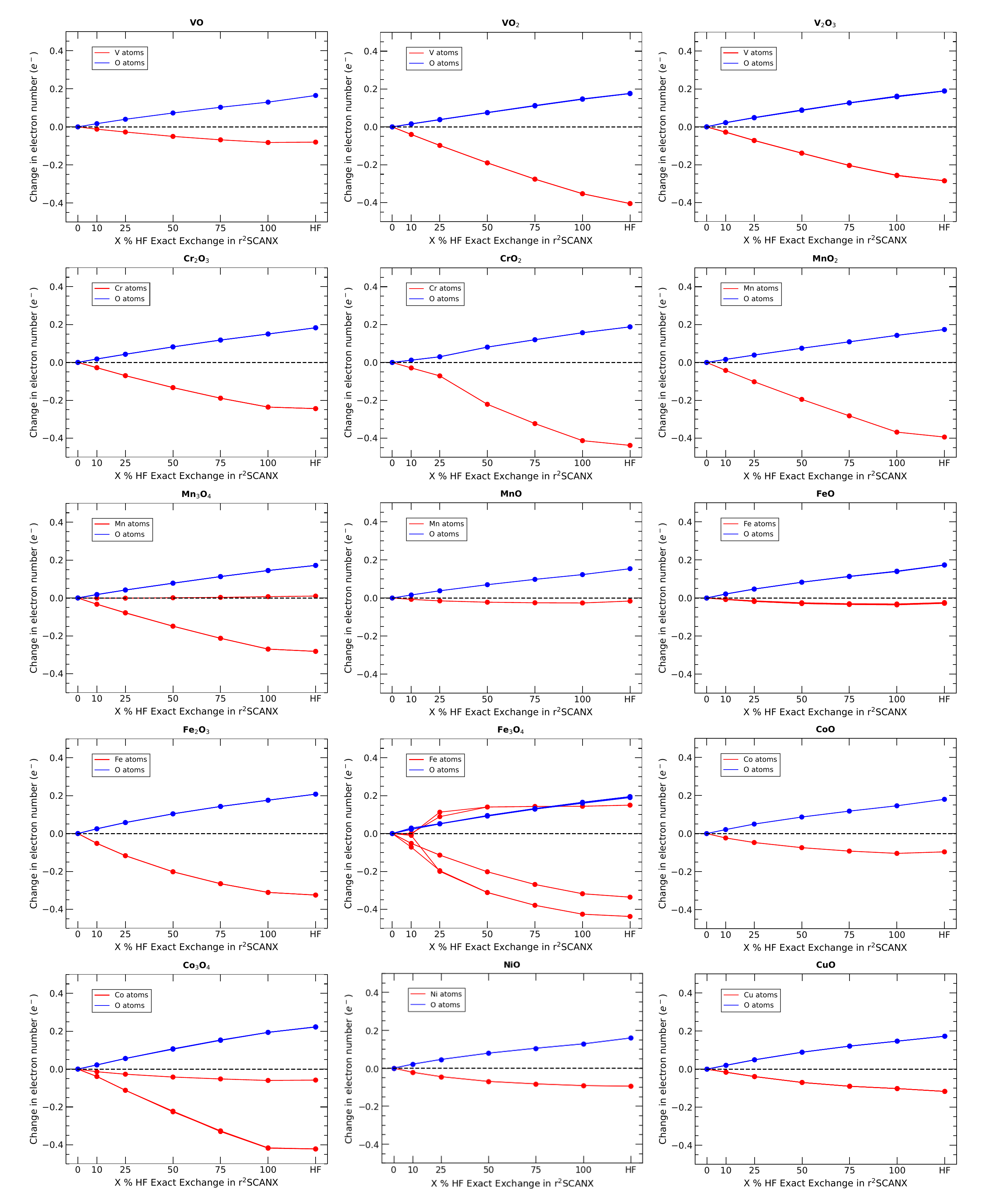}
\caption{Changes in electron numbers, in open-shell systems, as functions of the percentage X of exact exchange in r$^{2}$SCANX.}\label{fig:s4}
\end{figure*}

\newpage
\begin{figure*}[ht]
\centering
 \includegraphics[width=0.7\textwidth]{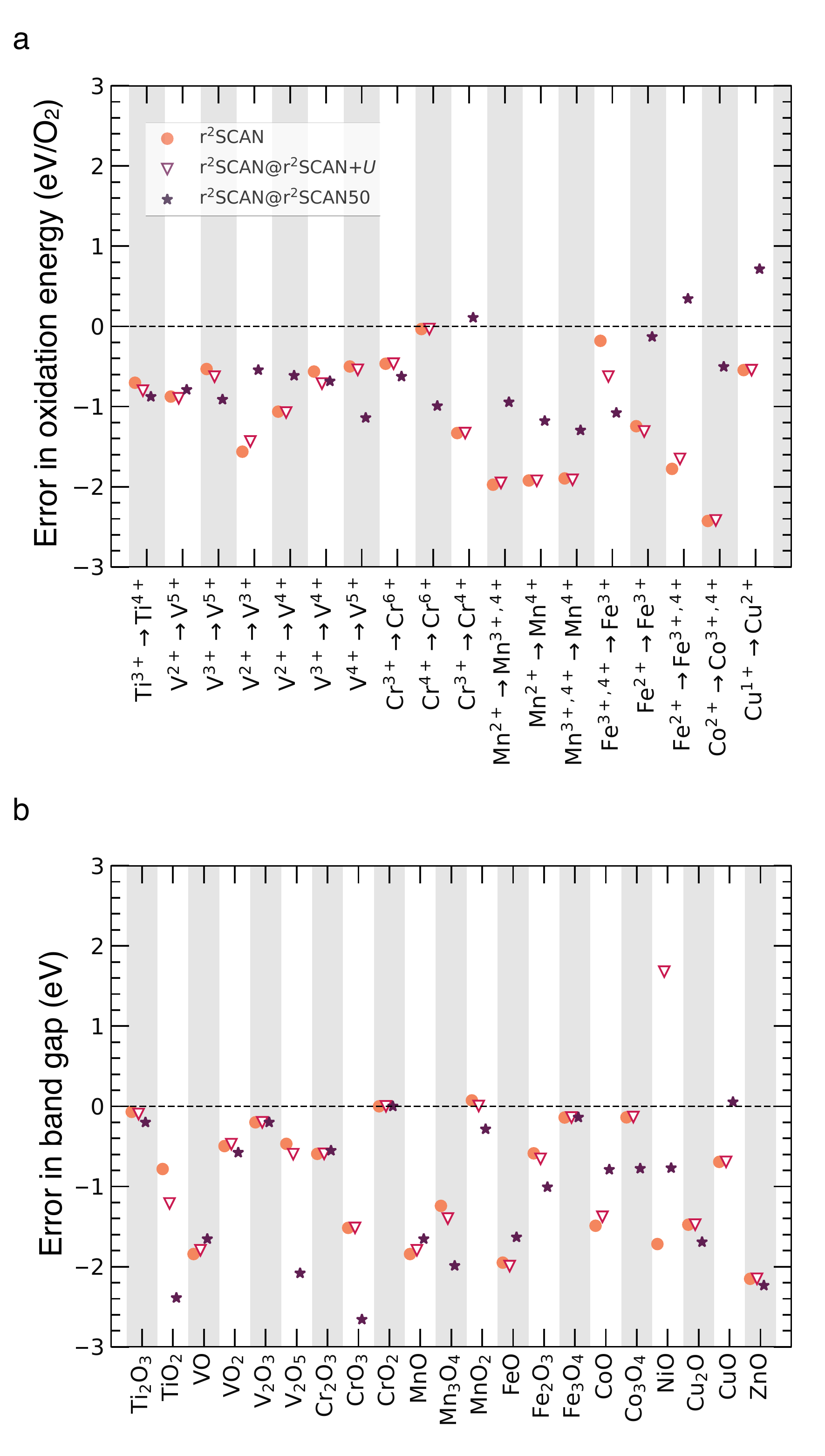}
\caption{Error in oxidation energy and band gap of M$_\mathrm{i}$O$_\mathrm{j}$s with r$^{2}$SCAN, r$^{2}$SCAN@r$^{2}$SCANX+\emph{U}, and r$^{2}$SCAN@r$^{2}$SCAN50.}\label{fig:s5}
\end{figure*}

\newpage
\begin{figure*}[ht]
\centering
 \includegraphics[width=0.7\textwidth]{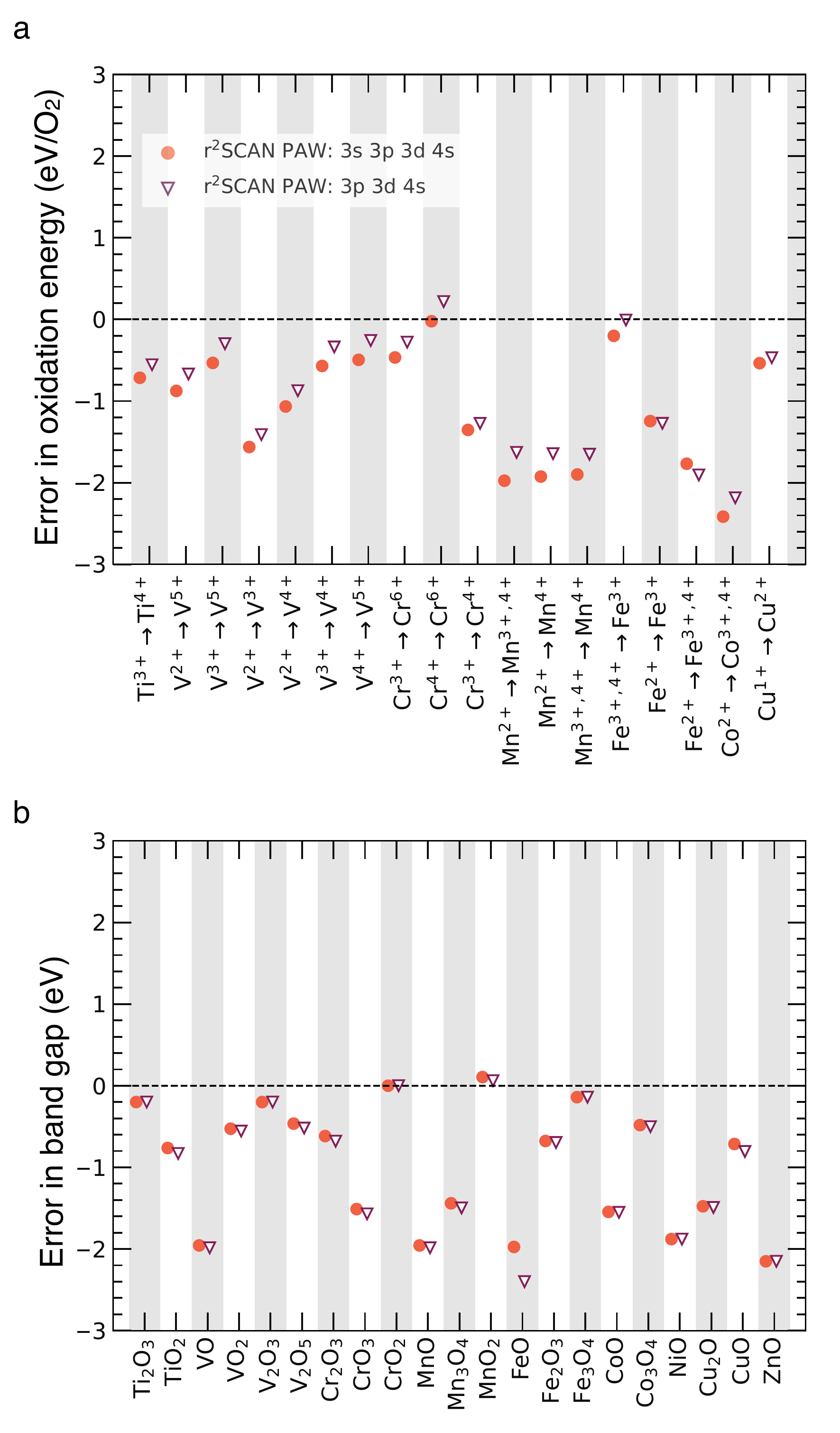}
\caption{Error in oxidation energy and band gap of M$_\mathrm{i}$O$_\mathrm{j}$ with r$^{2}$SCAN using different PAW potentials, 3\emph{s} 3\emph{p} 3\emph{d} 4\emph{s} and 3\emph{p} 3\emph{d} 4\emph{s}.}\label{fig:s6}
\end{figure*}

\clearpage
\newpage
\begin{table*}[!ht]
\caption{$U$ values are applied to the d-orbitals of transition metal atoms in the r$^2$SCAN+\emph{U} calculations performed in this study. These values are taken from Ref.~\cite{swathilakshmi_performance_2023}
Except for Ce, for which the $U$ values, that were fitted to SCAN+\emph{U}, are taken from Ref.~\cite{sai_gautam_evaluating_2018}}
\label{tab:s5}

\centering
\begin{minipage}{0.6\textwidth} 
\centering
\begin{ruledtabular}
\begin{tabular}{lc}

 {\bf Transiton metal atom} & {\bf $\bm{U}$ value (eV)}  \\ 

 Ti & 2.3 \\
\hline 
 V & 1.0 \\
\hline
 Mn & 1.8 \\
\hline 
 Fe & 3.1 \\ 
\hline 
 Co & 1.8 \\
 \hline 
 Ni & 2.1 \\
 \hline 
  Ce & 2.0 \\

\end{tabular}
\end{ruledtabular}
\end{minipage}
\end{table*}

\newpage
\begin{table*}[!ht]
\tiny
\caption{Experimental {formation} enthalpies (${\Delta \mathrm{H}_f}$ in eV/f.u.) of M$_\mathrm{i}$O$_\mathrm{j}$s.}
\label{tab:s1}
\begin{ruledtabular}
\begin{tabular}{lcc}
{\bf System}  &  {\bf Expt.\ $\pmb{\Delta \mathrm{H}_f}$ }  & {\bf Ref.}  \\ 
\hline

\textbf{TiO\textsubscript{2}} (\(P4_{2}/mnm\))\cite{sugiyama_crystal_1991} & --9.7409 & \cite{thomas_c_allison_nist-janaf_2013}\\

\hline 

\textbf{Ti\textsubscript{2}O\textsubscript{3}} (\(R\overline3c\))\cite{abrahams_magnetic_1963}& --12.6318& \cite{thomas_c_allison_nist-janaf_2013}\\

\hline 

\textbf{VO} (\(Fm\overline{3}m\))\cite{kumarakrishnan_cation_1985} & --4.4751 & \cite{Kubaschewski1979} \\

\hline 

\textbf{V\textsubscript{2}O\textsubscript{3}} (\(I2/a\))\cite{dernier_crystal_1970} & --12.58& \cite{Kubaschewski1979}\\

\hline 

\textbf{VO\textsubscript{2}} (\(P2_1/c\))\cite{Rogers_1993} & --7.3935& \cite{Kubaschewski1979}\\

\hline 

\textbf{V\textsubscript{2}O\textsubscript{5}} (\(Pmmn\))\cite{enjalbert_refinement_1986}& --16.0705 & \cite{Kubaschewski1979}\\
\hline 

\textbf{Cr\textsubscript{2}O\textsubscript{3}} (\(R\overline{3}c\))\cite{hill_crystallographic_2010}&  -11.7081& \cite{Kubaschewski1979} \\
\hline 

\textbf{CrO\textsubscript{3}} (\(C2cm\))\cite{stephens_crystal_1970}& --6.0058& \cite{Kubaschewski1979}\\
\hline 

\textbf{CrO\textsubscript{2}} (\(P4_2/mnm\))\cite{porta_chromium_1972}& --6.0362 & \cite{Kubaschewski1979}\\
\hline 

\textbf{MnO} (\(Fm\overline{3}m\))\cite{sasaki_x-ray_1979}& --3.9925 & \cite{Kubaschewski1979} \\

\hline 

\textbf{MnO\textsubscript{2}} (\(P4_2/mnm\))\cite{curetti_high-temperature_2021}& --5.3896& \cite{Kubaschewski1979}\\
\hline 
\textbf{Mn\textsubscript{3}O\textsubscript{4}} (\(I4_{1}/amd\))\cite{jarosch_crystal_1987}& --14.3706 &  \cite{Kubaschewski1979}\\

\hline 

\textbf{FeO} (\(Fm\overline{3}m\))\cite{yamamoto_modulated_1982}& --2.7406& \cite{Kubaschewski1979} \\
\hline 

\textbf{Fe\textsubscript{2}O\textsubscript{3}} (\(R\overline{3}c\))\cite{gokhan_unlu_structure_2019}& --8.5122 & \cite{Kubaschewski1979} \\

\hline 
\textbf{Fe\textsubscript{3}O\textsubscript{4}} (\(Fd\overline{3}m\))\cite{wright_charge_2002} & --11.5737 &  \cite{Kubaschewski1979}\\

\hline 

\textbf{CoO} (\(Fm\overline{3}m\))\cite{sasaki_x-ray_1979}& --2.4672& \cite{thomas_c_allison_nist-janaf_2013} \\

\hline 

\textbf{Co\textsubscript{3}O\textsubscript{4}} (\(Fd\overline{3}m\))\cite{picard_croissance_1980}& --9.3523& \cite{thomas_c_allison_nist-janaf_2013} \\

\hline 

\textbf{CuO} (\(C2/c\))\cite{soejima_structural_2000}& --1.5941 & \cite{thomas_c_allison_nist-janaf_2013} \\

\hline 

\textbf{Cu\textsubscript{2}O} (\(Pn\overline{3}m\))\cite{restori_charge_1986}& --1.7510 &  \cite{thomas_c_allison_nist-janaf_2013} \\

\hline 

\textbf{CeO\textsubscript{2}} (\(Fm\overline{3}m\))\cite{CeO2_struc_WOLCYRZ1992409}& --11.2963 &  \cite{Kubaschewski1979}\\

\hline 

\textbf{Ce\textsubscript{2}O\textsubscript{3}} (\(P\overline{3}m1\))\cite{Ce2O3_struc_BARNIGHAUSEN1985385}& --18.8807 &  \cite{Kubaschewski1979}\\

\end{tabular}
\end{ruledtabular}
\end{table*}

\clearpage
\begin{turnpage}
\begin{table*}[!ht]
\tiny
\caption{On-site magnetic moments (in $\mu_\mathrm{B}$) of M$_\mathrm{i}$O$_\mathrm{j}$s calculated with LSDA (PW92), PBE, r$^2$SCAN, and r$^2$SCANX hybrid functionals and compared to experimentally reported values.
}\label{tab:s2}
\begin{ruledtabular}
\begin{tabular}{llccccccccccc}
{\bf System}  &  {\bf Expt.} & 	{\bf PW92} & {\bf PBE} & {\bf LAK} & {\bf r$^2$SCAN} & {\bf r$^2$SCAN+\emph{U}} & {\bf r$^2$SCAN10} & {\bf r$^2$SCAN25} & {\bf r$^2$SCAN50} & {\bf r$^2$SCAN75} & {\bf r$^2$SCAN100}  & {\bf HF} \\ 
\hline

\textbf{TiO\textsubscript{2}} (\(P4_{2}/mnm\))\cite{sugiyama_crystal_1991} &  0.00	& 0.00 &	0.00 &	0.00 &	0.00 &	0.00 &	0.00 &	0.00 &	0.00 &	0.00	& 0.00	 & 0.00\\

\hline 

\textbf{Ti\textsubscript{2}O\textsubscript{3}} (\(R\overline3c\))\cite{abrahams_magnetic_1963} & 0.03 \cite{moon_absence_1969} & 
0.00 &	0.00 &	0.00 &	0.00 & 	0.00 & 	0.00 &	0.00 &	0.00 &	0.00 &	0.00 &0.00
\\

\hline 

\textbf{VO} (\(Fm\overline{3}m\))\cite{kumarakrishnan_cation_1985} & N/A & 	2.02&	2.22	&2.47	&2.45&	2.55&	2.52&	2.59&	2.66&	2.69	&2.72	&2.76 \\

\hline 

\textbf{V\textsubscript{2}O\textsubscript{3}} (\(I2/a\))\cite{dernier_crystal_1970} & 1.2-2.37\cite{moon_antiferromagnetism_1970,shin_observation_1992} & 
1.25 &	1.47&	1.74	&1.70&	1.80	&1.78	&1.83&	1.86	&1.87	&1.87&	1.90 \\
\hline 

\textbf{VO\textsubscript{2}} (\(P2_1/c\))\cite{Rogers_1993}& 0.00\cite{goodenough_VO2} & 0.46 & 0.77 & 0.95 & 0.93 & 0.98 & 0.97 & 0.99 & 1.00 & 1.00 & 1.00 & 1.02 \\
\hline 

\textbf{V\textsubscript{2}O\textsubscript{5}} (\(Pmmn\))\cite{enjalbert_refinement_1986}& 0.00 &	0.00&	0.00	&0.00	&0.00&	0.00	&0.00&	0.00	&0.00&	0.00&	0.00	&0.00\\ 

\hline

\textbf{Cr\textsubscript{2}O\textsubscript{3}} (\(R\overline{3}c\))\cite{hill_crystallographic_2010}&  2.76\cite{corliss_magnetic_1965} &
2.37	&2.45&	2.61 &	2.58	 & – &	2.63 &	2.67& 	2.71	& 2.73	& 2.74 & 2.77
 \\
\hline 

\textbf{CrO\textsubscript{3}} (\(C2cm\))\cite{stephens_crystal_1970}& 0.00 &	0.00 &	0.00 & 	0.00 &	0.00 &	N/A	&0.00	 & 0.00 &	0.00 &	0.00 &	0.00 &	0.00\\
\hline 

\textbf{CrO\textsubscript{2}} (\(P4_2/mnm\))\cite{porta_chromium_1972}& 2.00\cite{coey_half-metallic_2002} & 
1.89 & 1.94 &	2.10	& 2.06 &	N/A & 2.13 &	2.21 &	1.99 &	1.95 &	1.94 &	1.98
 \\
\hline 

\textbf{MnO} (\(Fm\overline{3}m\))\cite{sasaki_x-ray_1979}& 4.58\cite{cheetham_magnetic_1983} & 
4.08 &	4.15 &	4.35 &	4.3 &	4.42 &	4.36 &	4.42 &	4.48 &	4.52 &	4.54 & 	4.58 
 \\

\hline 

\textbf{MnO\textsubscript{2}} (\(P4_2/mnm\))\cite{curetti_high-temperature_2021}&2.35\cite{regulski_incommensurate_2003} & 
2.27	& 2.42	& 2.67	& 2.62	& 2.77	& 2.71	& 2.82	& 2.93	& 2.99	& 3.00 &	3.06 
 \\
\hline 
\multirow{2}{*}{\textbf{Mn\textsubscript{3}O\textsubscript{4}}  (\(I4_{1}/amd\))\cite{jarosch_crystal_1987}} &
   4.34,                                                          &  3.86; & 3.98; &	4.24;	& 4.19; & 4.35; &4.28; & 4.36;& 4.43;& 4.48;& 4.51;	& 4.54; \\
  &  3.25--3.64\cite{jensen_magnetic_1974} & 3.29 & 3.35 &  3.55  & 3.51 & 3.65 & 3.57  & 3.63 & 3.69 & 3.70 & 3.71 & 3.75 \\
\hline 

\textbf{FeO} (\(Fm\overline{3}m\))\cite{yamamoto_modulated_1982}& 3.32--4.2\cite{roth_magnetic_1958,battle_magnetic_1979} & 
3.23	& 3.30 & 	3.46&	3.42& 	3.54& 	3.47&	3.53&	3.59&	3.63&	3.65& 	3.67    \\
\hline 

\textbf{Fe\textsubscript{2}O\textsubscript{3}} (\(R\overline{3}c\))\cite{gokhan_unlu_structure_2019}& 4.9\cite{shull_neutron_1951} & 
3.24	& 3.45 & 3.95 & 3.86	& 4.12 &4.01 &4.16 & 4.31 &4.41 &4.47& 4.51 
 \\

\hline 
\multirow{2}{*}{\textbf{Fe\textsubscript{3}O\textsubscript{4}} (\(Fd\overline{3}m\))\cite{wright_charge_2002}} & \multirow{2}{*}{4.44,4.1\cite{wright_charge_2002}} &
3.19;  & 3.36;  & 3.83; 	& 3.73;	& 4.02; 	& 3.89; &	4.04;  & 	4.22;  & 	4.35; 	 & 4.43;	&4.46; 
 \\
& & 3.33; 3.33& 3.46; 3.46 & 3.76; 3.76 &  3.69;3.69 & 3.58; 4.12 &   3.74; 3.77 & 3.62; 4.13& 3.58; 4.32 & 3.62; 4.42 & 3.64; 4.49 & 3.66;4.52 \\
\hline 

\textbf{CoO}(\(Fm\overline{3}m\))\cite{sasaki_x-ray_1979}& 3.35--3.8\cite{roth_magnetic_1958,khan_magnetic_1970} & 2.31 & 2.39 & 2.57 & 2.54 & 2.62 & 2.61 & 2.67 & 2.73 & 2.76 & 2.78 & 2.79 \\
\hline 

\textbf{Co\textsubscript{3}O\textsubscript{4}} (\(Fd\overline{3}m\))\cite{picard_croissance_1980}& 3.02\cite{roth_magnetic_1964} & 2.07 & 2.21 & 2.51 & 2.45 & 2.57 & 2.55 & 2.62 & 2.69 & 2.73 & 2.75 & 2.76 \\
\hline 

\textbf{NiO} (\(Fm\overline{3}m\))\cite{Deng2024} & 1.64/1.90\cite{cheetham_magnetic_1983,anisimov_band_1991} & 1.19 & 1.38 & 1.59 & 1.63 & 1.69 & 1.67 & 1.75 & 1.81 & 1.85 & 1.87 & 1.89  \\
\hline 

\textbf{CuO} (\(C2/c\))\cite{soejima_structural_2000}& 0.68\cite{yang_neutron_1988} & 0.00 & 0.33 & 0.61 & 0.56 & N/A & 0.65 & 0.74 & 0.82 & 0.87 & 0.90 & 0.91 \\
\hline 

\textbf{Cu\textsubscript{2}O} (\(Pn\overline{3}m\))\cite{restori_charge_1986} & 0.00 &	0.00 &	0.00 & 	0.00 &	0.00 &	N/A	&0.00	 & 0.00 &	0.00 &	0.00 &	0.00 &	0.00\\
\hline 

\textbf{ZnO} (\(P6_{3}mc\))\cite{Abrahams1969} & 0.00 &	0.00 &	0.00 & 	0.00 &	0.00 &	N/A	&0.00	 & 0.00 &	0.00 &	0.00 &	0.00 &	0.00\\
\hline 

\textbf{CeO\textsubscript{2}} ( \(Fm\overline{3}m\))\cite{CeO2_struc_WOLCYRZ1992409} & 0.00 &	0.00 &	0.00 & 	0.00 &	0.00 &	0.00	&0.00	 & 0.00 &	0.00 &	0.00 &	0.00 &	0.00\\
\hline 

\textbf{Ce\textsubscript{2}O\textsubscript{3}} (\(P\overline{3}m\))\cite{Ce2O3_struc_BARNIGHAUSEN1985385} & 1.08\cite{Ce2O3_mag_PINTO198281} & 0.62 & 0.79 & 0.98 & 0.94 & 0.97 & 0.96 & 0.97 & 0.97 & 0.98 & 0.98 & 0.98 \\

\end{tabular}
\end{ruledtabular}
\end{table*}
\end{turnpage}

\clearpage
\begin{turnpage}
\begin{table*}[!ht]
\tiny
\caption{ Band gaps (in eV) of M$_\mathrm{i}$O$_\mathrm{j}$s calculated with LSDA (PW92), PBE, r$^2$SCAN, and r$^2$SCANY@r$^2$SCANX and compared to experimentally (Exp.) reported values. Band gaps in {\bf Table~1} are computed on a \emph{k}-grid with a uniform density of 48 \emph{k}-points per \AA$^{-1}$. However, band gaps shown in this table and in {\bf Supplementary Table~\ref{tab:s4}}, except for r$^2$SCAN+\emph{U}, are calculated on a sparser \emph{k}-grid with a uniform density of 700 \emph{k}-points per reciprocal atom.
}\label{tab:s3}
\begin{ruledtabular}
\begin{tabular}{llccccccccc}
{\bf System}  &  {\bf Exp.} & 	{\bf PW92} & {\bf PBE} & {\bf r$^2$SCAN} & {\bf r$^2$SCAN+\emph{U}} & {\bf r$^2$SCAN10} & {\bf r$^2$SCAN10@r$^2$SCAN} & {\bf r$^2$SCAN@r$^2$SCAN50} & {\bf r$^2$SCAN10@r$^2$SCAN50} & {\bf r$^2$SCAN@HF}   \\ 
\hline

\textbf{TiO\textsubscript{2}} (\(P4_{2}/mnm\))\cite{sugiyama_crystal_1991} &  3.0\cite{serpone_is_2006} & 1.76 & 1.85 & 2.22 & 2.51 & 3.07 & 3.41 & 0.61 & 1.85 & 0.00 \\
\hline 

\textbf{Ti\textsubscript{2}O\textsubscript{3}} (\(R\overline3c\))\cite{abrahams_magnetic_1963}&  0.2\cite{uchida_charge_2008} & 0.13 & 0.13 & 0.13 & 0.59 & 0.20 & 0.15 & 0.00 & 0.33 & 0.00 \\
\hline 

\textbf{VO} (\(Fm\overline{3}m\))\cite{kumarakrishnan_cation_1985} & N/A & 0.07 & 0.57 & 1.67 & 2.35 & 3.10 & 2.99 & 1.98 & 3.37 & 2.41 \\
\hline 

\textbf{V\textsubscript{2}O\textsubscript{3}} (\(I2/a\))\cite{dernier_crystal_1970} & 0.2\cite{shin_observation_1992}  & 0.00 & 0.00 & 0.00 & 0.68 & 1.01 & 0.00 & 0.00 & 1.19 & 0.36 \\
\hline 

\textbf{VO\textsubscript{2}} (\(P2_1/c\))\cite{Rogers_1993}& 0.70\cite{shin_vacuum-ultraviolet_1990} & 0.00 & 0.04 & 0.20 & 0.69 & 1.49 & 1.36 & 0.12 & 0.85 & 0.00 \\
\hline 

\textbf{V\textsubscript{2}O\textsubscript{5}} (\(Pmmn\))\cite{enjalbert_refinement_1986}& 2.5\cite{kumar_structural_2008} & 1.60 & 1.70 & 2.03 & 2.14 & 2.90 & 3.22 & 0.42 & 1.67 & 0.00 \\
\hline 

\textbf{Cr\textsubscript{2}O\textsubscript{3}} (\(R\overline{3}c\))\cite{hill_crystallographic_2010}&  3.2\cite{abdullah_structural_2014}  & 1.20 & 1.64 & 2.61 & N/A & 4.00 & 3.93 & 2.65 & 4.03 & 2.42 \\
\hline 

\textbf{CrO\textsubscript{3}} (\(C2cm\))\cite{stephens_crystal_1970}& 3.8\cite{misho_preparation_1989}  & 1.97 & 2.05 & 2.28 & N/A & 3.12 & 3.31 & 1.14 & 2.26 & 0.00 \\
\hline 

\textbf{CrO\textsubscript{2}} (\(P4_2/mnm\))\cite{porta_chromium_1972}&  0.00\cite{coey_half-metallic_2002}  & 0.00 & 0.00 & 0.00 & N/A & 0.00 & 0.00 & 0.00 & 0.00 & 0.00 \\\hline 

\textbf{MnO} (\(Fm\overline{3}m\))\cite{sasaki_x-ray_1979}& 3.6--3.8\cite{messick_direct_1972} & 0.89 & 1.00 & 1.86 & 2.13 & 2.90 & 2.87 & 2.05 & 3.03 & 2.21 \\
\hline 

\textbf{MnO\textsubscript{2}} (\(P4_2/mnm\))\cite{curetti_high-temperature_2021}& 0.27--0.3\cite{druilhe_electron_1967,islam_studies_2005}  & 0.01 & 0.00 & 0.36 & 0.74 & 1.48 & 1.68 & 0.00 & 0.33 & 0.00 \\
\hline 

\textbf{Mn\textsubscript{3}O\textsubscript{4}}  (\(I4_{1}/amd\))\cite{jarosch_crystal_1987} & 2.3--2.5\cite{xu_chemical_2006} & 0.15 & 0.45 & 1.16 & 1.39 & 2.21 & 2.38 & 0.41 & 1.70 & 0.00 \\
\hline 

\textbf{FeO} (\(Fm\overline{3}m\))\cite{yamamoto_modulated_1982}& 2.20\cite{bowen_electrical_1975} & 0.02 & 0.00 & 0.45 & 1.58 & 1.90 & 1.72 & 0.77 & 2.04 & 0.99 \\
\hline 

\textbf{Fe\textsubscript{2}O\textsubscript{3}} (\(R\overline{3}c\))\cite{gokhan_unlu_structure_2019}&  2.20\cite{droubay_structure_2007} & 0.41 & 0.64 & 1.61 & 2.71 & 2.96 & 2.95 & 1.19 & 2.64 & 1.23 \\
\hline 

\textbf{Fe\textsubscript{3}O\textsubscript{4}} (\(Fd\overline{3}m\))\cite{wright_charge_2002} & 0.14\cite{park_charge-gap_1998} & 0.03 & 0.00 & 0.00 & 0.23 & 0.00 & 0.00 & 0.00 & 0.00 & 0.00 \\
\hline 

\textbf{CoO} (\(Fm\overline{3}m\))\cite{sasaki_x-ray_1979}&  2.40\cite{wei_insulating_1994,van_elp_electronic_1991} & 0.01 & 0.01 & 0.91 & 2.13 & 2.67 & 2.57 & 1.61 & 3.26 & 2.37 \\
\hline 

\textbf{Co\textsubscript{3}O\textsubscript{4}} (\(Fd\overline{3}m\))\cite{picard_croissance_1980}& 1.60\cite{van_elp_electronic_1991} & 0.38 & 0.68 & 1.46 & 1.94 & 2.88 & 2.86 & 0.82 & 2.37 & 0.29 \\
\hline 

\textbf{NiO} (\(Fm\overline{3}m\))\cite{Deng2024} & 4.3\cite{sawatzky_magnitude_1984} & 0.58 & 1.11 & 2.58 & 3.52 & 4.29 & 4.18 & 3.53 & 4.62 & 3.63 \\
\hline 

\textbf{CuO} (\(C2/c\))\cite{soejima_structural_2000}& 1.40\cite{ghijsen_electronic_1988} & 0.00 & 0.00 & 0.71 & N/A & 2.16 & 1.80 & 1.45 & 2.91 & 1.86 \\
\hline 

\textbf{Cu\textsubscript{2}O} (\(Pn\overline{3}m\))\cite{restori_charge_1986}& 2.17--2.4\cite{ghijsen_electronic_1988,zhang_symmetry-breaking_2020} & 0.58 & 0.58 & 0.81 & N/A & 1.68 & 1.67 & 0.59 & 1.58 & 0.03 \\
\hline 

\textbf{ZnO} (\(P6_{3}mc\))\cite{Abrahams1969}& 3.4\cite{Reynolds1999} & 0.77 & 0.82 & 1.25 & N/A & 2.16 & 2.22 & 1.17 & 2.13 & 1.14  \\
\hline 

\textbf{CeO\textsubscript{2}} ( \(Fm\overline{3}m\))\cite{CeO2_struc_WOLCYRZ1992409} & 6\cite{CeO2_bg_PhysRevLett.53.202} & 1.95 & 2.02 & 2.28 & 2.35 & 3.18 & 3.84 & 0.00 & 0.89 & 0.00 \\
\hline 

\textbf{Ce\textsubscript{2}O\textsubscript{3}} (\(P\overline{3}m\))\cite{Ce2O3_struc_BARNIGHAUSEN1985385} & 2.3\cite{Ce2O3_bg_PROKOFIEV199641} & 0.00 & 0.00 & 0.54 & 1.90 & 2.00 & 1.97 & 1.39 & 2.56 & 2.26 \\

\end{tabular}
\end{ruledtabular}
\end{table*}
\end{turnpage}

\newpage
\begin{turnpage}
\begin{table*}[!ht]
\tiny
\caption{Band gaps (in eV) of M$_\mathrm{i}$O$_\mathrm{j}$s calculated, with LAK and r$^2$SCANX@LAK and compared with experimentally (Exp.) reported data.
}\label{tab:s4}
\begin{ruledtabular}
\begin{tabular}{llccc}
{\bf System}  &  {\bf Exp.} & 	{\bf LAK}	& {\bf r$^2$SCAN@LAK}  & 	{\bf r$^2$SCAN10@LAK}   \\ 
\hline

\textbf{TiO\textsubscript{2}} (\(P4_{2}/mnm\))\cite{sugiyama_crystal_1991} &   3.0\cite{serpone_is_2006} & 2.38 & 1.91 & 3.12 \\
\hline 

\textbf{Ti\textsubscript{2}O\textsubscript{3}} (\(R\overline3c\))\cite{abrahams_magnetic_1963}& 0.2\cite{uchida_charge_2008} & 0.13 & 0.13 & 0.15 \\
\hline 

\textbf{VO} (\(Fm\overline{3}m\))\cite{kumarakrishnan_cation_1985} & N/A & 2.06 & 1.72 & 3.06 \\
\hline 

\textbf{V\textsubscript{2}O\textsubscript{3}} (\(I2/a\))\cite{dernier_crystal_1970} & 0.2\cite{shin_observation_1992} & 0.00 & 0.00 & 0.04 \\
\hline 

\textbf{VO\textsubscript{2}} (\(P2_1/c\))\cite{Rogers_1993}& 0.70\cite{shin_vacuum-ultraviolet_1990} & 0.42 & 0.20 & 1.36 \\
\hline 

\textbf{V\textsubscript{2}O\textsubscript{5}} (\(Pmmn\))\cite{enjalbert_refinement_1986}& 2.5\cite{kumar_structural_2008}  & 2.19 & 1.69 & 2.90 \\
\hline 

\textbf{Cr\textsubscript{2}O\textsubscript{3}} (\(R\overline{3}c\))\cite{hill_crystallographic_2010}&  3.2\cite{abdullah_structural_2014} & 2.96 & 2.60 & 3.97 \\
\hline 

\textbf{CrO\textsubscript{3}} (\(C2cm\))\cite{stephens_crystal_1970}& 3.8\cite{misho_preparation_1989} & 2.35 & 2.01 & 3.05 \\
\hline 

\textbf{CrO\textsubscript{2}} (\(P4_2/mnm\))\cite{porta_chromium_1972}& 0.00\cite{coey_half-metallic_2002}  & 0.00 & 0.00 & 0.00 \\
\hline 

\textbf{MnO} (\(Fm\overline{3}m\))\cite{sasaki_x-ray_1979}& 3.6--3.8\cite{messick_direct_1972} & 2.20 & 1.91 & 2.94 \\
\hline 

\textbf{MnO\textsubscript{2}} (\(P4_2/mnm\))\cite{curetti_high-temperature_2021}& 0.27-0.3\cite{druilhe_electron_1967,islam_studies_2005} & 0.61 & 0.14 & 1.45 \\
\hline 

\textbf{Mn\textsubscript{3}O\textsubscript{4}} (\(I4_{1}/amd\))\cite{jarosch_crystal_1987}& 2.3--2.5\cite{xu_chemical_2006} & 1.26 & 1.00 & 2.25 \\
\hline 

\textbf{FeO} (\(Fm\overline{3}m\))\cite{yamamoto_modulated_1982}& 2.20\cite{bowen_electrical_1975} & 0.50 & 0.44 & 1.71 \\
\hline 

\textbf{Fe\textsubscript{2}O\textsubscript{3}} (\(R\overline{3}c\))\cite{gokhan_unlu_structure_2019}& 2.20\cite{droubay_structure_2007} & 1.99 & 1.51 & 2.90 \\
\hline 

\textbf{Fe\textsubscript{3}O\textsubscript{4}}  (\(Fd\overline{3}m\))\cite{wright_charge_2002} & 0.14\cite{park_charge-gap_1998}& 0.00 & 0.00 & 	0.00 \\
\hline 

\textbf{CoO} (\(Fm\overline{3}m\))\cite{sasaki_x-ray_1979}& 2.40\cite{wei_insulating_1994,van_elp_electronic_1991} & 1.06 & 0.91 & 2.58 \\
\hline 

\textbf{Co\textsubscript{3}O\textsubscript{4}} (\(Fd\overline{3}m\))\cite{picard_croissance_1980}& 1.60\cite{van_elp_electronic_1991} & 1.69 & 1.46 & 2.86 \\
\hline 

\textbf{NiO} (\(Fm\overline{3}m\))\cite{Deng2024} & 4.3\cite{sawatzky_magnitude_1984} & 3.00 & 2.58 & 4.18 \\
\hline 

\textbf{CuO} (\(C2/c\))\cite{soejima_structural_2000}& 1.40\cite{ghijsen_electronic_1988} & 1.04 & 0.71 & 1.93 \\
\hline 

\textbf{Cu\textsubscript{2}O} (\(Pn\overline{3}m\))\cite{restori_charge_1986}& 2.17--2.4\cite{ghijsen_electronic_1988,zhang_symmetry-breaking_2020} & 0.99 & 0.82 & 1.70 \\
\hline 

\textbf{ZnO} (\(P6_{3}mc\))\cite{Abrahams1969}& 3.4\cite{Reynolds1999} & 1.47 & 1.25 & 2.22 \\
\hline 

\textbf{CeO\textsubscript{2}} ( \(Fm\overline{3}m\))\cite{CeO2_struc_WOLCYRZ1992409} & 6\cite{CeO2_bg_PhysRevLett.53.202} & 2.37 & 2.29 & 3.84 \\
\hline 

\textbf{Ce\textsubscript{2}O\textsubscript{3}} (\(P\overline{3}m\))\cite{Ce2O3_struc_BARNIGHAUSEN1985385} & 2.3\cite{Ce2O3_bg_PROKOFIEV199641} & 0.80 & 0.54 & 1.97 \\

\end{tabular}
\end{ruledtabular}
\end{table*}
\end{turnpage}

\clearpage

\clearpage

\begin{turnpage}
\begin{table*}[!ht]
\caption{Oxidation energies in eV/O\textsubscript{2} of 2Ti\textsubscript{2}O\textsubscript{3} + O$_2$ $\rightarrow$ 4TiO$_2$ with r\textsuperscript{2}SCAN, r2SCAN+\emph{U} and proposed r\textsuperscript{2}SCANY@r\textsuperscript{2}SCANX methods. 
Chemical accuracy of ~3 kcal/mol $\approx$ 130.2 meV/O\textsubscript{2}.
}\label{tab:s_encut1000}
\centering
\begin{minipage}{0.9\textwidth} 
\centering
\begin{ruledtabular}
\begin{tabular}{lccc}
{\bf Method}  & {\bf ENCUT=700} & {\bf ENCUT=1000} & {\bf Deviation in meV/O{\textsubscript{2}}}\\ 
\hline

r\textsuperscript{2}SCAN & -8.3199 & -8.3209 & 1.0\\
r\textsuperscript{2}SCAN10 & -8.638 & -8.6362 & 1.8\\
r\textsuperscript{2}SCAN10@r\textsuperscript{2}SCAN & -8.6571 & -8.6568 & 0.3 \\
r\textsuperscript{2}SCAN@r\textsuperscript{2}SCAN50 & -8.4935 & -8.4906 & 2.9\\
r\textsuperscript{2}SCAN10@r\textsuperscript{2}SCAN50 & -8.6801 & -8.6767 & 3.4\\
r\textsuperscript{2}SCAN+\emph{U} & -8.8425 & -8.8442 & 1.7\\

\end{tabular}
\end{ruledtabular}
\end{minipage}
\end{table*}
\end{turnpage}
\clearpage

\end{document}